\documentclass[acmsmall]{acmart}

\usepackage{algorithm2e}[noline]
\usepackage{listings}
\usepackage{multirow}
\usepackage{wrapfig}
\usepackage[makeroom]{cancel}

\usepackage{xr}

\usepackage{xfp}
\newcommand\SupplementaryMaterials{%
  \xdef\presupfigures{\arabic{figure}}%
  \xdef\presupsections{\arabic{section}}%
  \xdef\presupeqs{\arabic{equation}}%
  \xdef\presuptables{\arabic{table}}%
  \renewcommand\thefigure{S\fpeval{\arabic{figure}-\presupfigures}}
  \renewcommand\thesection{S\fpeval{\arabic{section}-\presupsections}}
  \renewcommand\theequation{S\fpeval{\arabic{equation}-\presupeqs}}
  \renewcommand\thetable{S\fpeval{\arabic{table}-\presuptables}}
}

\definecolor{codegreen}{rgb}{0,0.6,0}
\definecolor{codegray}{rgb}{0.5,0.5,0.5}
\definecolor{codepurple}{rgb}{0.58,0,0.82}
\definecolor{backcolour}{rgb}{0.95,0.95,0.92}

\lstdefinestyle{mystyle}{
    backgroundcolor=\color{backcolour},   
    commentstyle=\color{codegreen},
    keywordstyle=\color{magenta},
    numberstyle=\tiny\color{codegray},
    stringstyle=\color{codepurple},
    basicstyle=\ttfamily\footnotesize,
    breakatwhitespace=false,         
    breaklines=true,                 
    captionpos=b,                    
    keepspaces=true,                 
    numbers=left,                    
    numbersep=5pt,                  
    showspaces=false,                
    showstringspaces=false,
    showtabs=false,                  
    tabsize=2
}

\AtBeginDocument{%
  }

\setcopyright{acmlicensed}
\acmJournal{PACMCGIT}
\acmYear{2025} \acmVolume{8} \acmNumber{1} \acmArticle{14} \acmMonth{5}\acmDOI{10.1145/3728292}

\begin{document}

\title{Improved Stochastic Texture Filtering Through Sample Reuse}

\author{Bartlomiej Wronski}
\orcid{0009-0005-0806-2307}
\affiliation{%
  \institution{NVIDIA}
  \city{Brooklyn}
  \country{USA}
}
\email{bwronski@nvidia.com}

\author{Matt Pharr}
\orcid{0000-0002-0566-8291}
\affiliation{%
  \institution{NVIDIA}
  \city{San Francisco}
  \country{USA}
}
\email{matt@pharr.org}

\author{Tomas Akenine-M\"oller}
 \orcid{0000-0001-6226-3170}
 \affiliation{%
 \institution{NVIDIA}
 \city{Lund}
 \country{Sweden}}

\begin{abstract}
	Stochastic texture filtering (STF) has re-emerged as a technique
	that can bring down the cost of texture filtering of advanced
	texture compression methods, e.g., neural texture compression.
	However, during texture magnification, the swapped order of filtering and shading with STF
	can result in aliasing. The inability to smoothly interpolate material properties stored in textures,
	such as surface normals, leads to potentially undesirable appearance changes.
	We present a novel method to improve the quality of stochastically-filtered magnified textures
	and reduce the image difference compared to traditional texture filtering.
	When textures are magnified, nearby pixels filter similar sets of texels
    and we introduce techniques for sharing texel values among pixels with only a small increase in cost (0.04--0.14~ms per frame).
	We propose an improvement to weighted importance sampling that guarantees
	that our method never increases error beyond single-sample stochastic texture filtering.
	Under high magnification, our method has >10 dB higher PSNR than single-sample STF.
	Our results show greatly improved image quality both with and without spatiotemporal denoising.
\end{abstract}



\begin{teaserfigure}
	\centering
	\setlength{\tabcolsep}{0.75pt}
	\newcommand{\myheight}{19.0mm}
	\renewcommand{\arraystretch}{0.5}	
	\hspace*{-1.4mm}
	{\scriptsize
		\begin{tabular}{ccccccc}
			& ground truth & STF & STF denoised & our & our denoised \\
			full image & \textbf{PSNR}($\uparrow$)/\textbf{FLIP}($\downarrow$):& 
			30.3 / 0.042 & 
			35.9 / 0.030 &
			42.6 / 0.021 &
			\textbf{48.1 / 0.017} \\
			\multirow{2}{*}[18.00mm]{
			\includegraphics[height=38.42mm]{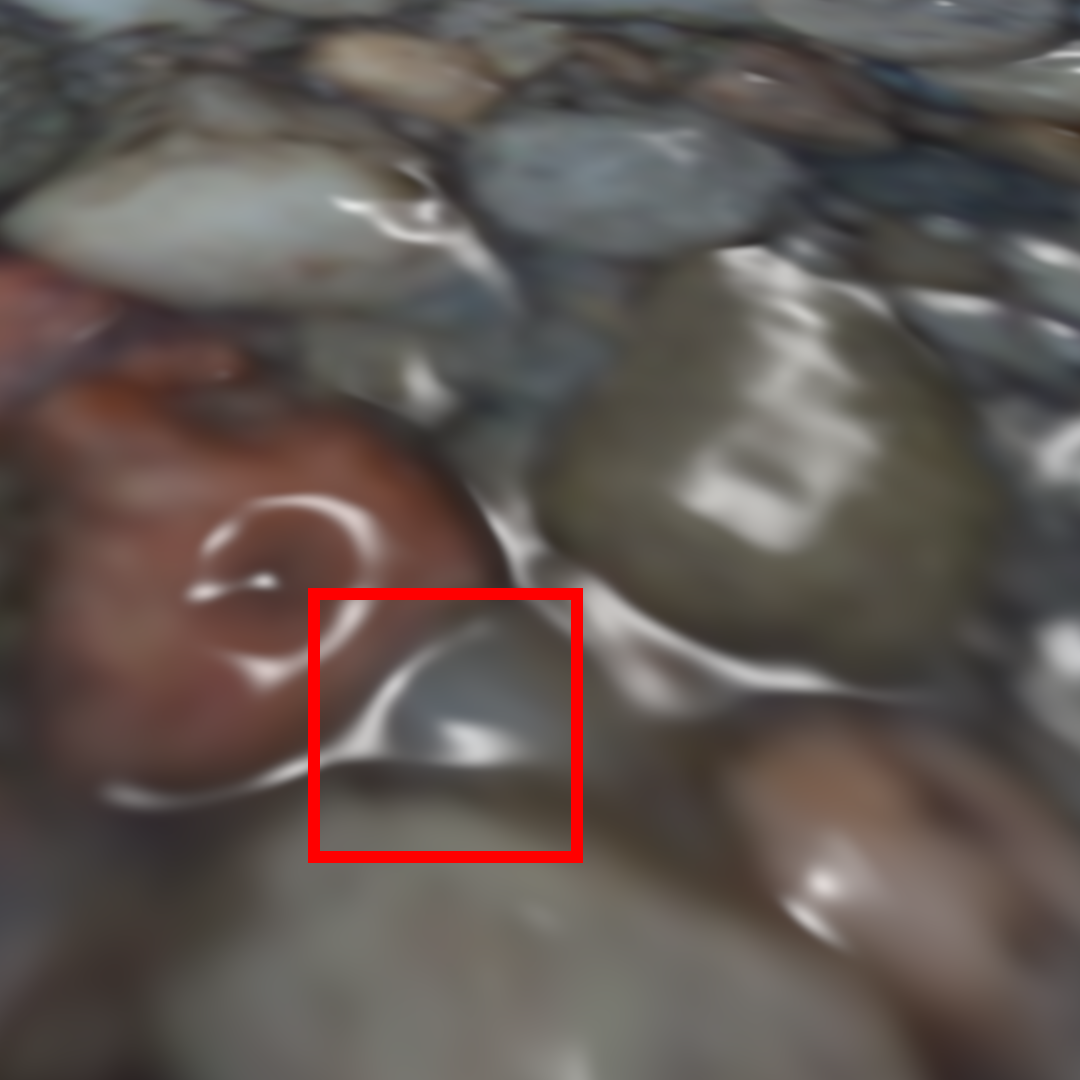}}&
			\includegraphics[height=\myheight]{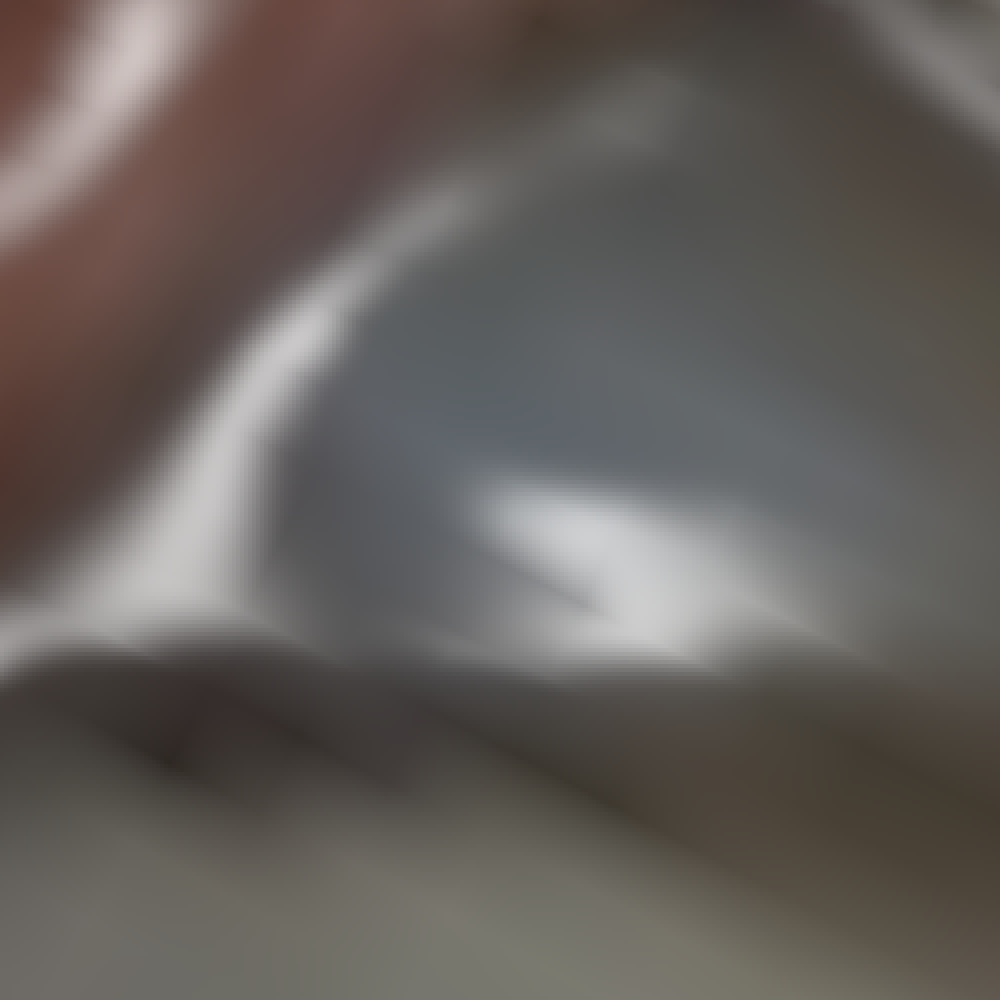}&
			\includegraphics[height=\myheight]{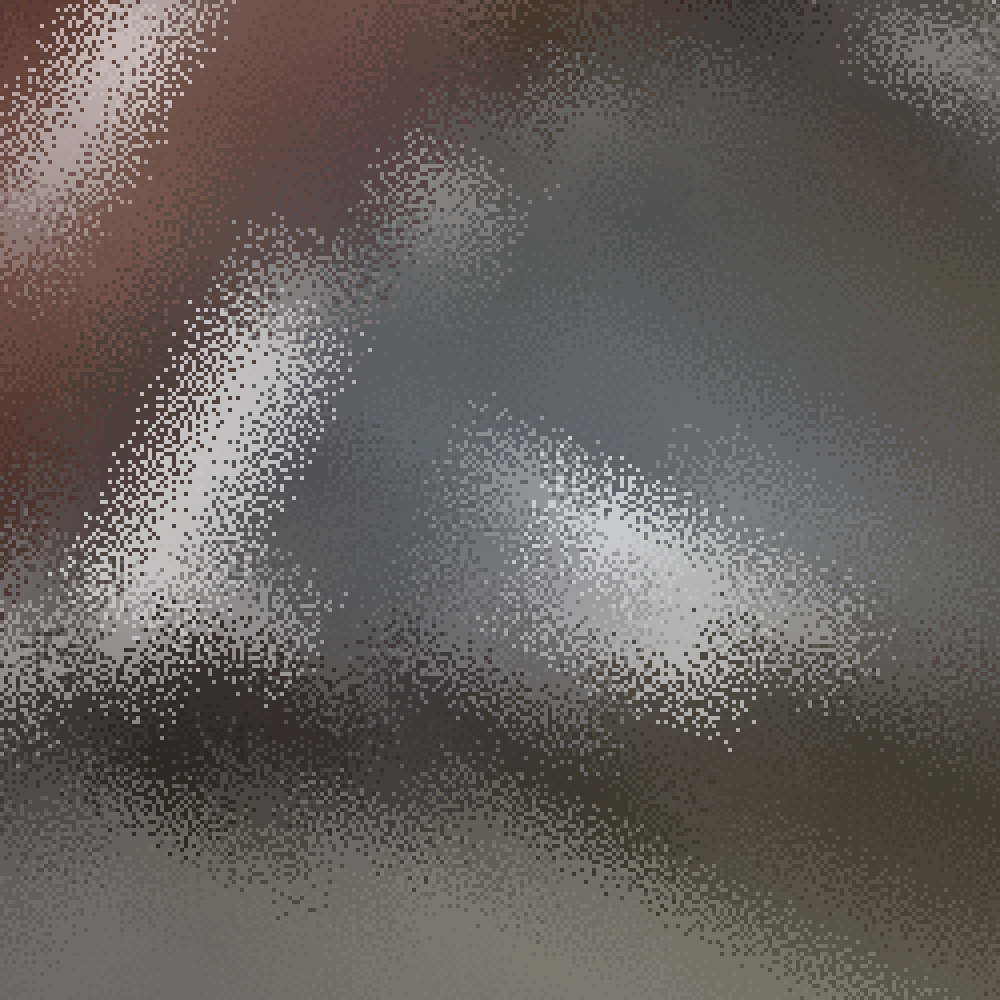}&
			\includegraphics[height=\myheight]{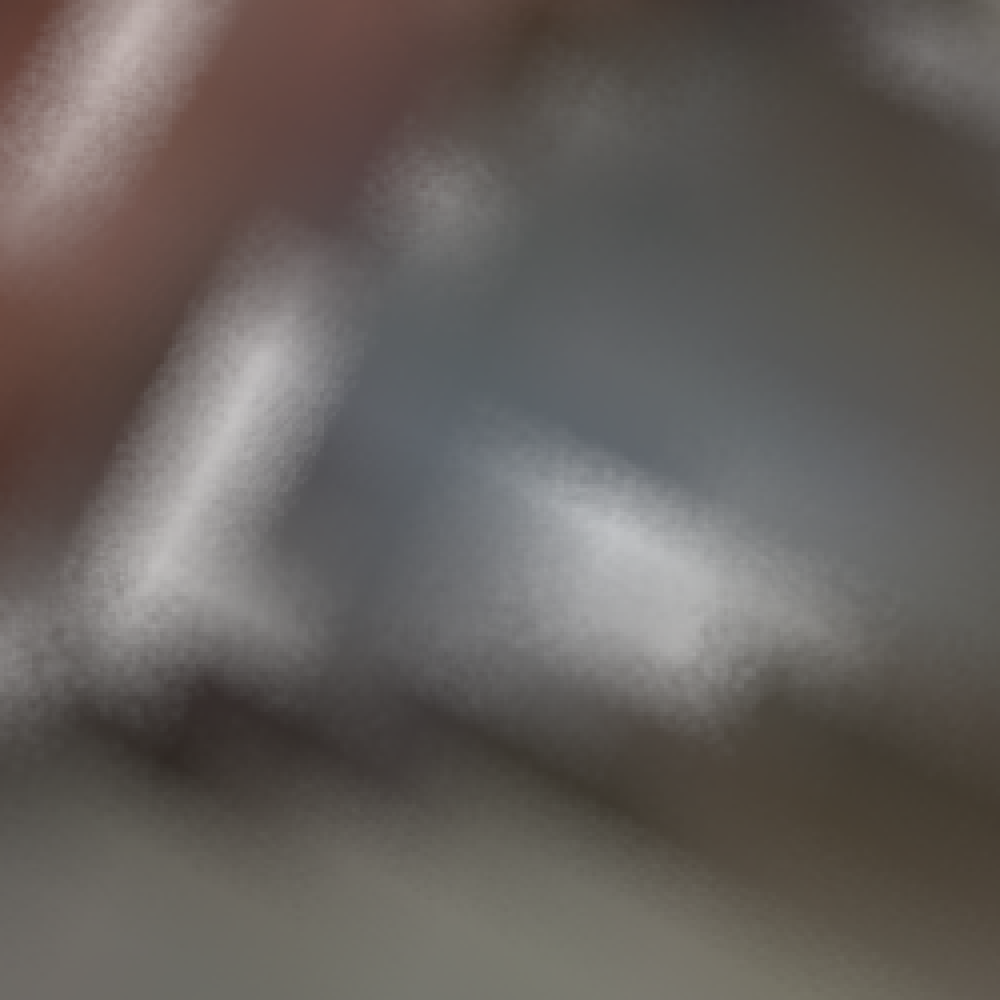}&
			\includegraphics[height=\myheight]{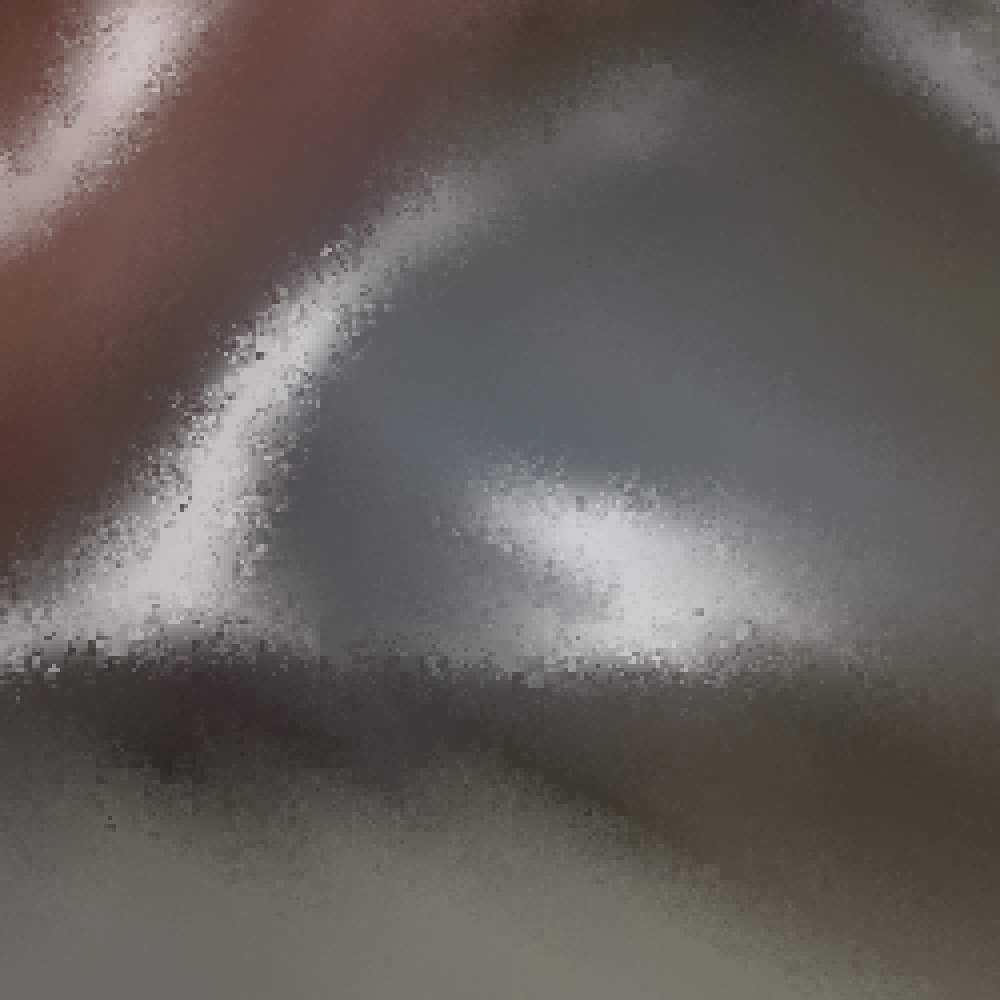}&
			\includegraphics[height=\myheight]{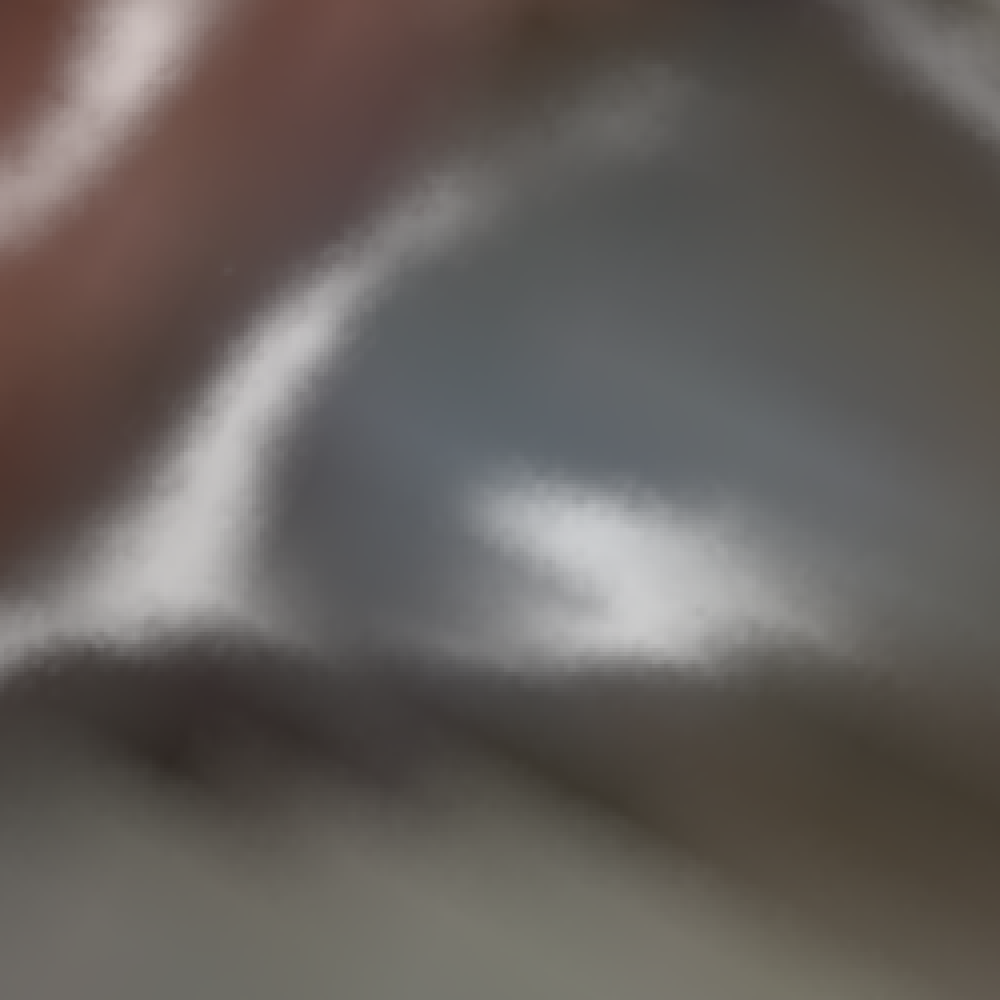} &
			\rotatebox{90}{\hspace{5mm}\textbf{bilinear}}
			\\
			&
			\includegraphics[height=\myheight]{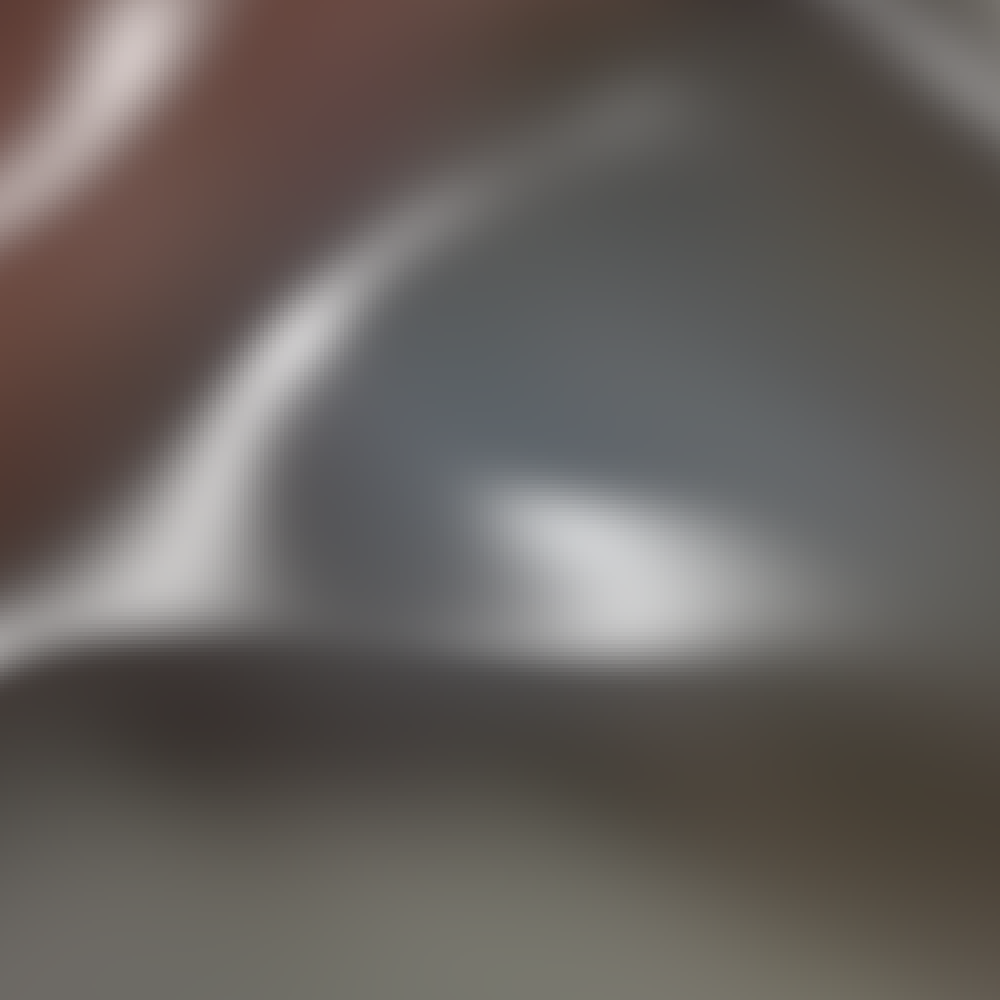}&
			\includegraphics[height=\myheight]{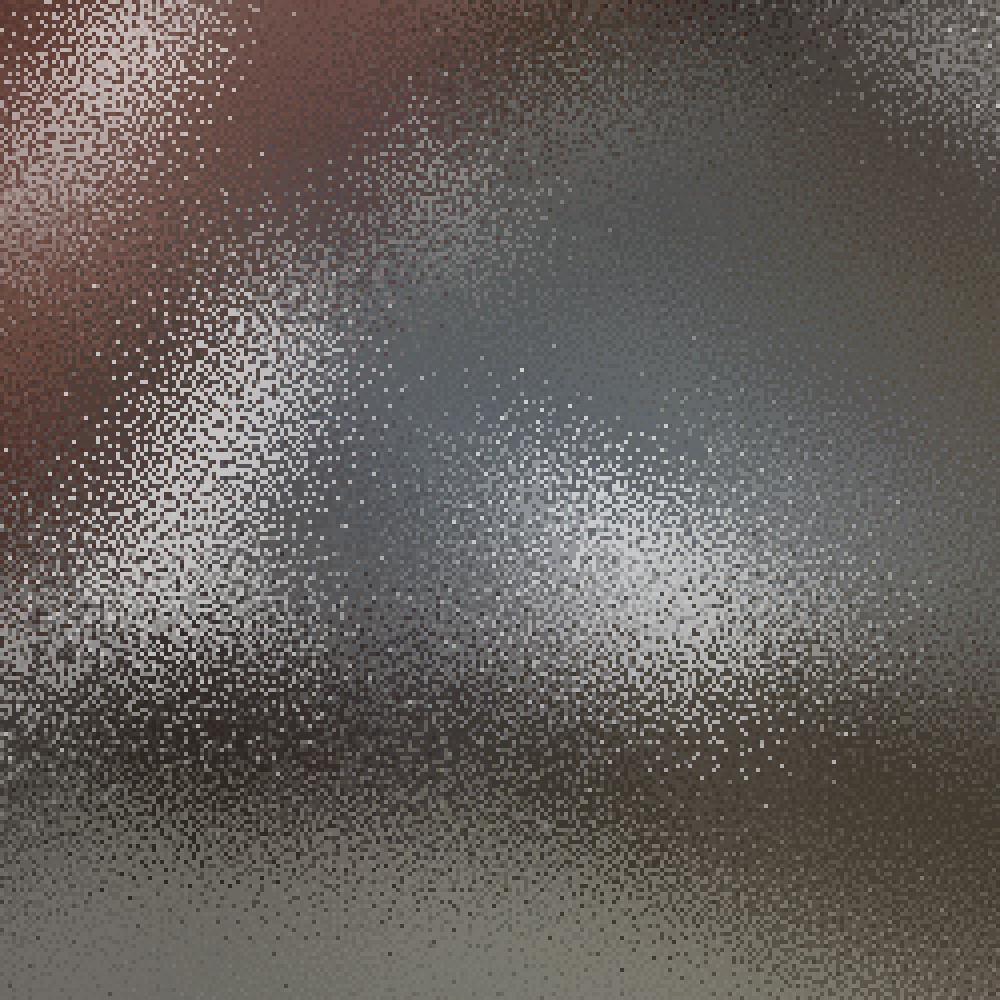}&
			\includegraphics[height=\myheight]{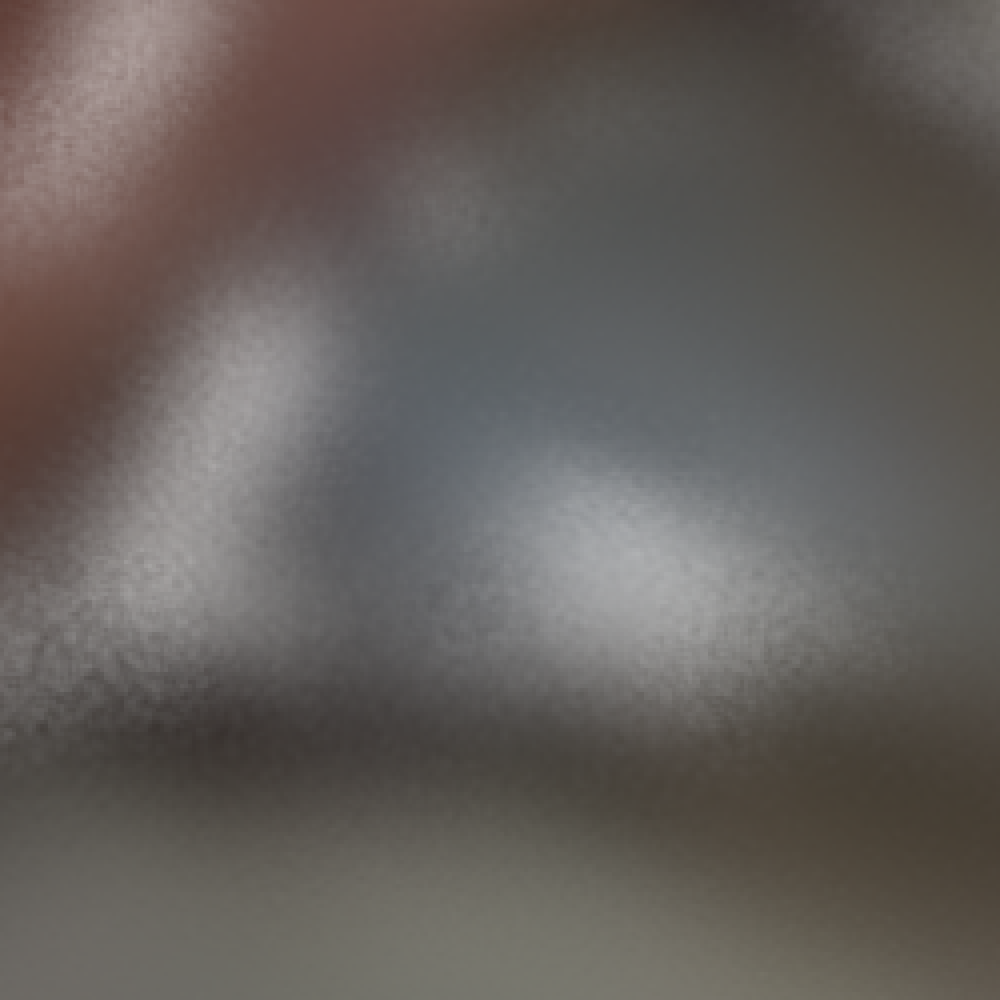}&
			\includegraphics[height=\myheight]{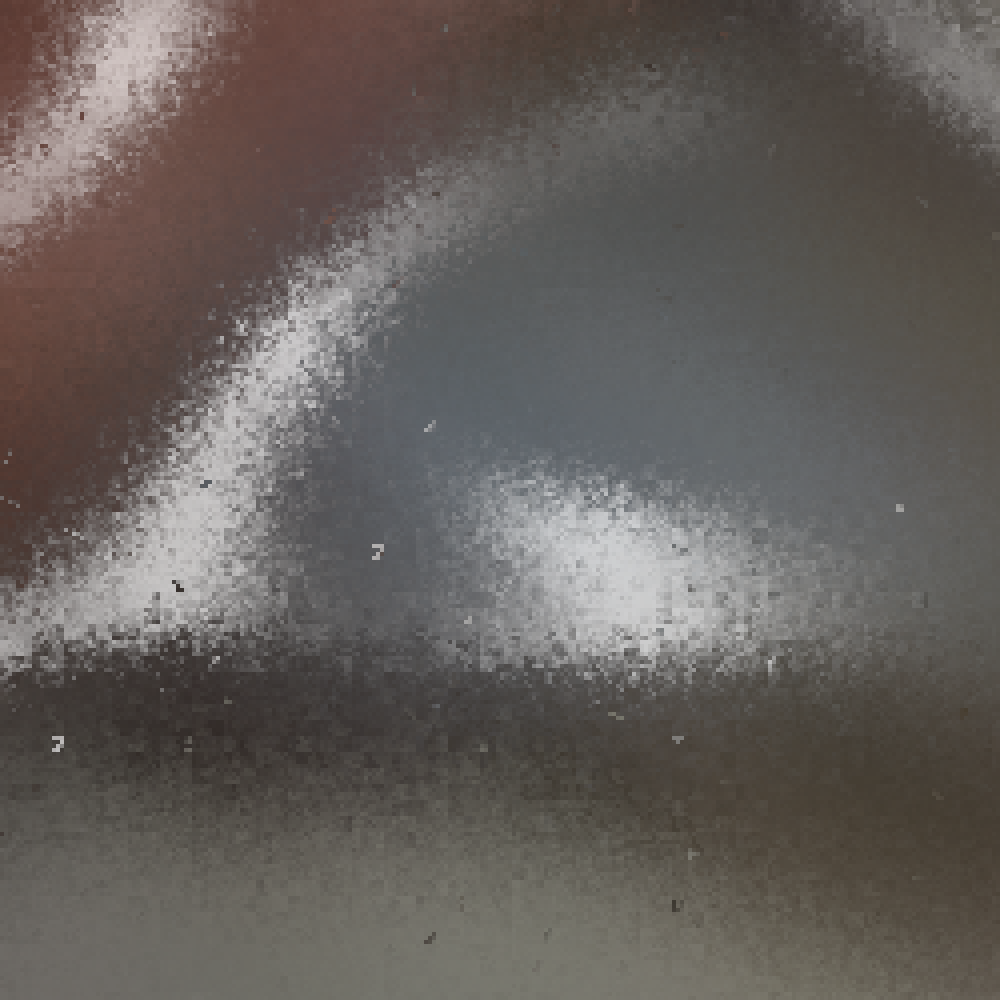}&
			\includegraphics[height=\myheight]{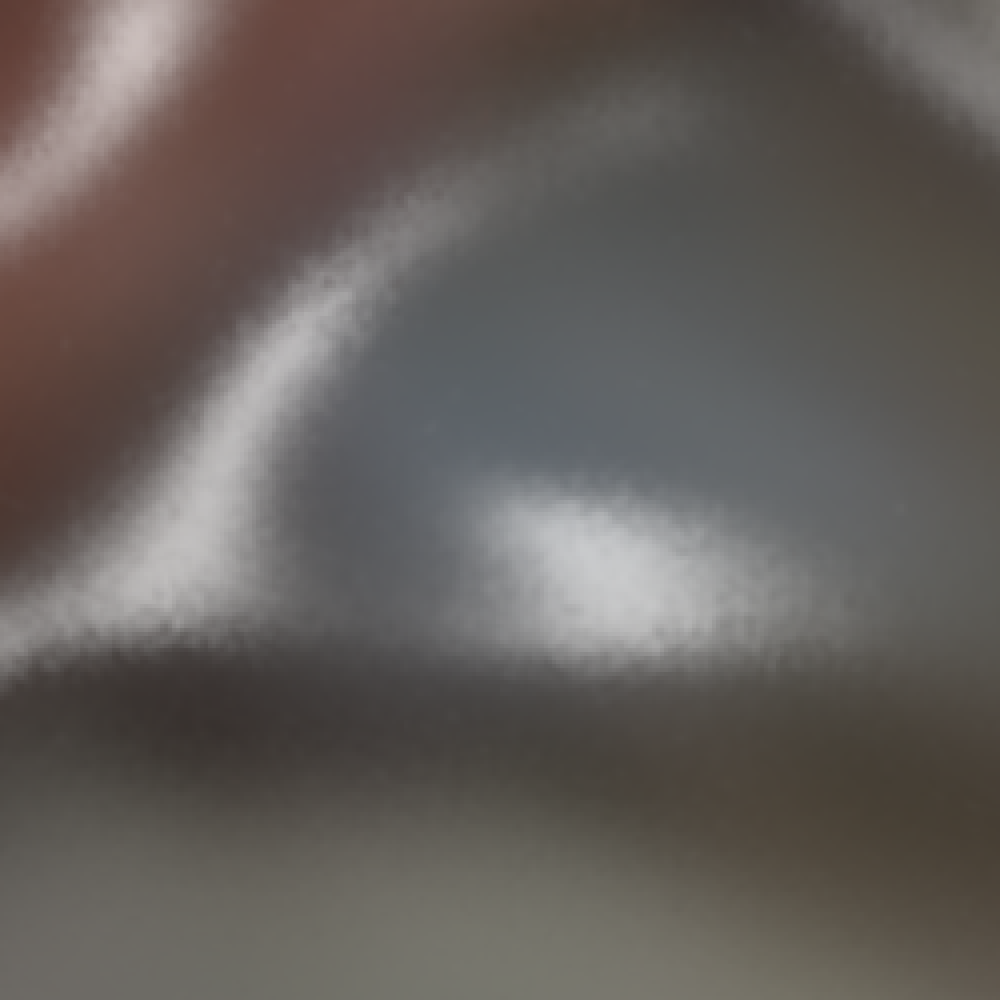}& 
			\rotatebox{90}{\hspace{1mm}\textbf{bicubic B-spline}}\\
			& \textbf{PSNR}($\uparrow$)/\textbf{FLIP}($\downarrow$):& 
			28.3 / 0.054 & 
			33.8 / 0.038 &
			38.2 / 0.032 &
			\textbf{44.7 / 0.020}			
		\end{tabular}
	}
	\caption{We present an improved stochastic texture filtering (STF) method
		with no additional texture lookup cost compared to standard STF.
		Above, the images from our method have between $9.9$ and $12.3$~dB higher PSNR
		and lower FLIP errors for both bilinear and bicubic B-spline filtering compared to standard STF. DLSS was used for denoising.
	}
	\Description{figure description}
	\label{fig_teaser}
\end{teaserfigure}

\maketitle

\section{Introduction}
\label{sec_intro}

The gap between computation and memory bandwidth in microprocessors has been increasing
for nearly 50 years~\cite{Dongarra2022}.
GPUs are not immune to this trend:
over the past 20 years, the computation available on GPUs (measured by FP32 GFLOPS) has increased by
a factor of $\sim\!2,750\times$, while the peak off-chip memory bandwidth has increased by a factor of just
$\sim\!60\times$.\footnote{We compare the NVIDIA GeForce FX 5800 Ultra, 30 FP32 GFLOPS, 16 GB/s bandwidth, released March 2003,
to the NVIDIA 4090 RTX, 82.6 TFLOPS, 1.01 TB/s bandwidth, released October 2022;
GPUs from other manufacturers share similar relative ratios.}
If specialized tensor operations in hardware is included,
the growth in available computation is up to an order of magnitude greater, depending on the precision.
This trend has motivated the development of new compression techniques that trade off increased computation to save bandwidth;
examples include neural texture compression (NTC)~\cite{Vaidyanathan2023} and NeuralVDB~\cite{Kim2024neuralvdb}.
However, these techniques are not compatible with current GPU texture filtering hardware and may be too expensive even for a simple bilinear filter decoding $2\times 2$ texels.
Building on earlier work by Hofmann et al.~\cite{Hofmann2021}, Vaidyanathan et al.~\cite{Vaidyanathan2023}, and others,
Pharr et al.~\cite{Pharr2024} thus introduced a family of Monte Carlo stochastic texture filtering (STF) techniques to approximate
the effects of traditional texture filtering without increasing the texture sampling cost beyond a single sample.

STF has the additional advantage that it allows efficient implementation of \emph{filtering after shading} (FAS), where
the antialiasing filter is applied to the final shaded value rather than the texture inputs to a shader~\cite{Pharr2024}.
When the shading function is linear or an affine combination of the input texture values, both filtering before shading (FBS) and FAS converge to the same result.
When it is non-affine, results differ:
FAS is generally more accurate than FBS when textures are minimized,
but it has a number of disadvantages when textures are magnified~\cite[Section~3.2]{Pharr2024}:
\begin{itemize}
\item FAS can introduce aliasing under magnification.
\item FAS is unable to reproduce smooth interpolation of some properties (e.g., surface curvature).
\item FAS may produce results different than those intended by the author of the art assets.
\end{itemize}
Further, spatiotemporal reconstruction techniques like TAA~\cite{Yang:2020:Survey,Karis:2014:High} and DLSS~\cite{Liu:2022:DLSS} may fail to converge and produce residual noise when used to integrate STF samples~\cite[Section~6]{Pharr2024}.

In this work, we address these limitations and improve the quality of stochastic texture magnification
without increasing the cost beyond a single texture lookup and without affecting the favorable properties of FAS for minification.
We base our approach on a simple insight: under magnification, adjacent screen pixels tend to filter the same set of texels.
Thus, if each pixel continues to take a single STF texel sample and shares this sample with nearby pixels,
then each pixel may be able to compute a more accurate estimate of the true filtered value.
This sharing can be performed efficiently by taking advantage of the SIMT/SIMD nature of GPU execution to communicate between shader threads
without any VRAM memory traffic.
By averaging multiple texel values, our method performs some filtering before shading.
Intuitively, it can be seen as a hybrid between filtering after shading and filtering before shading.

Our contributions include:
\begin{itemize}
\item An algorithm for efficient STF sample reuse and sharing using wave intrinsics (Section~\ref{sec_algorithms}) including practical implementation details (Section~\ref{sec_method_details}).
\item An in-depth analysis of the error of the traditional STF estimator (Section~\ref{sec:stferror}) including the impact of 
  filtering after shading of non-affine rendering functions. We further explain in the supplementary material (Section~\ref{sec:app-nonlinear-taylor})
  the relationship between the estimator variance and the bias introduced by reordering filtering and shading.
\item A new Monte Carlo estimator based on weighted importance sampling with a lower error than previous estimators when used for stochastic texture filtering (Section~\ref{sec:estimators}). 
\item An algorithm that generates optimized \emph{sharing footprints} introducing variation, in which pixels share texel values with each other and distribute error as blue noise in the rendered images (Section~\ref{sec_filter_footprints}).
\item An algorithm to generate custom blue noise patterns optimized to account for this sharing (Section~\ref{sec_blue_noise}).
\end{itemize}
Under magnification, these techniques provide much better image quality
than traditional STF, both visually and using error metrics; see Figure~\ref{fig_teaser}.
We validate our claims and analyze results with a number of Monte Carlo estimators, 
different pseudo-random sequences and using different spatiotemporal denoisers in Section~\ref{sec_results}.

\section{Background and Previous Work}
\label{sec_prevwork}

Before introducing our algorithm, we first provide relevant background on texture filtering, Monte Carlo sampling, and the GPU quad and wave intrinsics we use
for efficient communication between nearby pixels.

\subsection{Texture Filtering and Representations}

A texture $t(u,v)$ is defined by a set of texels $\mathbf{T}_{u_i, v_i}$ defined 
at integer coordinates $(u_i, v_i)$ on a grid, scaled by translated Dirac delta functions:\footnote{%
Without loss of generality, we focus on filtering 2D textures and use this assumption throughout the text,
but the described techniques apply to any texture dimensionality.}
\begin{equation}
  t(u, v) = \sum_i^n \, \delta(u-u_i) \delta(v-v_i) \, \mathbf{T}_{u_i,v_i}.
  \label{eq:texture-definition}
\end{equation}
With this notation, $n$ is the total number of texels in the texture.
To construct a continuous texture function, it is necessary to specify a reconstruction filter $f_\mathrm{r}$ and convolve it with the texture function:
\begin{equation}
  t_\mathrm{r}(u,v) = t \otimes f_\mathrm{r} = \int \! \! \! \! \int t(u', v') \, f_\mathrm{r}(u-u', v-v') \, \mathrm{d}u' \, \mathrm{d}v',
\label{eq:texture-continuous-integral}
\end{equation}
where $t \otimes f_\mathrm{r}$ is a convolution. %
In turn, the filtered value at a point $(u,v) \in \mathbb{R}^2$ can be written as a sum of weighted
texel values.  Given a $(u,v)$ lookup point, we will generally write Equation~\ref{eq:texture-continuous-integral} as
\begin{equation}
  t_\mathrm{r}(u,v) = \sum_i f_\mathrm{r}(u-u_i, v-v_i) \, \mathbf{T}_{u_i, v_i}.
\label{eq:texture-continuous-sum}
\end{equation}

The bilinear (tent) and bicubic filters are commonly used for texture
reconstruction.  Given such a filter with limited spatial extent, these
sums only need to be over a few weighted texels.

\subsection{Monte Carlo Estimators}
\label{sec_prevwork_mc}

The standard importance sampling Monte Carlo (MC) estimator samples a random
variable $X$ from a probability distribution function (PDF) $p$, $X \sim p$
and gives the estimate
\begin{equation}
  F = \frac{f(X)}{p(X)} \approx \int f(x) \, \mathrm{d}x.
  \label{eq:mc-estimator}
\end{equation}
This estimator is unbiased if $p(x)>0$ everywhere when $f(x)>0$~\cite{Pharr2023}.

A continuous Monte Carlo integral estimator like Equation~\ref{eq:mc-estimator} can
  be converted to apply to a discrete sum by introducing Dirac delta functions
  centered at the points where the sum is being evaluated.
  PDFs are then comprised of corresponding Dirac deltas multiplied by discrete \emph{probability mass functions} (PMFs) that sum to~1.
  Such an approach was effectively taken
in previous applications of STF, where one sample was taken from the sum in Equation~\ref{eq:texture-continuous-sum}
based on a PMF proportional to the filter weight.
In turn, the importance sampling Monte Carlo estimator approximates
the filtered value $t(u,v)$ with a single unweighted texel value; we will
call this approach \emph{one-tap STF} in the following.
See Section~\ref{app:stf-mc-connection} for a derivation showing how the continuous
  estimator of Equation~\ref{eq:mc-estimator} was applied for one-tap STF.

The Monte Carlo literature offers extensive analysis of convergence of not only estimators,
but also different sources of randomness, such as different pseudo- and quasi-random sequences.
Most of this analysis focuses on convergence properties with sample counts growing toward infinity.
In contrast, real-time rendering uses spatiotemporal filtering~\cite{Yang:2020:Survey,Karis:2014:High}
and can realistically average only a small number of past samples, sometimes combined with %
spatial filtering~\cite{Liu:2022:DLSS}.
Therefore, real-time rendering focuses on minimizing the perceptual error for a minimal number of samples,
assuming small radius spatial filtering and an exponentially moving average temporal filter.
The most common randomness sources in real-time rendering are blue noise dithering masks~\cite{georgiev:2016:blue},
such as the spatiotemporal blue noise masks~\cite{Wolfe2022} used by Pharr et al.~\cite{Pharr2023}.
Wolfe et al.~\cite{Wolfe2022} proposed two different algorithms for generating those masks---a scalar version based on
the void and cluster algorithm, and a vector version 
based on
simulated annealing.
In Section~\ref{sec_blue_noise}, we discuss the impact of those masks on the quality of our algorithm as well as compare them with
the most recent spatiotemporal blue noise mask generation advancements~\cite{donnelly:2023:filter}.

\subsection{Wave Intrinsics}
\label{sec_wave_intrinsics_background}
\begin{wrapfigure}{R}{0.45\textwidth}
	\centering
	\includegraphics[width=0.25\columnwidth]{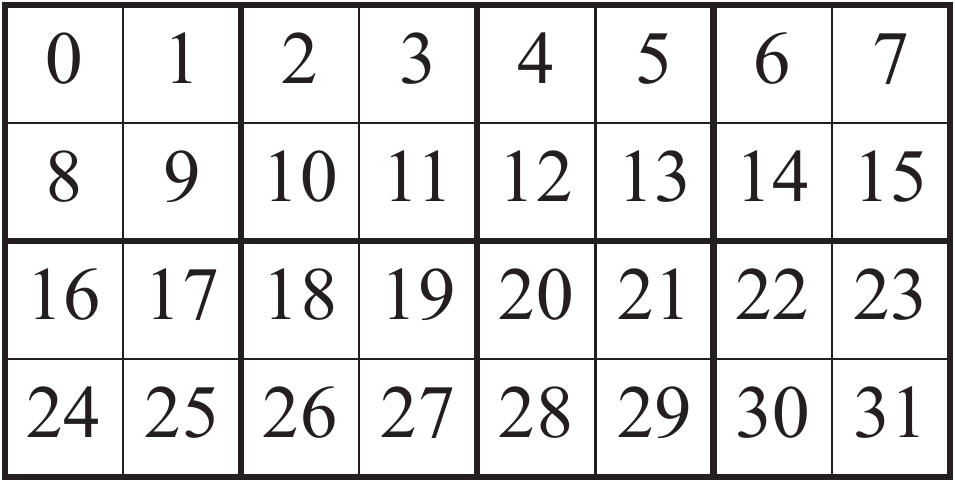}
	\caption{An example of a 32 lane wave: eight $2\times 2$ quads
	configured as $8\times 4$ pixels. The numbers
	are lane ids, configured for linear row ordering.
	Each pixel corresponds to a
	lane in the wave, and it can access a value of the other wave lanes
	using \texttt{WaveReadLaneAt()}.
	}
	\label{fig_wave}
\end{wrapfigure}
We make extensive use of \textit{wave intrinsics} to share samples between pixels;
they were introduced in DirectX HLSL Shader Model 6.0~\cite{Microsoft2021}.
Normally in shader programming, only a single thread of execution 
is exposed, meaning that each thread does not know what
its neighbors are doing.
Wave intrinsics allow threads to communicate.
HLSL 6.0 uses the term \textit{lane} for a single thread of execution,
and \textit{wave}, sometimes called a warp, for a set of lanes that
are executed together in parallel.
Wave sizes are hardware dependent and not guaranteed by the DirectX specification, but common sizes are 8, 16, 32, and 64 lanes.
All our evaluation (Section~\ref{sec_results}) is done on NVIDIA hardware,
where recent GPUs have a wave size of 32; %
our work
generalizes to arbitrary wave sizes. %
A lane can access a value from any lane in its wave using 
\texttt{WaveReadLaneAt(value, laneId)}; see Figure~\ref{fig_wave}.
Wave intrinsics are an attractive method for inter-lane communication since they are
typically implemented as swap or shuffle instructions within a
wave~\cite{Microsoft2021}, and thus cost only one instruction with no
memory bandwidth impact.
Previous work exploited pixel shader quad derivatives to perform in-place screen-space filtering (such as bilateral or convolution filters)~\cite{penner:2011:shader,mcguire:2012:scalable}.
We build upon those ideas using modern wave intrinsics and generalize beyond screen-space effects.

\begin{wrapfigure}{L}{0.39\textwidth}
	\centering
	\includegraphics[width=0.25\textwidth]{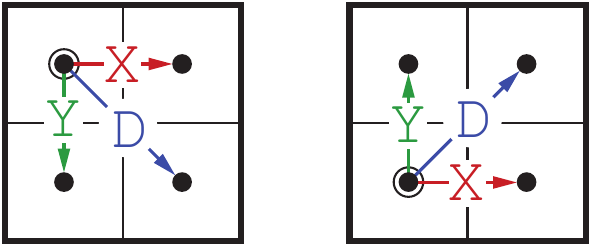}
	\caption{Quads with the current pixel marked with
		a circle. The values accessed by \texttt{QuadReadAcrossX/Y/Diagonal()}
		are marked \texttt{X}, \texttt{Y}, and \texttt{D}.}
	\label{fig_quad}
\end{wrapfigure}
Under magnification, adjacent screen pixels sample adjacent source texels and we aim to share
samples between adjacent pixels to improve filtering quality.
The zero memory cost of wave intrinsics is highly beneficial for this purpose.
Before we describe and motivate our approach in detail, we discuss how wave lanes are mapped to screen pixels.
We will show in Section~\ref{sec_filter_footprints} that this mapping affects the possible sample sharing configurations
and filtering quality.

For compute shaders, it is the programmer's responsibility to map the dispatch indices (and, as a consequence, the wave lane indices)
to processed elements, such as pixels, vertices, or rays.
Figure~\ref{fig_wave} organizes 32 lanes as $8\times 4$ lanes but other configurations
such as $16\times 2$ lanes are possible.

With pixel shaders, there is only a guarantee of $2\times 2$ quad granularity
with the upper left pixel of each quad located at even $(x, y)$ coordinates on the screen.
Two quads are shown in Figure~\ref{fig_quad}.
HLSL provides special helper functions for quad communication.
For a single pixel in a $2\times 2$ quad, one can access computed values in
the other pixels in the quad using the 
\texttt{QuadReadAcrossX()},
\texttt{QuadReadAcrossY()},
and \texttt{QuadReadAcrossDiagonal()} functions (Figure~\ref{fig_quad}).
It is still possible to access other wave lanes using the
\texttt{WaveReadLaneAt()} function in pixel shaders, but
they can map to arbitrary screen pixels, inactive lanes, and depend on the rasterizer behavior.

\section{Error Analysis of Stochastic Texture Filtering}\label{sec:stferror}
We observe that the error and noise characteristics of STF do not resemble traditional Monte Carlo rendering noise.
A single sample of the rendering equation MC integral is typically extremely noisy, even given a smooth signal.
For this reason, most of the Monte Carlo literature analyzes the average error and the error convergence.
By comparison, many pixels produced by STF are not noisy, even without temporal integration.
See Figure~\ref{fig_teaser} as an example.
Aiming to improve STF, we analyze the error of a single STF sample and show that it is not only bounded, but is zero in many common cases.

We begin by observing that all valid texture filters have weights that sum to one.
Without loss of generality, we assume a filter with nonnegative weights.\footnote{A similar analysis applies to filters with negative lobes sampled with positivization~\cite{Pharr2024}, rescaling the convex hull bounds.}
The one-tap STF importance sampling estimator effectively selects one of the samples of the texture where the texture filter is nonzero (one element of the sum in Equation~\ref{eq:texture-continuous-sum}).
To analyze the maximum error of a single sample, we can ignore the sampling probabilities and look at the individual selected samples.
The STF estimator has a geometric interpretation---the filtered value lies
inside the convex hull of texel values, and STF selects one of the texels.
The one-tap STF estimator is thus guaranteed to never leave the range of the contributing texels' values.
This property is crucial for texture filtering, as textures often encode physical properties with a defined valid range.
The highest error case occurs when the selected texel differs the most from the filtered value, which can be arbitrarily close to another texel.
In other words, the maximum error of any interpolating texture filter with nonnegative weights is
\begin{equation}\label{eq:stf_error}
	\max_{i} \mathbf{T}_{u_i,v_i} - \min_{i} \mathbf{T}_{u_i,v_i},
\end{equation}
with $i$ ranging over the contributing texels.

From Equation~\ref{eq:stf_error} and the geometrical interpolation interpretation, 
it is clear that when all filtered texel values are the same,
the one-tap STF estimator error is zero.
Natural and computer-generated images have a low-frequency spectral bias~\cite{Reinhard2001natural} and adjacent texel values often change slowly.
While this is not the case for all scenes, many rendering texture assets contain large flat and low-frequency regions.
In those regions, the STF estimator error is close or equal to zero.

Given the low error and favorable properties of the existing STF estimator, we set the following goals for the improved estimator:
\begin{enumerate}
	\item It should not produce any error where the one-tap STF estimator yields zero error.
	\item It should not increase the error in common low-frequency regions.
	\item It should reduce the error and variance in high-frequency regions, e.g., edges, where the maximum and minimum local values differ significantly.
\end{enumerate}
To achieve properties 1 and 2, we argue that a texture filtering estimator must produce results within the convex hull of the texel values.
In Section~\ref{sec_results_estimators}, we show how violating those requirements can lead to poor image quality and introduction of new
errors as compared to one-tap STF.

\section{A Stochastic Texture Filtering Algorithm with Sample Reuse}
\label{sec_algorithms}

In this section, we present our entire algorithm for STF with
texel reuse across neighboring pixels.

\subsection{Evaluation of Existing Estimators}
\label{sec_existing_estimators}
With our approach, each lane starts by sampling a single texel according to its texture filter,
just like one-tap STF.
After sample sharing, each lane has $n$ sampled texels $x_i$, each drawn from a PMF $p_i$ corresponding to a distinct lane's texture filter.
Our task is to compute an estimate of the texture filtering Equation \ref{eq:texture-continuous-integral} using these samples.
We start by considering the use of standard MC estimators.

\begin{wrapfigure}{L}{0.37\textwidth}
	\centering
	\includegraphics[width=0.36\textwidth]{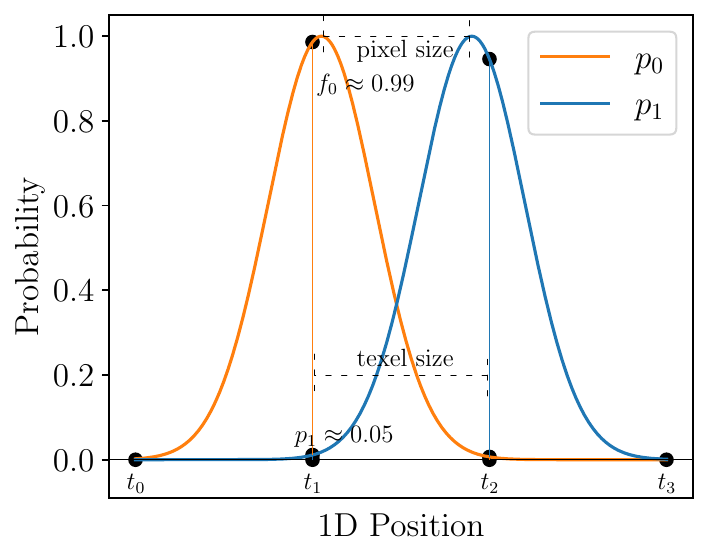}
	\caption{Reconstruction filters for two adjacent pixels in 1D.
	Texel locations are on the $x$-axis. When pixel 1 samples texel $t_1$ from the distribution $p_1$ according to
	its texture reconstruction filter and then shares it with pixel 0, the standard MC importance sampling estimator weights this texel
	$f_0 / p_1 \approx 20$ causing a variance spike, even for constant filtered signals.
	}
	\label{fig_pdf_mismatch}
\end{wrapfigure}

The standard importance sampling estimator computes the filtered value as
the weighted texel values divided by the probabilities $p_i$ of sampling each
$x_i$:
\begin{equation}
  \frac{1}{n} \sum_i^n \frac{f(x_i)}{p_i(x_i)}.
  \label{eq_basic_mc}
\end{equation}
Although importance sampling requires that $p_i > 0$ whenever $f > 0$; this may not be
the case with sample sharing since other lanes' filters (and thus PMFs), generally differ from the current lane's.
If the importance sampling estimator is still used, the error may be unbounded.
Further, even if a PMF $p_i$ is valid, the estimator may still produce arbitrarily high error---in
the context of STF, consider a pixel where the filter weight for a texel $\mathbf{T}_{u_i,v_i}$ is very small;
if it samples that texel and shares it with a neighboring pixel where the filter weight is much
larger, the ratio $f/p$ may be arbitrarily large, leading to high variance.
This is illustrated in Figure~\ref{fig_pdf_mismatch}.

One way to ensure that the PMFs are valid is by using \emph{defensive importance sampling},
where for example a constant PMF over the entire domain is mixed with the regular sampling
PMF~\cite{Hesterberg1995WeightedAI}.
However, we have found that this may lead to pixels sometimes having \emph{zero} texels that are inside their own filter,
making this approach unsuitable.
\emph{Multiple importance sampling} (MIS)~\cite{Veach1995MIS} provides an
estimator that reduces the variance when some PMFs are a better match to the function than others and
further requires that only one of the PMFs be nonzero when $f>0$.
The MIS balance heuristic (and Veach's other heuristics) have the disadvantage
of $n^2$ PMF evaluations.
This cost can be noticeable when sharing samples across multiple pixels.
It can be avoided with
Bitterli's \emph{pairwise MIS}, which only requires $2n$ PMF evaluations~\cite{Bitterli2022}.

We have also considered a \emph{regression estimator} that is based on computing weights for a control variate based on the sample values~\cite[Equation~9.11]{Owen2013mcbook}. It is:
\begin{equation}
\qquad\qquad\qquad\qquad\qquad\qquad\qquad\quad
  \frac{1}{n} \sum_i^n \frac{f(x_i)}{p_i(x_i)} - \beta \left( \frac{w_i(x_i)}{p_i(x_i)} - 1 \right),
  \label{eq_mc_regression}
\end{equation}  
where $w_i(x_i)$ is the filter weight for texel $x_i$ and
\begin{equation}
\qquad\qquad\qquad\qquad\qquad\qquad\qquad\quad
  \beta = \frac{\sum_i^n \left(w_i(x_i) - \bar{w}\right) f(x_i) w_i(x_i)}{\sum_i^n (w_i(x_i) - \bar{w})^2},
\end{equation}
with $\bar{w} = (1/n) \sum_i^n w_i(x_i)$.

One problem with all the approaches in this section
is that they do not fulfill the first two goals identified in Section~\ref{sec:stferror}
and may yield results outside the convex hull of texel values.
One way to work around this issue is to clamp the filtered texture value to lie withing the bounds of texel values used.
This introduces bias but significantly reduces the error as we show in our evaluation (Section~\ref{sec_results_estimators}).

\subsection{Our Estimator}
\label{sec:estimators}

\emph{Weighted importance sampling} (WIS) estimators (also known as \emph{self-normalized importance sampling}) provide an alternative that provide
results bounded by the texel values.
WIS began with the ``weighted uniform'' estimator, suggested by Handscomb~\cite{Handscomb1964}
and later analyzed by Powell and Swann~\cite{Powell1966weighted}.
Spanier~\cite{Spanier1979new} suggested and analyzed two successive generalizations,
leading to sampling from one PDF $x_i \sim p_1$ and then reweighting using a second:%
\footnote{The reweighting in WIS bears some similarity to importance resampling
~\cite{talbot2005resampling,bitterli2020spatiotemporal}, though
those methods are used when it is inexpensive to generate candidate samples
and where selecting a subset of those samples for evaluation is desired;
this is not the case for texture filtering. See Bitterli
et al.~\cite[Appendix~B]{bitterli2020spatiotemporal} for further discussion
of connections between WIS and resampled importance sampling as well as to
Monte Carlo ratio estimators.}
\begin{equation}
  \frac{\sum_{i}^{n} f(x_i)/p_1(x_i)}{\sum_{i}^{n} p_2(x_i)/p_1(x_i)}.
  \label{eq_spanier_wis}
\end{equation}
These WIS estimators both have a small bias but are \emph{consistent}.
In graphics, WIS is commonly used when filtering pixel samples in path tracers, for
example~\cite[Section~5.4.3]{Pharr2023}.
Other prior applications of WIS in rendering include Monte Carlo
radiosity~\cite{bekaert2000weighted} and a photon mapping
technique~\cite{SzirmayKalos2023wisphoton}.

We propose a generalization of Equation~\ref{eq_spanier_wis} that allows
taking samples from $n$ different PDFs $x_i \sim p_i$, as is the case with
texel sharing.
Assuming for now we have a continuous PDF $p_\mathrm{c}$ that is nonzero where
$f(x)>0$, and additional PDFs $p_i(x)$ that are nonzero when $p_\mathrm{c}$ is,
then our estimator is
\begin{equation}
  \frac{\sum_{i}^{n} f(x_i)/p_i(x_i)}{\sum_{i}^{n} p_\mathrm{c}(x_i)/p_i(x_i)},
  \label{eq_bart_weighted_estimator}
\end{equation}
which we believe is novel.

This estimator is \emph{consistent}, i.e., it converges to the value of the integral $\mu = \int f(x) \mathrm{d}x$ with probability 1.
We can see this by first writing it in the equivalent form:
\begin{equation}
  \frac{\frac{1}{n} \sum_i^n f(x_i)/p_i(x_i)}{\frac{1}{n} \sum_i^n p_\mathrm{c}(x_i)/p_i(x_i)}.
\end{equation}
Following Owen~\cite[Theorem~9.2]{Owen2013mcbook}, since $p_i>0$ where
$f>0$, the numerator of Equation~\ref{eq_bart_weighted_estimator} is the average of independent
random variables that each are unbiased estimates of $\mu$.
In turn, the numerator converges to $\mu$.
Next, because $p_\mathrm{c}$ is a PDF, it integrates to 1. We can use the same argument to show
that the denominator converges to $1$ as $n\rightarrow\infty$.
Thus, their ratio converges to $\mu$ as well.
Like other WIS estimators, our estimator is biased, but the bias is small in practice.

To apply this continuous estimator to the texture filtering sum,
  we take the reconstructed texture function $t_\mathrm{r}$ of Equation~\ref{eq:texture-continuous-integral} for $f$
  and define corresponding PDFs with the product of Dirac delta functions at $(u_i,v_i)$ and the reconstruction filters.
  (See Section~\ref{app:stf-mc-connection} in the supplemental for a derivation showing how
  Pharr et al. effectively did this with one-tap STF~\cite{Pharr2023}.)
  We would like to estimate the filtered texture value at a point $(u,v)$,
  where each lane has sampled a texel location $(u_i,v_i)$ from its texture filter $f_i$'s PMF $p_i$.
  If we write the current lane's texture reconstruction filter as $f_\mathrm{c}(u',v')=f_\mathrm{r}(u-u',v-v')$ 
  with an associated PMF $p_\mathrm{c}$, the estimator is:
\begin{equation}
  \frac{\sum_i^n f_\mathrm{c}(u_i,v_i) \, \mathbf{T}_{u_i, v_i}/p_i(u_i,v_i)}%
{\sum_i^n p_\mathrm{c}(u_i,v_i) / p_i(u_i,v_i)}.
\label{eq:estimator-for-stf}
\end{equation}
If the texture reconstruction filter is positive and normalized then
$p_\mathrm{c}(u_i,v_i)=f_\mathrm{c}(u_i,v_i)$
and similarly $p_i=f_i$, giving
\begin{equation}
  \frac{\sum_i^n f_\mathrm{c}(u_i,v_i) \, \mathbf{T}_{u_i, v_i}/f_i(u_i,v_i)}{\sum_i^n f_\mathrm{c}(u_i,v_i)/f_i(u_i,v_i)}.
\end{equation}
If we define weights $w_i = \frac{f_\mathrm{c}(u_i,v_i)}{f_i(u_i,v_i)}$ and simplify, we have:
\begin{equation}
  \sum_i^n \frac{w_i}{\sum_j^n w_j} \, \mathbf{T}_{u_i, v_i}.
  \label{eq:estimator-convex}
\end{equation}
Thus, the weights are normalized and sum to one, guaranteeing that our estimator always
produces a convex combination of the filtered texels.
As a result, constant texture regions have zero error.

However, unlike the general case, we do not necessarily have $p_i>0$ where $f>0$ and $p_\mathrm{c}>0$ and
thus convergence is not guaranteed.
We find in practice that the error due to this is small.
Crucially, unlike the standard importance sampling estimator, ours does not have the risk of returning
very large or small values due to a mismatch between $f$ and the $p_i$.
On the other hand, a significant $f$ and $p_i$ mismatch can still cause a visual error in high-contrast regions.
The resulting high weight is normalized with other contributing weights but causes our algorithm to effectively select one of
the texels from the other lanes.
This can result in pixels with a firefly-like appearance in highly specular regions (see Figure~\ref{fig_teaser}).

\subsubsection{Exact Filtering}
\label{sec:exact_filtering}
\begin{figure}[t]
	\centering
	\includegraphics[width=0.85\columnwidth]{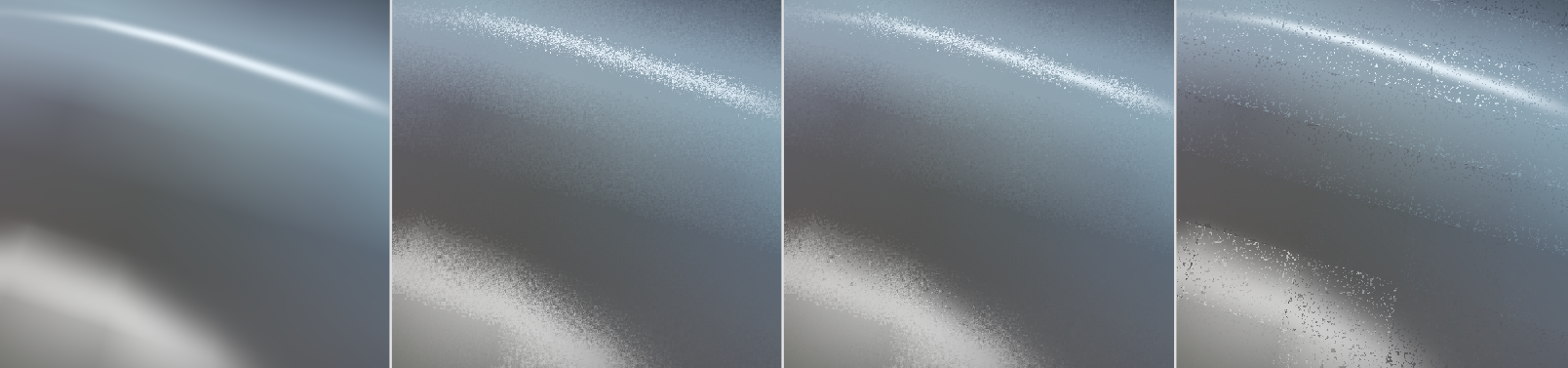}
	\caption{
		From left to right: reference bilinear filtering, our WIS estimator, our WIS estimator with \textit{exact filtering},
		and exact filtering with equal sampling probabilities of texels inside of the filter footprint.
		The rightmost image shows a much higher percentage of pixels using exact filtering at the cost of severe visual artifacts,
		mostly visible at bilinear filter footprint boundaries.
	}
	\label{fig_exact_filtering}
\end{figure}

Some simple filters, such as bilinear, require only four distinct texels for exact reconstruction.
We observe that using high-quality, anti-correlated random number generators, such as blue noise,
there is a high probability that all four distinct texels will be sampled in the pixel neighborhood.
In such cases, we can return the filtered value directly without using our estimator.
We call this extension \textit{exact filtering}, and show an example of its efficiency in Figure~\ref{fig_exact_filtering}
(two middle columns), and evaluate it in Section~\ref{sec_results}.
It introduces an additional small bias but reduces the error.
Exact filtering is practical with small filter footprints, such as bilinear.
In contrast, a bicubic filter requires 16 texels for perfect reconstruction and the probability of
sampling all those values even across an entire 32-element wave is low.

A limitation of exact filtering is that because each pixel samples a texel according to its filter PMF,
texels that only make small contributions to the pixels in a wave may be sampled only rarely.
We briefly explored strategies that change the sampling probabilities to be more uniform under strong magnification,
similarly to defensive importance sampling~\cite{Hesterberg1995WeightedAI}.
We found that while those strategies increase the number of pixels where exact filtering is possible, they can lead to severe grid-like 
visual artifacts (Figure~\ref{fig_exact_filtering} rightmost column).
The main reason is that a lane cannot access elements outside of the wave. For instance, the top left pixel
of a wave cannot access any elements to the left or above it, which might be needed for its exact bilinear filter
depending on how pixels' UV values align with the screen pixel grid.

\subsection{Sharing Footprints Within Waves}
\label{sec_filter_footprints}
With our approach, each lane in a wave has an associated \emph{texel sharing footprint} that specifies which other lanes it draws texel values from.
With quad intrinsics, these footprints are restricted to
fixed grid positions and access is allowed to only three neighbors
(Section~\ref{sec_wave_intrinsics_background}).
With wave intrinsics, one is free to use any lanes inside the wave and any number of them, giving a wider design space.
In the following, we assume a 32-wide wave, though the concepts apply to other wave sizes.
A wave with 32 lanes may be configured as $16\times 2$ or
$8\times 4$ pixels. We found that $8\times 4$ is preferable, since
it allows for larger, square-like sharing footprints, which tend to give lower error.

Our design criteria for sharing footprint size and placement were:
\begin{enumerate}
	\item\label{enum_ff_square} Prefer lanes close to the current lane to improve the chances of sharing a useful texel value.
	\item\label{enum_ff_same_number} Use the same sharing filter footprint size for all lanes inside a wave to avoid unnecessary thread divergence.
	\item\label{enum_ff_move1} Maximize the variety of shared texel values between lanes.
\end{enumerate}

\subsubsection{Square Footprints}
\label{sec_square_footprints}
\begin{figure}[t]
	\centering
	\includegraphics[width=0.7\columnwidth]{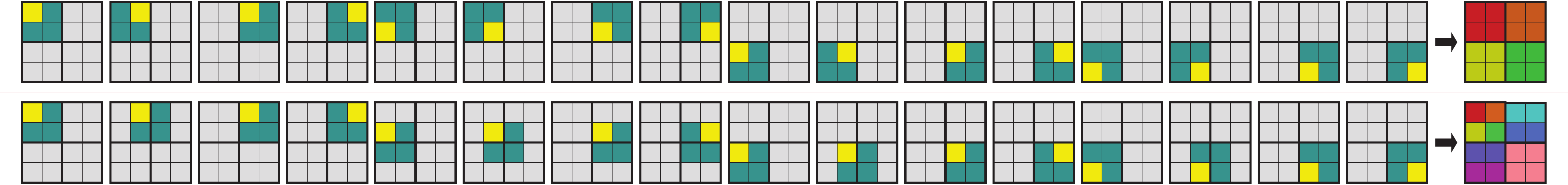}
	\caption{
		Examples of $2\times 2$ footprints for waves of size 16,
		configured as $4\times 4$ lanes.
		A texel sharing footprint consists of a yellow lane, which is the lane
		whose filtered value we want to compute, together with
		a set of texel samples from dark green lanes.
		\textbf{Top}: using quad intrinsics the footprints
		are restricted by the API to have the upper left coordinates of a
		quad be even in $x$ and $y$.		
		All lanes in each quad filter the same texel values.
		Under high magnification, $2\times 2$ regions produce almost identical
		color after stochastic filtering, illustrated by the colored pattern at the right.
		\textbf{Bottom}: using wave intrinsics, the $2\times 2$ footprints
		have more degrees of freedom in that they can be placed anywhere
		inside the wave. 
		The sharing pattern visualization at the right
		has fewer regions of similar color, resulting in less spatial correlation.
	}
	\label{fig_footprint2x2}
\end{figure}

Square sharing footprints are a natural choice that ensures that the closest possible lanes are used for sharing.
Figure~\ref{fig_footprint2x2} considers $2\times 2$ footprints.  The top half
shows the only possible footprint placement for quad intrinsics.
With quad intrinsics, the four lanes inside each quad end up with the same set of texel values after sharing.
Under high magnification, all will compute nearly the same filtered color introducing high spatial correlation.

In the bottom part of the figure, we show one possible layout for $2\times 2$
footprints when using wave intrinsics, which allows
the footprints to have more variety.
The false-colored pattern in the lower right shows that this method provides
smaller regions with the same texel values used at adjacent lanes compared to quad intrinsics.
The error is not significantly reduced, but it is less spatially correlated,
which is preferable for spatiotemporal denoising.

An illustration of $3\times 3$ and $4 \times 4$ footprints can be found
in our supplemental material, and $3\times 3$ is also visualized at the top
of Figure~\ref{fig_stochastic_footprints}.

\subsubsection{Pseudorandom Sparse Footprints}
\label{sec_stochastic_footprints}
\begin{wrapfigure}{R}{0.475\textwidth}
	\centering
	\hspace*{6mm}
	\begin{tabular}{cc}
	\includegraphics[width=0.15\columnwidth]{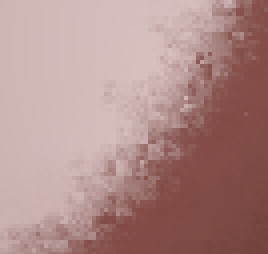} &
	\includegraphics[width=0.15\columnwidth]{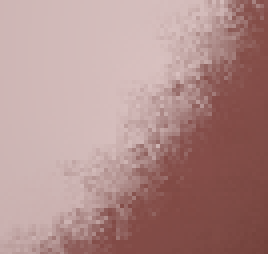}
    \end{tabular}
	\caption{Single frame crops of \textit{deterministic} (left) and \textit{sparse} (right) sharing footprints with 9 samples.
	The deterministic pattern yields visible square and edge discontinuities, while using the pseudorandom
	sparse pattern reduces the blocky appearance and resembles dithering.
	}
	\label{fig_structured_patterns}
\end{wrapfigure}
Square footprints 
are straightforward to implement and exhibit excellent reuse locality,
increasing the probability of successful reuse of texels between lanes.
However, with larger footprints, they can lead to visible structured square patterns,
such as those visible in the left part of Figure~\ref{fig_structured_patterns}.
The lanes at the edges and corners of the wave
filter the same texel values as their neighbors, leading to correlation
artifacts that are difficult to remove with spatiotemporal denoisers.
When a denoiser averages correlated pixels, the variance is not reduced.

\begin{figure}[tb]
	\centering
	\includegraphics[width=0.7\columnwidth]{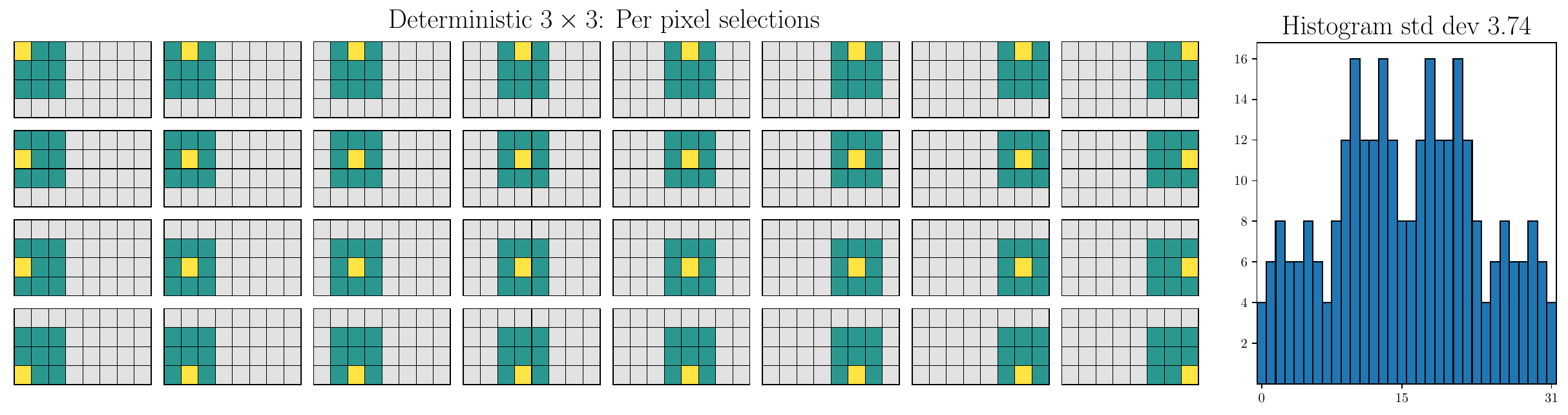}\\[1ex]
    \includegraphics[width=0.7\columnwidth]{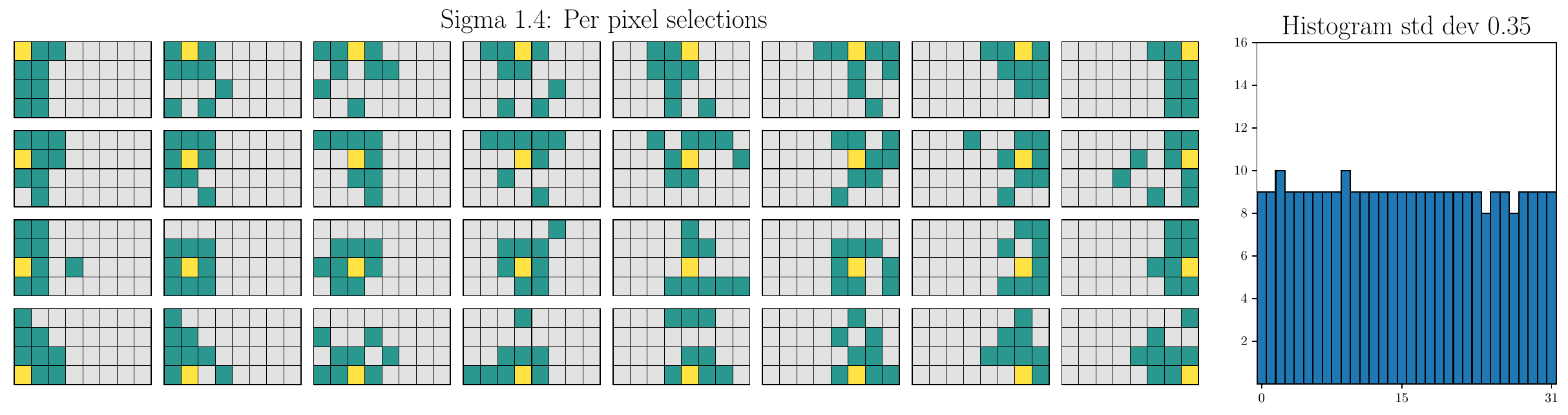}	
	\caption{\textbf{Top:} comparison of deterministic 9 sample sharing footprints ($3\times 3$ squares) and pseudorandom sparse footprints.
	The grid diagram (\textbf{left}) shows which wave elements samples (green) are reused by the given lane (yellow).
	The histogram (\textbf{right}) shows how many times a texel sample from the given lane is
	used by all the lanes in the wave.
	\textbf{Bottom:} sparse footprints balance sample reuse locality with uniform texel reuse contributions to break
	up visible structures.
	}
	\label{fig_stochastic_footprints}
\end{figure}

To alleviate this issue, we propose an alternative based on pseudorandom,
pre-generated sparse sharing footprints.
Figure~\ref{fig_stochastic_footprints} compares a regular $3\times 3$ square footprint
to pseudorandom, pre-generated sparse sharing footprints.
We balance lane locality (and in turn, locality of the UV coordinates at each lane),
where greater locality gives a higher chance of sharing samples, with dithering-like structures that
give a result that is easier to resolve with a spatiotemporal filter.
Furthermore, we reduce correlations between adjacent lanes by enforcing that none of the lanes are
overrepresented in the aggregate of the sharing footprints.
The right part of Figure~\ref{fig_stochastic_footprints} shows 
the histogram of each lane's usage for both a $3\times 3$ footprint and
a sparse footprint.
The $3\times 3$ footprint underrepresents the
texels sampled by the corner lanes and overrepresents the center ones, while
the sparse footprint provides higher uniformity.
This uniformity of texel use is beneficial as it avoids
repeated similar and correlated texels.

Figure~\ref{fig_structured_patterns} compares the resulting images of the two approaches.
Pseudorandom sparse footprints are more noisy, but the noise resembles dithering and is easier to
temporally resolve and denoise as we show in Section~\ref{sec_results_denoising}.
We describe a simple optimization-based method to generate those patterns in Section~\ref{sec:app-stochastic-footprint}.
In our implementation, we generate multiple different patterns that we cycle through over frames to further break up any visual structure.

\begin{figure}[t]
	\centering
	\includegraphics[width=\columnwidth]{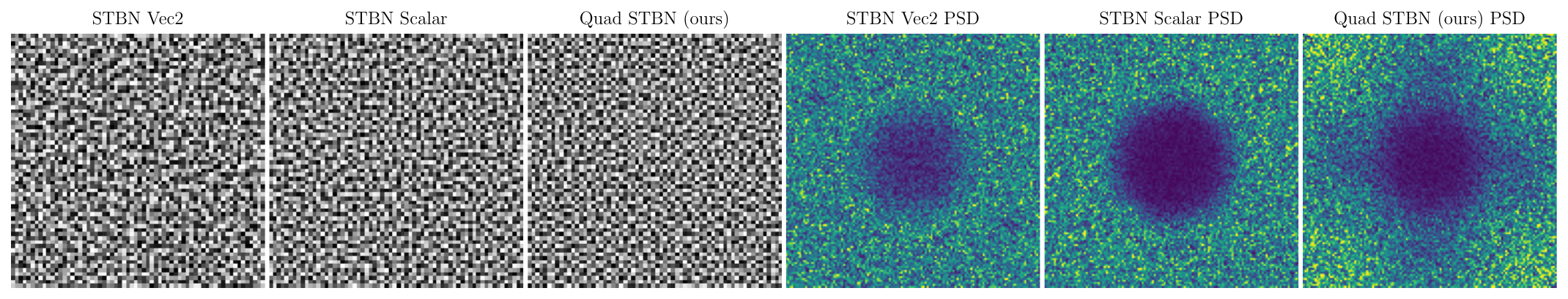}
	\caption{Comparison of the visual appearance and power spectra of various 
	STBN patterns--\texttt{Vec2} and scalar versions of the original STBN algorithm~\cite{Wolfe2022}
	as well as our proposed modification. The \texttt{Vec2} version has the worst blue noise
	properties and a significant amount of energy in the low frequencies.
	We observe in the rightmost power spectral density plot that our modification
	prioritizes diagonal frequencies and removes horizontal and vertical leftover patterns.}
	\label{fig_stbn_spectra}
\end{figure}

\subsection{Blue Noise Mask Improvements}
\label{sec_blue_noise}
The original STF technique recommended the use of spatiotemporal blue noise masks~\cite{Wolfe2022}.
Since our technique relies even more on neighboring sample diversity and because
of advancements in blue noise masks~\cite{donnelly:2023:filter}, we reevaluate their choice.

We found that the \texttt{Vec2} version of the STBN mask~\cite{Wolfe2022}
yields suboptimal results.
This can be explained by the algorithm used to generate those masks---simulated annealing~\cite{Wolfe2022}---rather than void and cluster,
yielding inferior blue noise properties.
We confirm this in the power spectra in Figure~\ref{fig_stbn_spectra}.
Our first modification is to simply concatenate independent scalar realizations from the void and cluster
algorithm.

Further, we adapt the STBN pattern to the unique needs of our algorithm.
While STBN masks optimize for blue noise value distribution and anti-correlation of a pixel
with all of its neighbors, we are reusing only some of the neighbors for wave sample sharing.
For instance, we can never sample across the wave boundaries.
In the specific case of $2 \times 2$ quad sharing, we always reuse samples between
neighbors on a fixed $2 \times 2$ grid.

Our proposed modification changes the void and cluster energy splatting function to
splat twice as much energy to the neighbors in each fixed $2 \times 2$ quad.
We visualize the resulting pattern and its power spectral density in Figure~\ref{fig_stbn_spectra}.
This modification removes some of the horizontal and vertical patterns and spectral
components and increases the noise energy in the spectrum diagonals.
We evaluate the impact of those modifications in Section~\ref{sec_results_blue_noise}.

\subsection{Implementation Details}
\label{sec_method_details}
Here we combine all introduced components of our algorithm in a GPU-friendly manner.
We focus on the more general and flexible wave intrinsics; see Figure~\ref{fig_implementation_code}

Our method starts with each lane sampling a single texel according to its filter,
following the standard STF approach.
Each lane then iterates over the lanes of its associated footprint.
In each loop iteration, we use {\texttt{WaveReadLaneAt()}} to fetch the information about texel samples from other lanes in the wave.
In particular, we obtain
texel values and their associated integer coordinates and the PMF value for sampling them.
The integer texel coordinates are needed to compute the current lane's filter sampling PMF.
The estimator introduced in Section~\ref{sec:estimators} is then used to compute the weight of each sample.
We accumulate the weighted texels as well as the weights, and divide the accumulated texels by the accumulated weights.
This can be expressed as follows.
\begin{figure}
\lstdefinestyle{mystyle}
{
	language = C,
	keywordstyle = [1]{\color{magenta}},
	keywordstyle = [2]{\color{violet}},
	morekeywords = [1]{float4, uint, int2},
	morekeywords = [2]{WaveReadLaneAt, GetFilterPMF},
	basicstyle=\scriptsize\ttfamily, %
	frame=single,
}
\begin{lstlisting}[language=C,  style=mystyle]
	// uv is in [0,1]^2 and txDim is the texture resolution.
	float texelFloatCoords = uv * txDim - float2(0.5f);
	// (...) One-tap STF sampling with access of one texel. 
	// texelFloatCoords are the original floating-point coordinates for filtering.
	// sampled_uv are the integer coordinates of the STF-fetched texel.
	// T is the fetched/decompressed single texel.
	float p_sampled_uv = GetFilterPMF(texelFloatCoords, sampled_uv);
	float4 T = texture[sampled_uv];
	
	float4 sum_w_i_T_i = 0.0f;
	float sum_w_i = 0.0f;
	for (uint i = 0; i < FOOTPRINT_SIZE; ++i) {
		uint laneIdx = waveLaneSet[currentLaneIdx][i];
		int2 uv_i = WaveReadLaneAt(sampled_uv, laneIdx);
		float p_i = WaveReadLaneAt(p_sampled_uv, laneIdx);
		float4 T_i = WaveReadLaneAt(T, laneIdx);
		float p_c = GetFilterPMF(texelFloatCoords, uv_i);
		float w_i = p_c / p_i;
		sum_w_i_T_i += w_i * T_i;
		sum_w_i += w_i;
	}
	return sum_w_i_T_i / sum_w_i;
\end{lstlisting}
\caption{Implementation of the core texel sharing algorithm with filtering
  using Equation~\ref{eq_bart_weighted_estimator}.}
\label{fig_implementation_code}
\end{figure}
In the code above, the texel sharing footprint is encoded as a lookup table in the \texttt{waveLaneSet} array.
If deterministic sharing footprints are used, the corresponding offsets may be computed at runtime. 
The number of loop iterations (\texttt{FOOTPRINT\_SIZE}) is based on the size of the sharing footprints---4, 9, or 16 in our experiments.
An example implementation of the {\textcolor{violet}{\texttt{GetFilterPMF}}} function is provided in the supplement (Section~\ref{app:filter-pdf}).

Our approach guarantees that the weight for the current lane is always equal to one due to the equality of the PMFs.
Conversely, if a texture sample of a neighbor falls outside of the original lane's filtering footprint,
the weight is zero, preventing incorrect contributions.
As a consequence, our method is robust when no wave sharing is possible:
if the magnification factor is insufficient or if wave neighbors sample texels that make no contribution to the current lane,
our algorithm returns exactly the same result as one-tap STF, since each lane is always included in its own footprint.

In our implementation, we repeat this loop for each sampled material texture for simplicity.
We note however, that in practice when multiple textures share the same UV coordinates and resolution,
the loop could be executed only once and the PMF and weight computations could be shared to minimize the amount of
shader code and to reduce the arithmetic instruction overhead.

\section{Results}
\label{sec_results}
We have implemented our technique in Falcor~\cite{Kallweit2022}.
Rendering starts with Falcor's \texttt{GBufferRaster}-pass, which outputs only depth and UVs,
similar to modern visibility buffer approaches~\cite{Haar:2015:gpu}.
The resulting buffer is fed to a custom compute shader pass, which
performs texture lookups and texel sharing before filtering texels and computing shading.
All results were rendered using an NVIDIA RTX 4090 GPU and all used Falcor
other than the performance measurements with neural texture compression in Section~\ref{sec_results_performance}.

\subsection{Comparison of Estimators}
\label{sec_results_estimators}
\begin{figure}[t]
	\centering
	\begin{minipage}{0.395\textwidth}
		\includegraphics[width=\columnwidth]{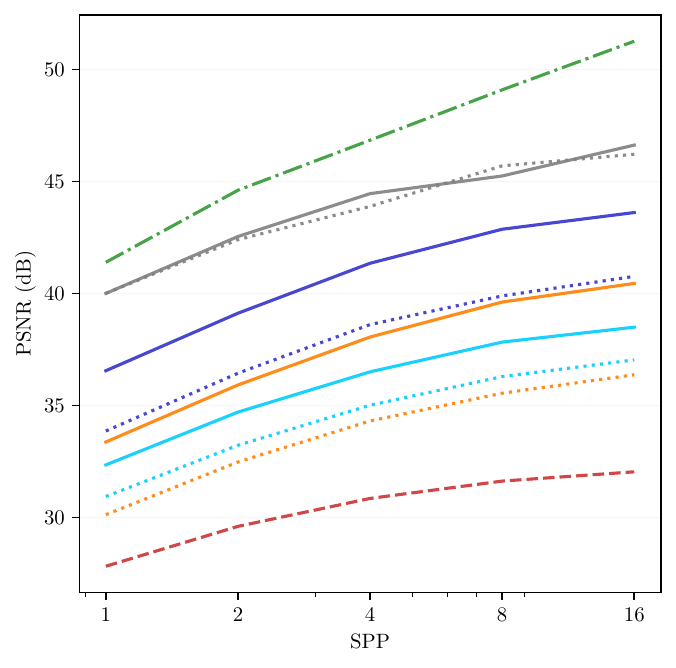}
	\end{minipage}	
	\begin{minipage}{0.595\textwidth}
		\newcommand{\myfactor}{0.0605}
		\setlength{\tabcolsep}{1.0pt}
		\includegraphics[width=\columnwidth]{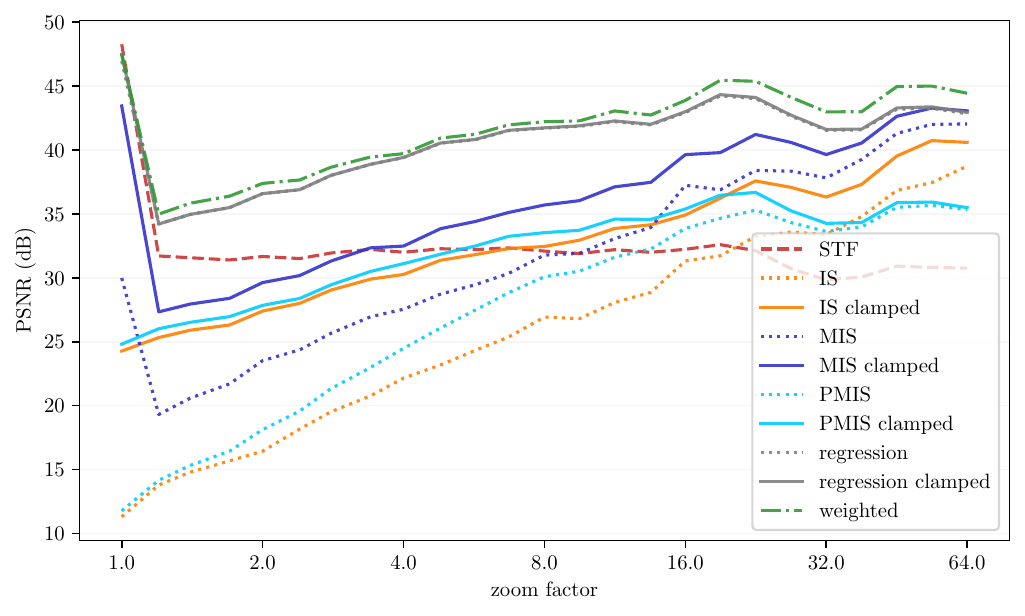}\\
		\hspace*{6mm}
		\begin{tabular}{cccccccccccccc}
			\includegraphics[height=\myfactor\textwidth]{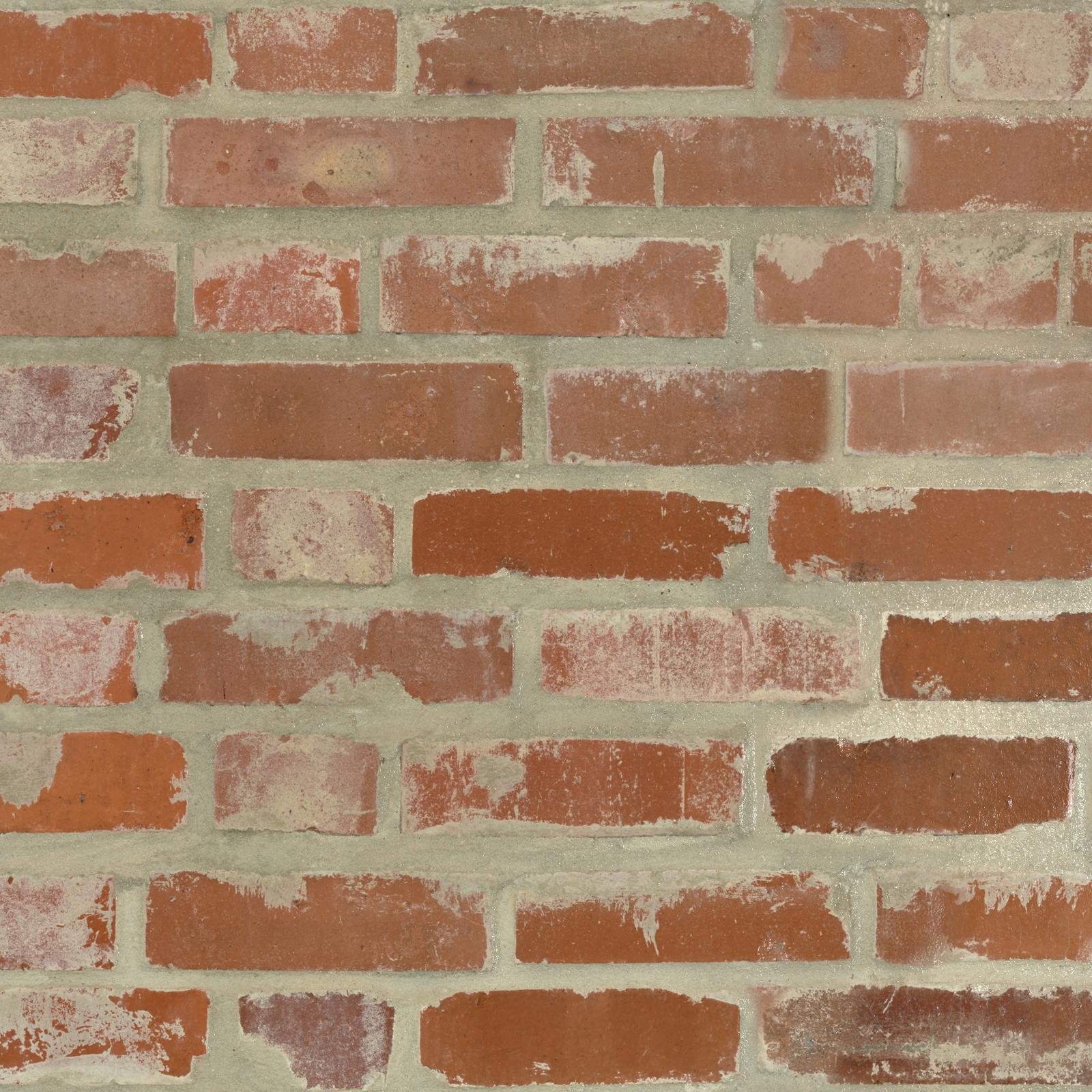} &
			\includegraphics[height=\myfactor\textwidth]{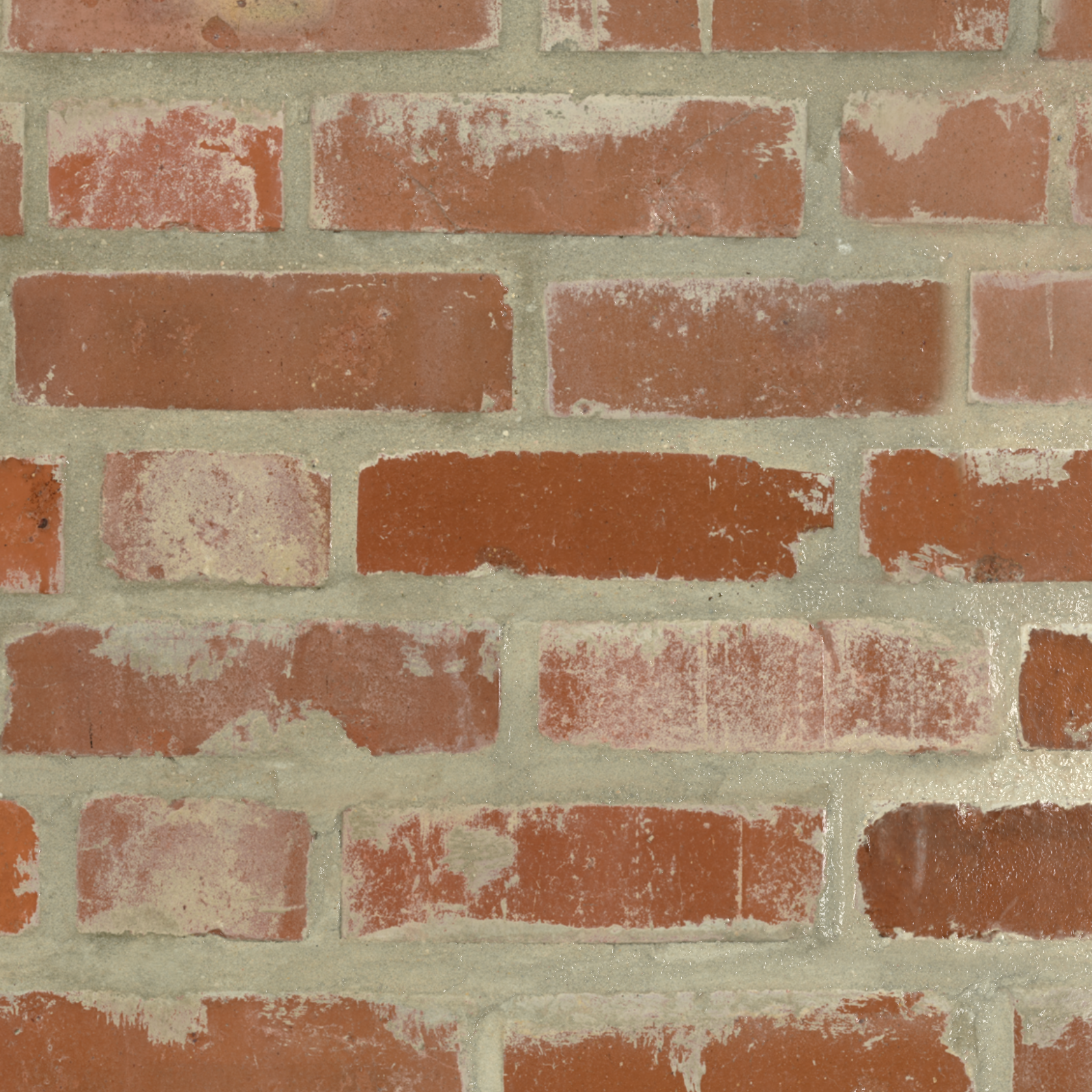} &
			\includegraphics[height=\myfactor\textwidth]{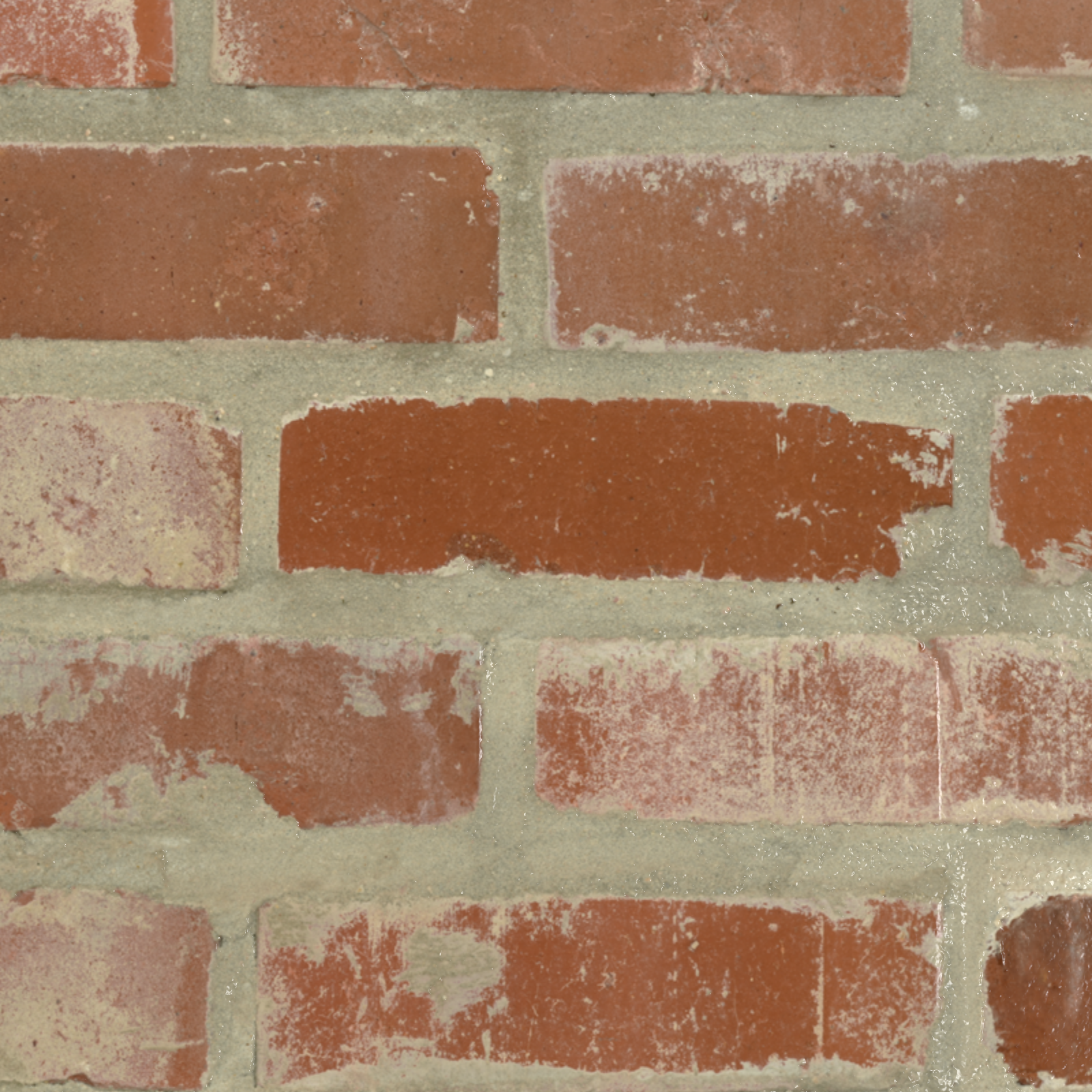} &
			\includegraphics[height=\myfactor\textwidth]{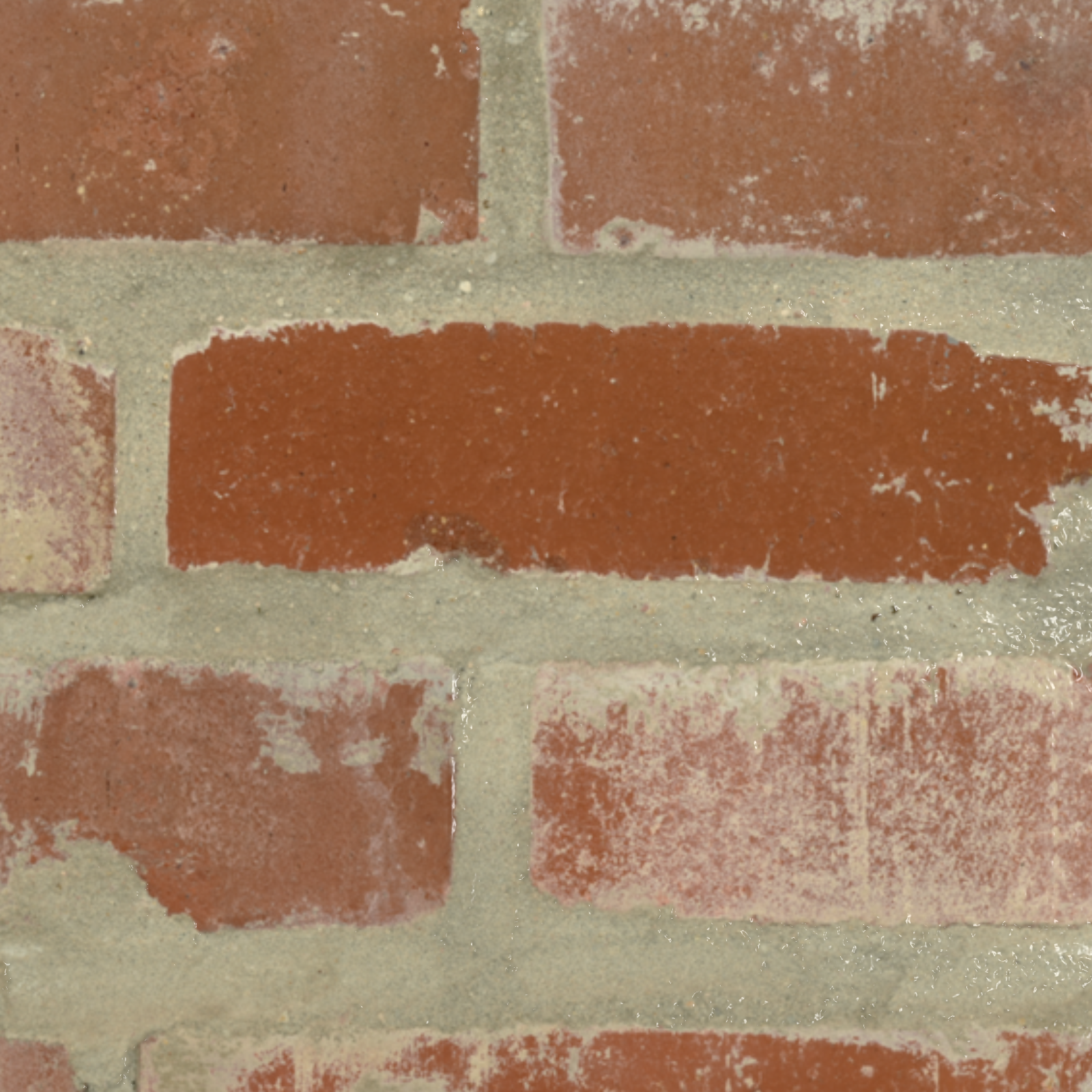} &
			\includegraphics[height=\myfactor\textwidth]{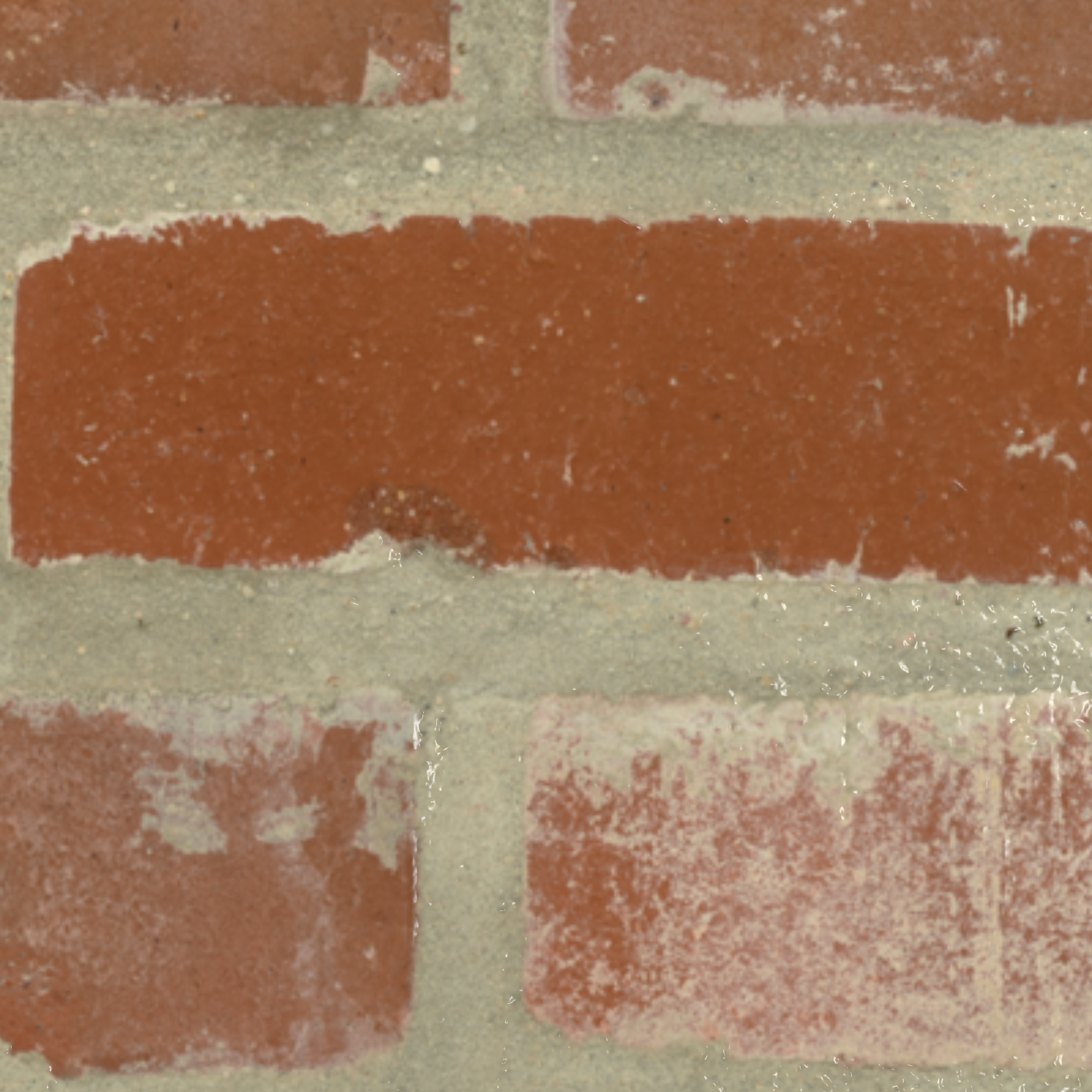} &
			\includegraphics[height=\myfactor\textwidth]{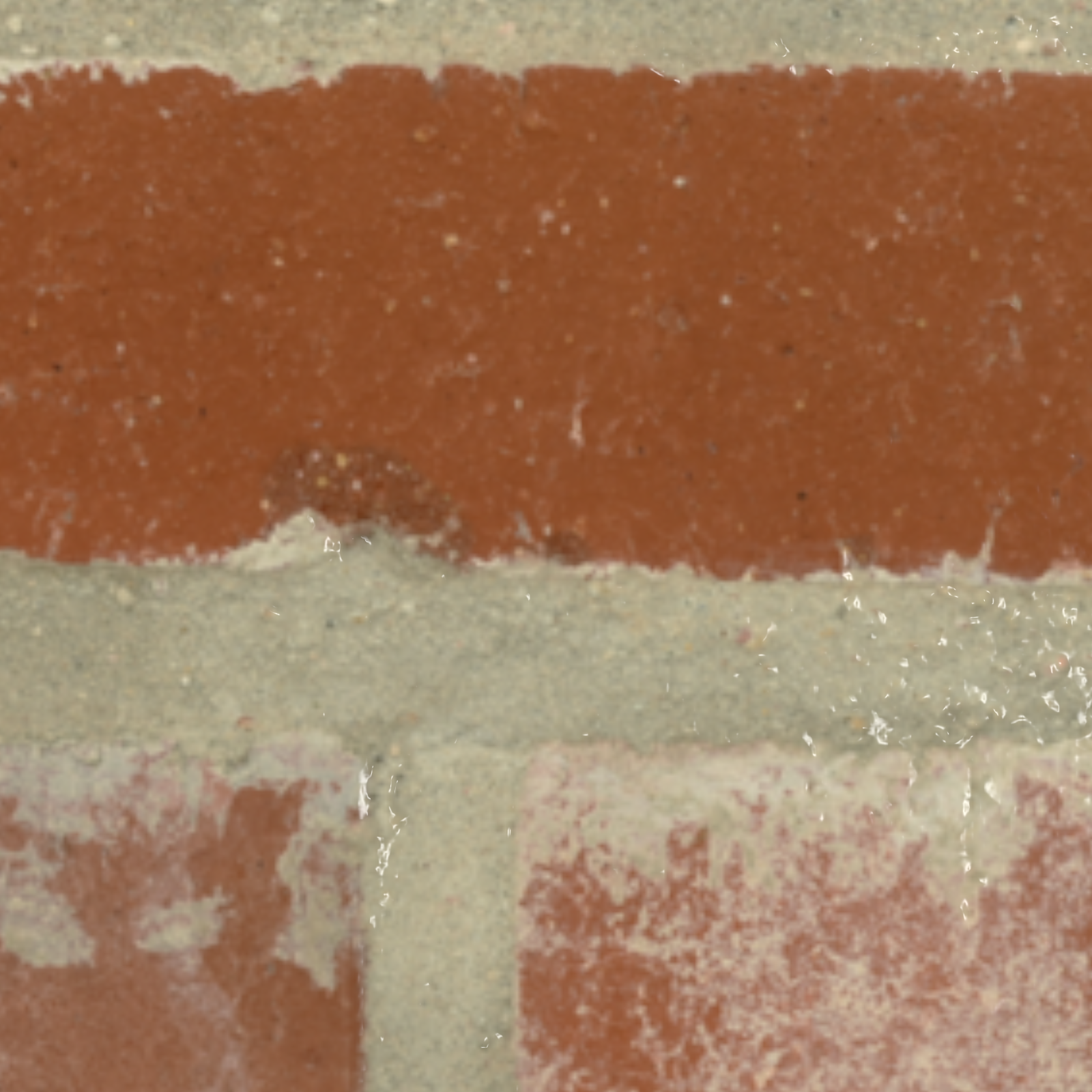} &
			\includegraphics[height=\myfactor\textwidth]{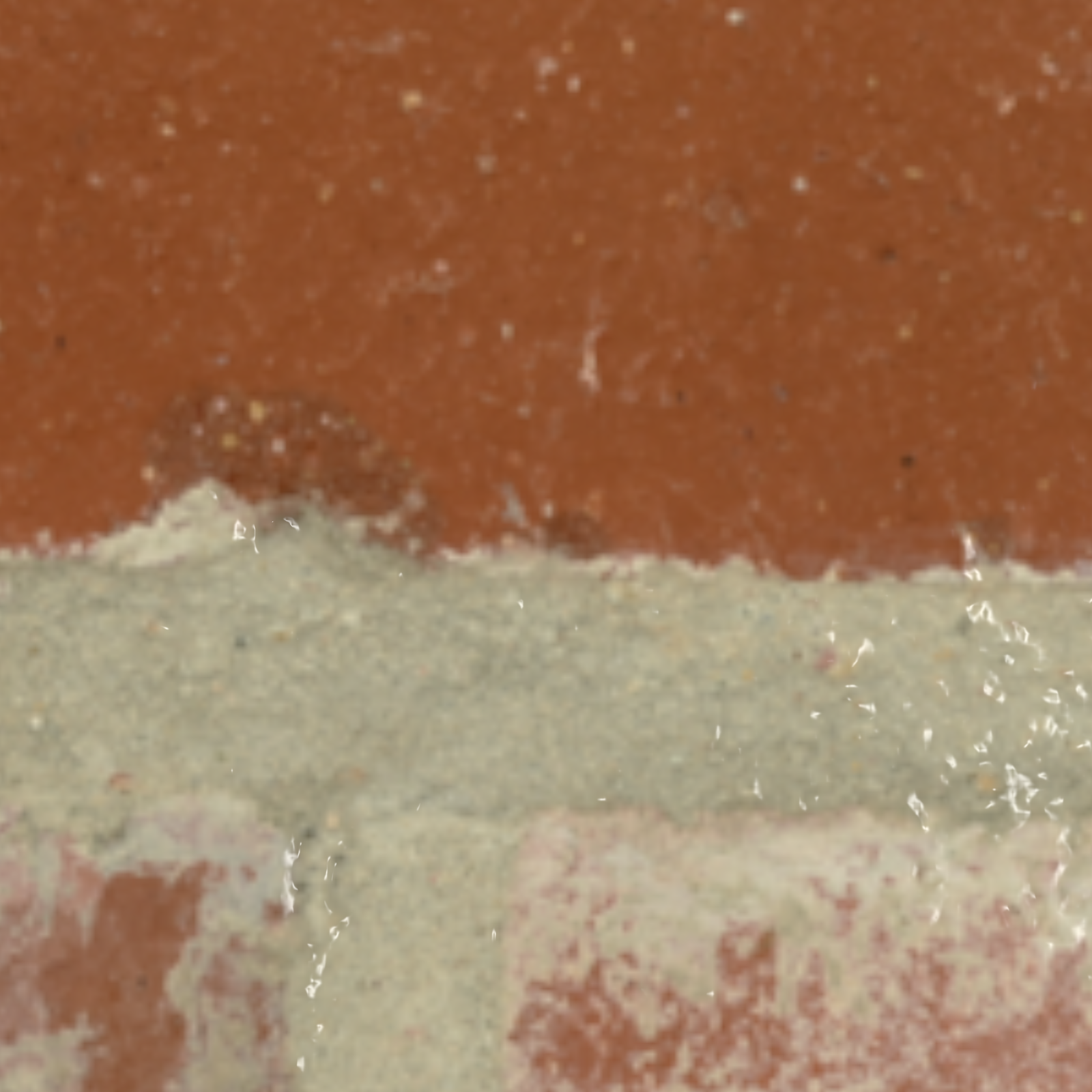} &
			\includegraphics[height=\myfactor\textwidth]{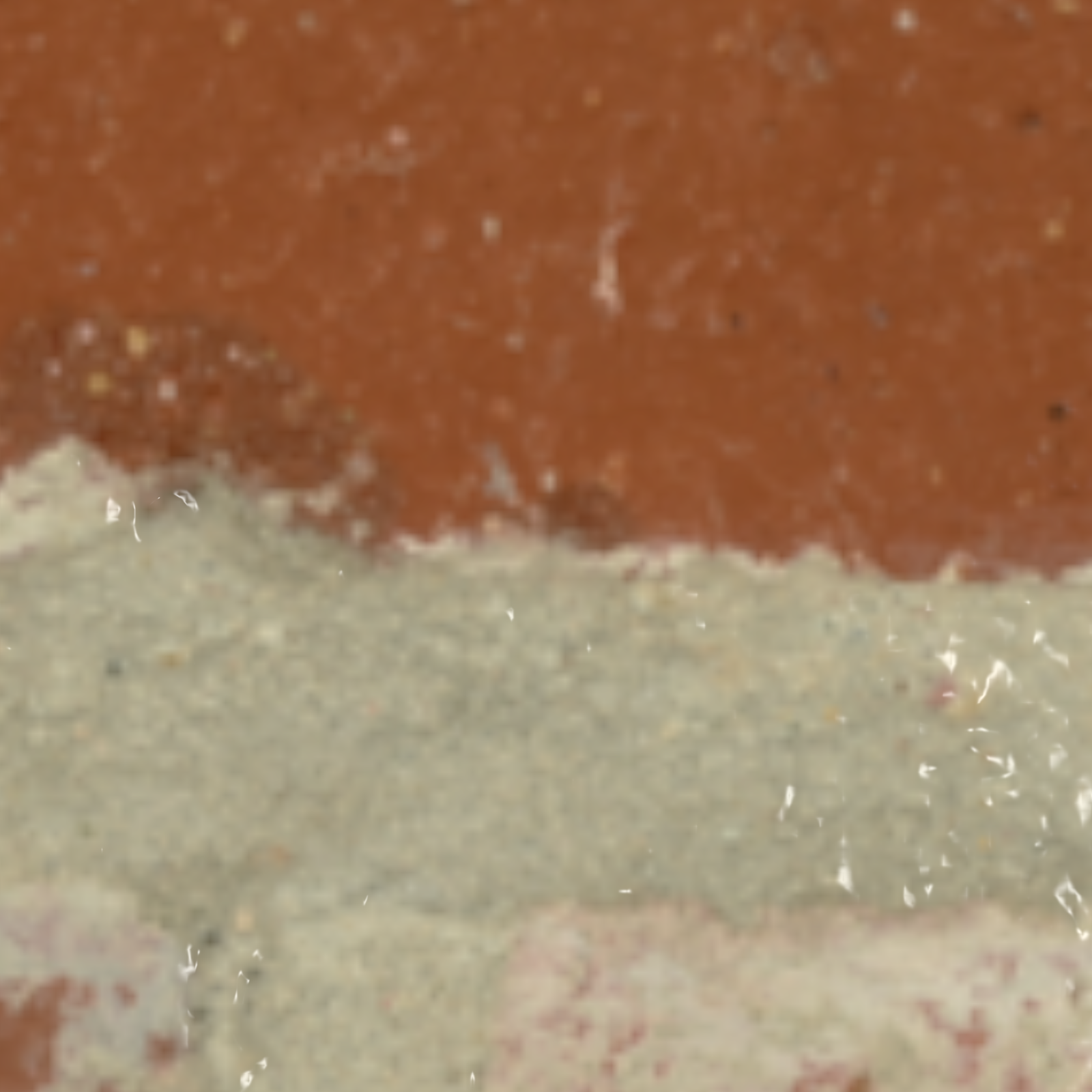} &
			\includegraphics[height=\myfactor\textwidth]{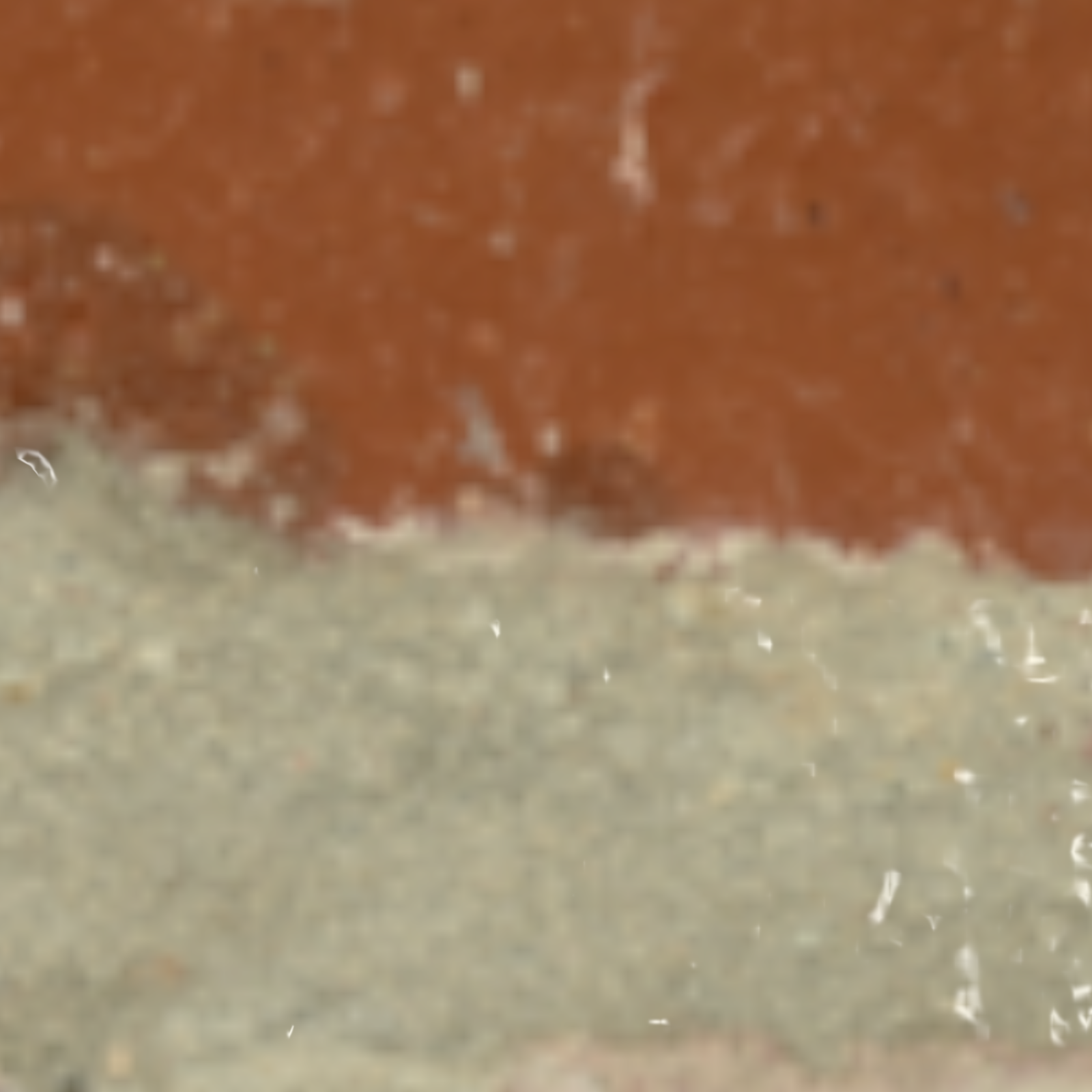} &
			\includegraphics[height=\myfactor\textwidth]{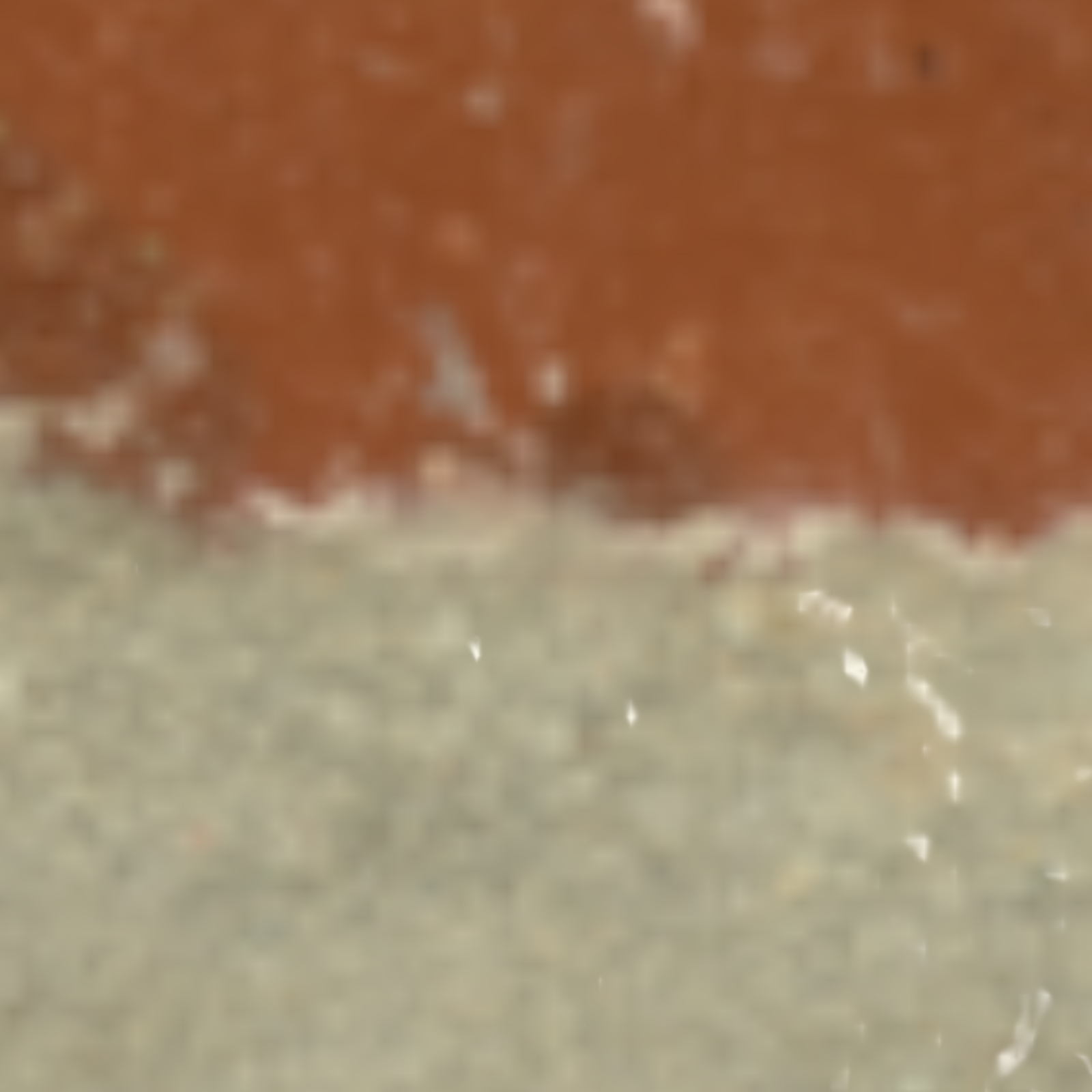} &
			\includegraphics[height=\myfactor\textwidth]{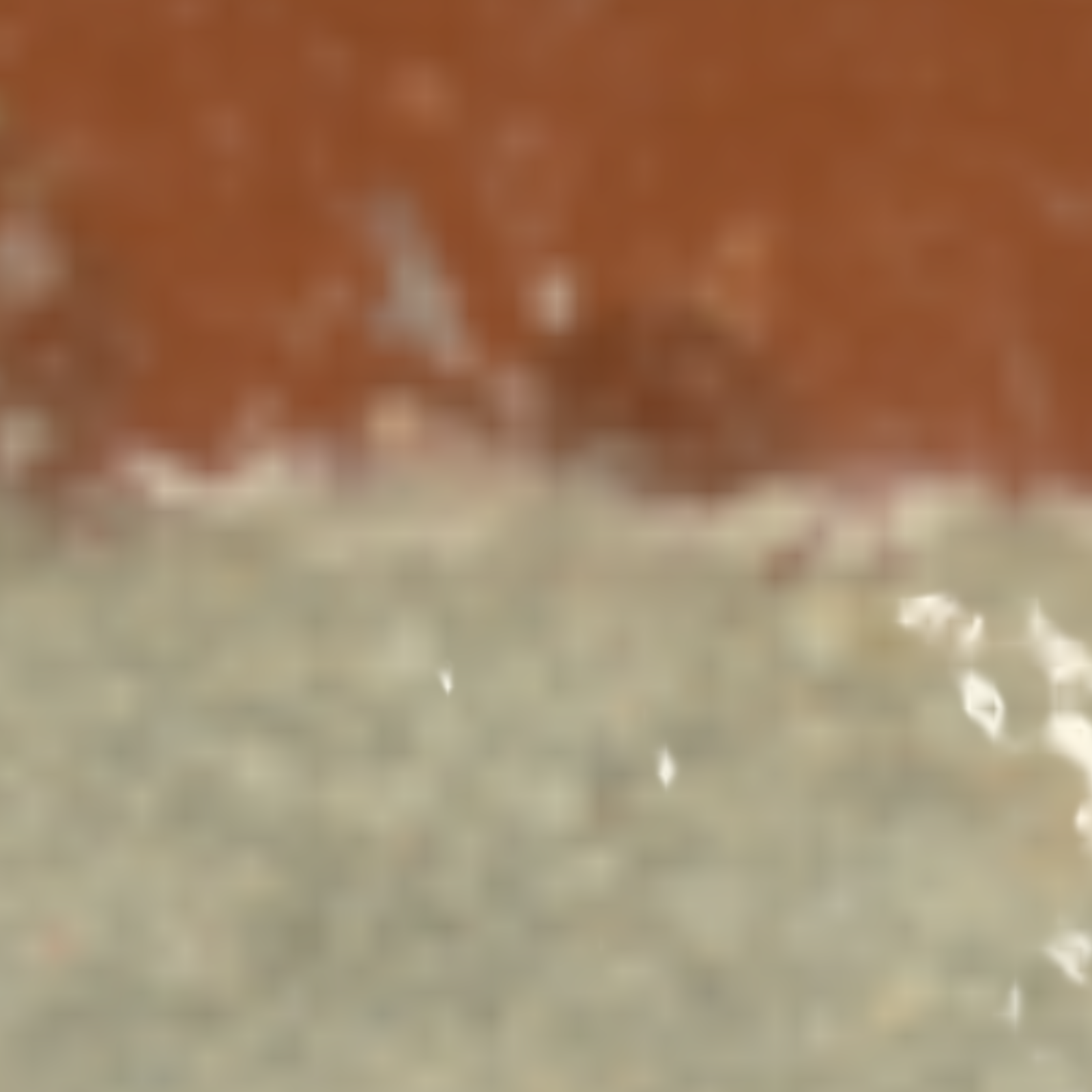} &
			\includegraphics[height=\myfactor\textwidth]{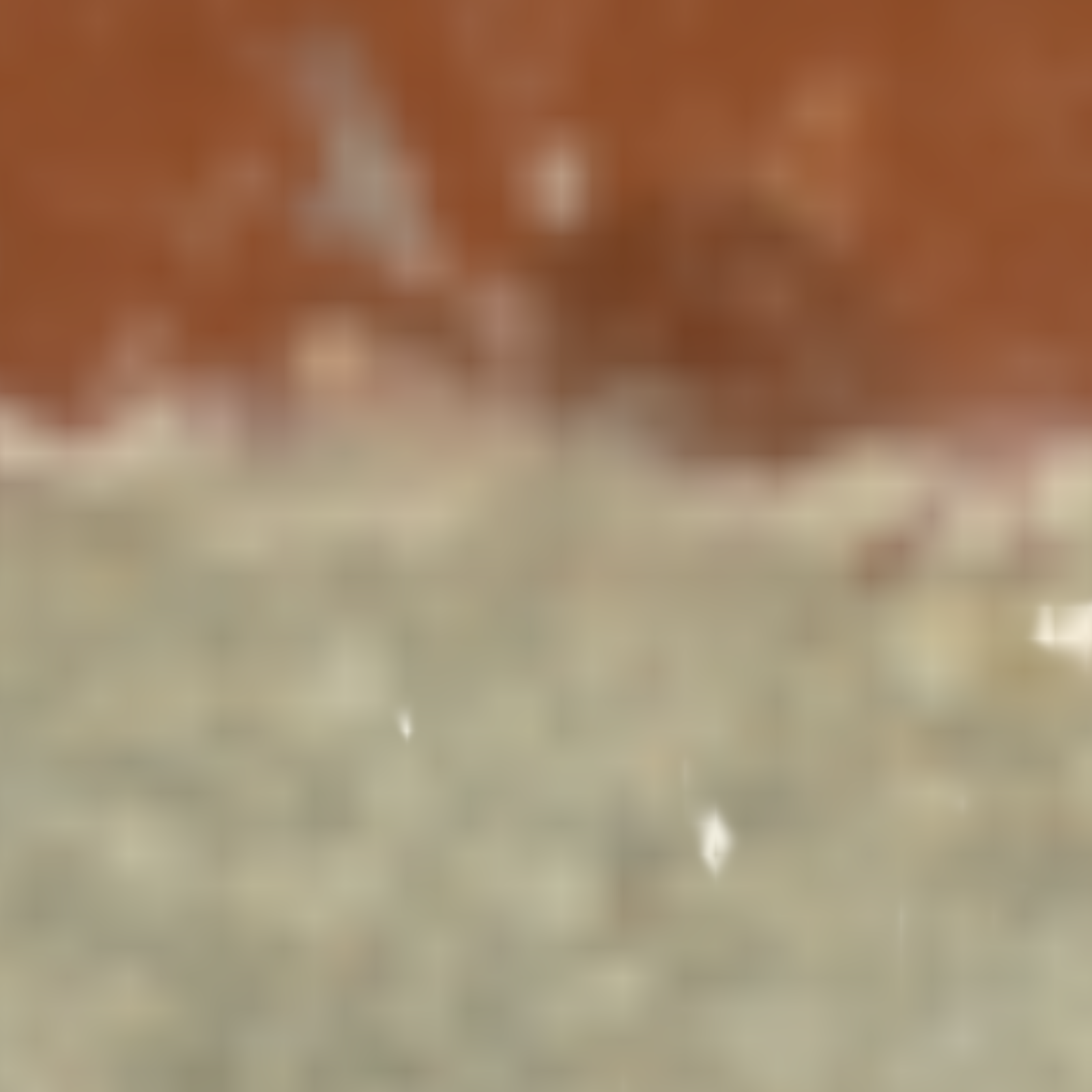} &
			\includegraphics[height=\myfactor\textwidth]{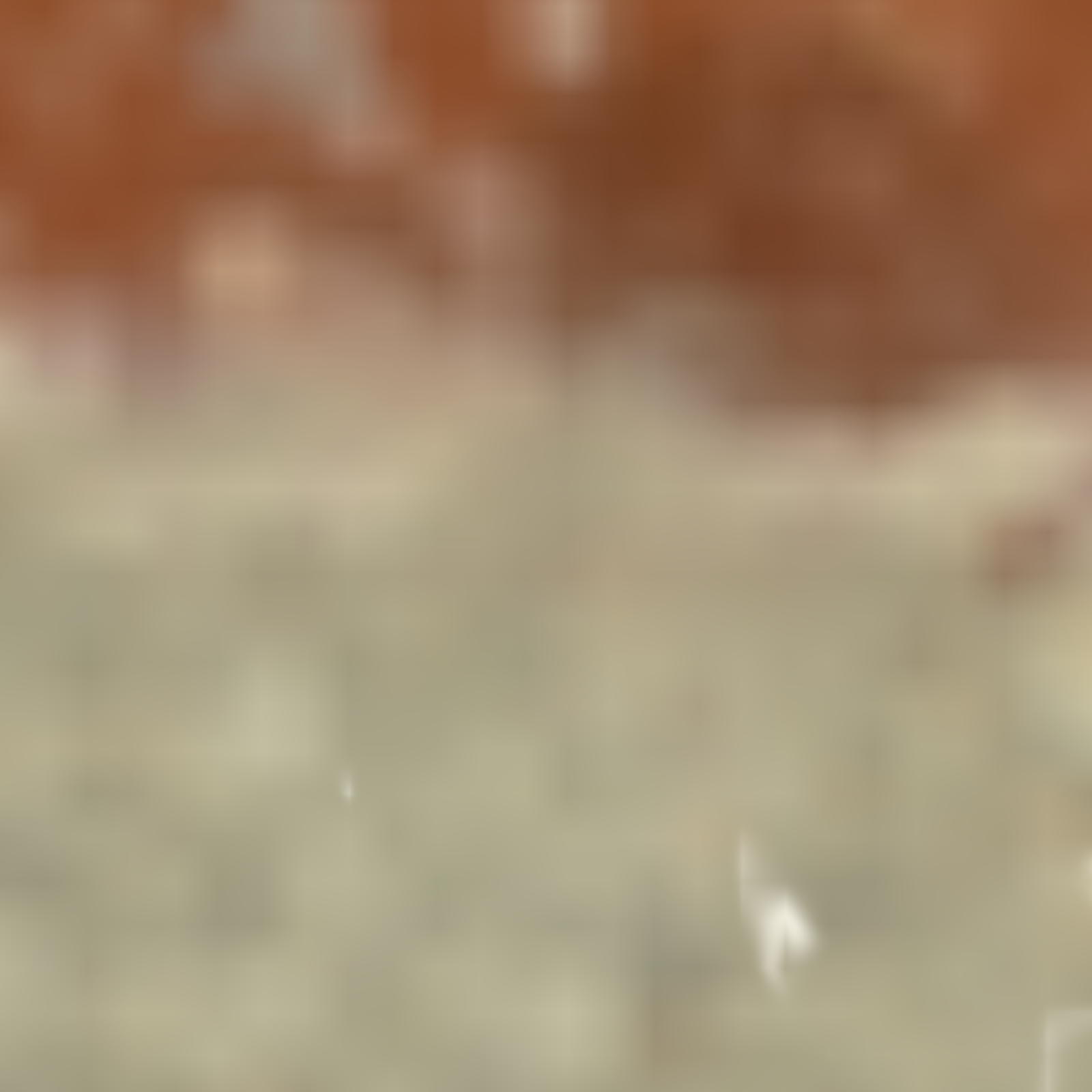}
		\end{tabular}
	\end{minipage}	
	\caption{\textbf{Left}: PSNR as a function of the number of samples per pixel (SPP)
		for the scene in Figure~\ref{fig_estimator_images}.
		The setup is high magnification, bilinear filtering,
		diffuse and specular normal mapping, and
		$3\times 3$ sharing footprints.
		The estimators we evalute include
		classic one-tap STF, clamped and non-clamped versions of
		standard importance sampling (IS), multiple importance sampling (MIS),
		pairwise MIS (PMIS), the regression estimator, and our new weighted estimator.
		Our weighted estimator provides the best results.
		\textbf{Right}: PNSR as a function of zoom factor with rendered images below.
		Note that a zoom factor of 1.0 gives a 1--1 mapping between texels and pixels, and a factor of 64 means a texel maps
		to $64\times 64$ pixels, for example.
	}
	\label{fig_estimators_vs_SPP}
\end{figure}

\begin{figure}[t]
	\newcommand{\myfactor}{0.16}
	\centering
	\setlength{\tabcolsep}{1.0pt}
	\renewcommand{\arraystretch}{0.7}
	\footnotesize{
		\begin{tabular}{cccccc}
		full image & STF & IS & MIS & PMIS & regression 
		\\
		\includegraphics[height=\myfactor\textwidth]{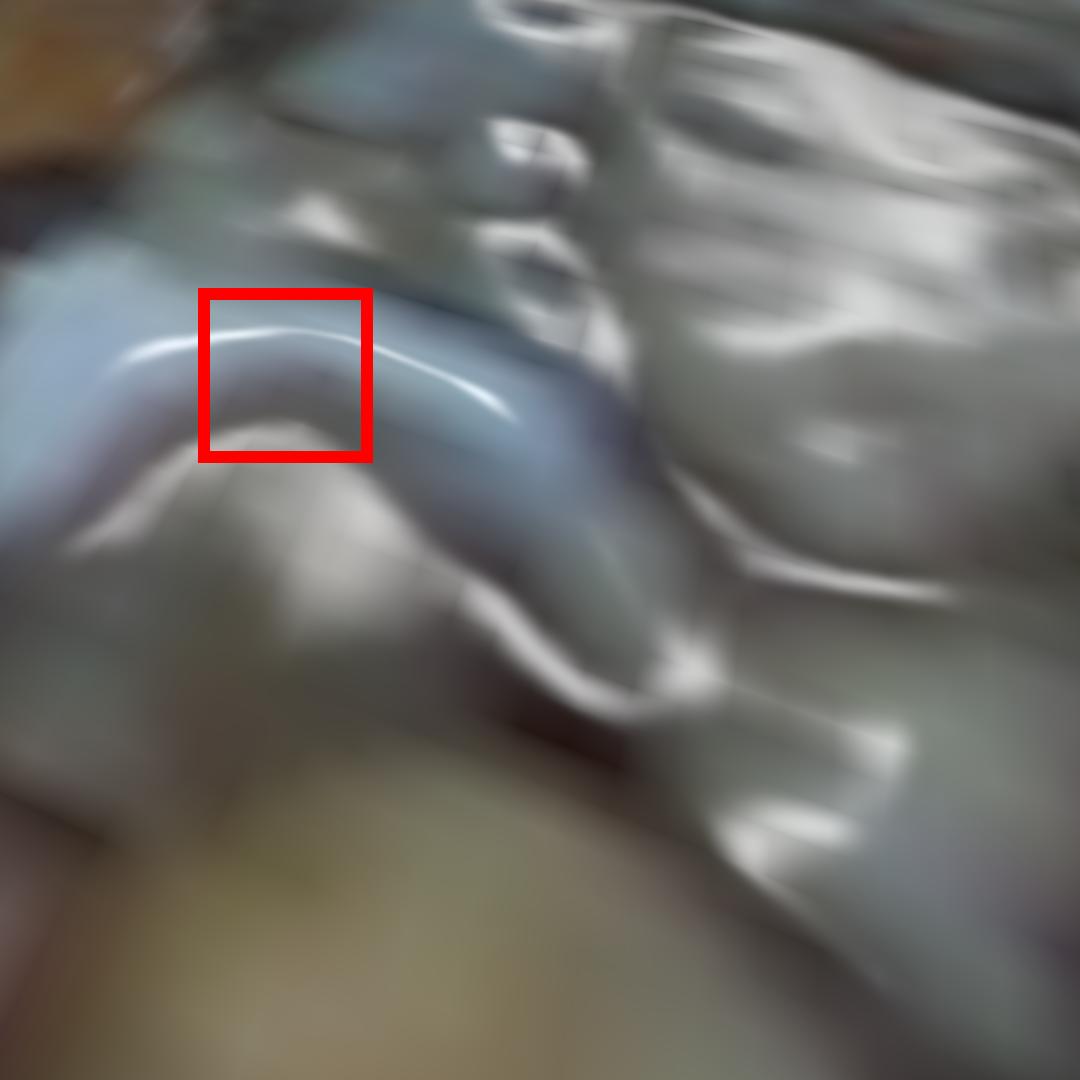} &
		\includegraphics[height=\myfactor\textwidth]{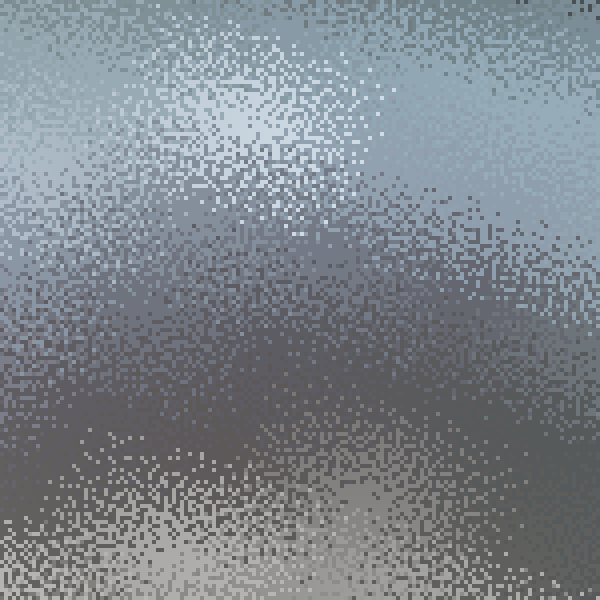} &
		\includegraphics[height=\myfactor\textwidth]{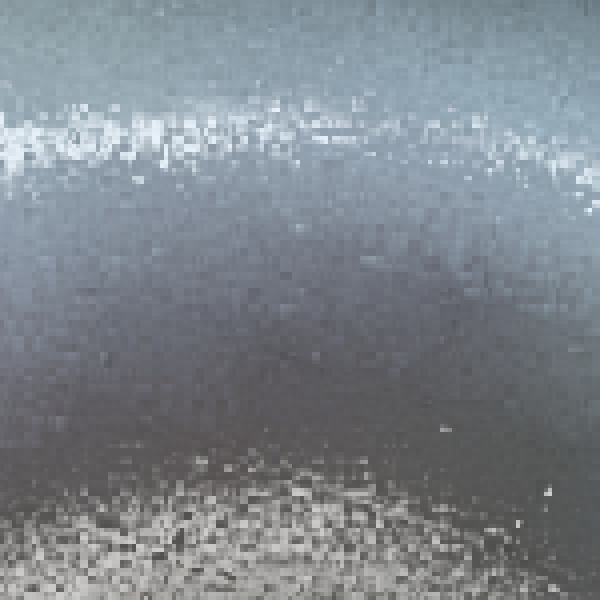} &
		\includegraphics[height=\myfactor\textwidth]{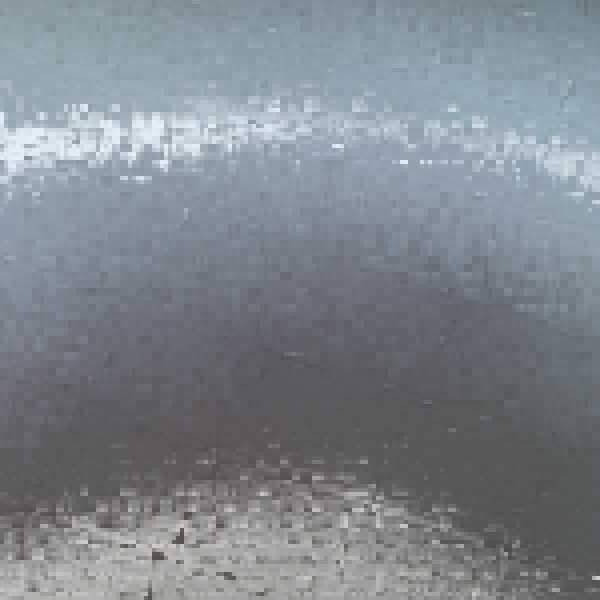} &
		\includegraphics[height=\myfactor\textwidth]{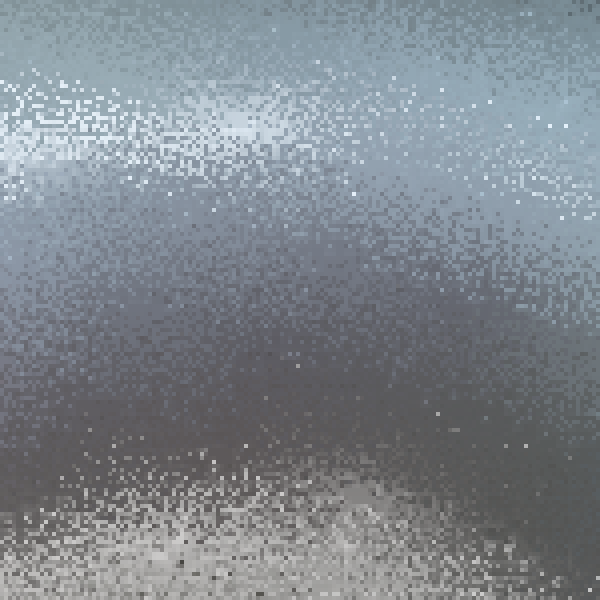} & 
		\includegraphics[height=\myfactor\textwidth]{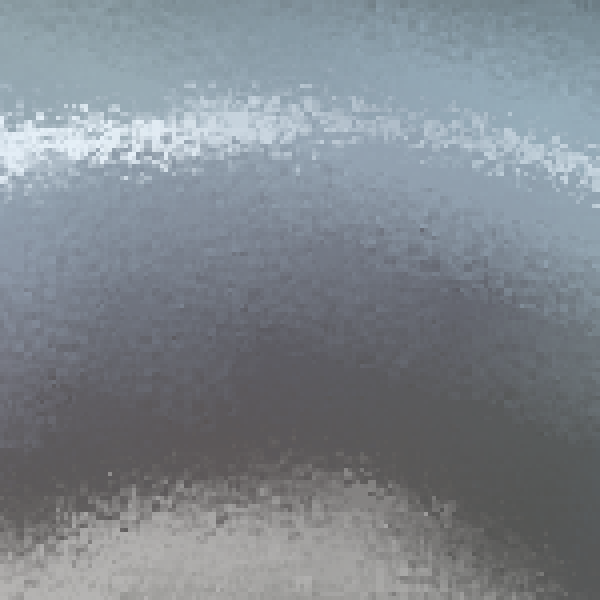} 
		\\
		\includegraphics[height=\myfactor\textwidth]{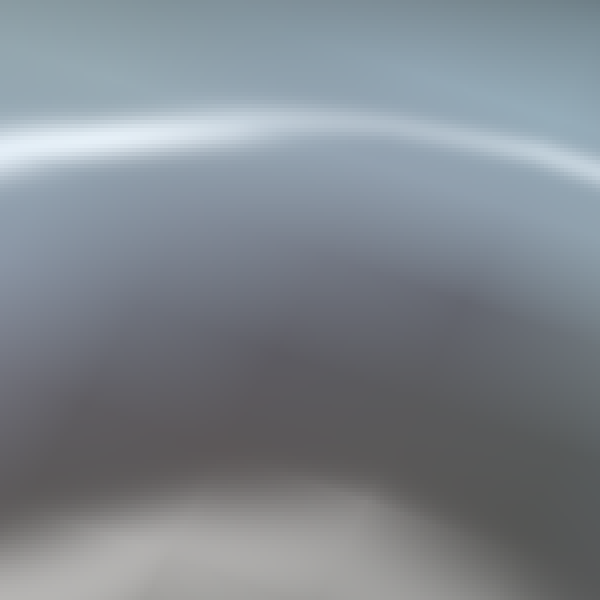} & 
		\includegraphics[height=\myfactor\textwidth]{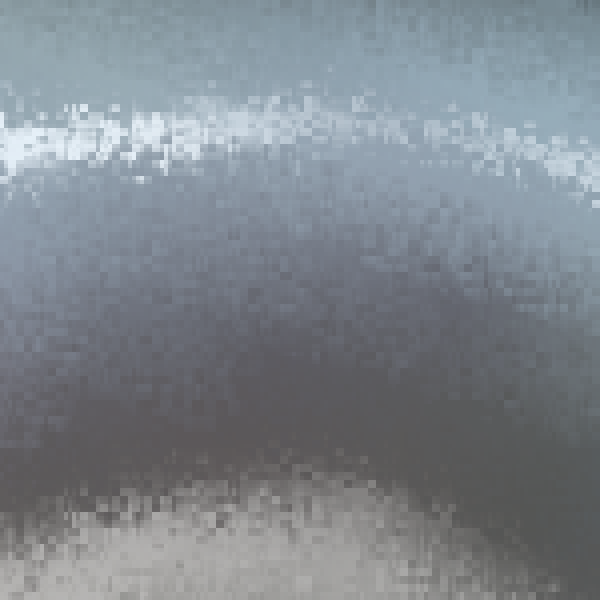} &
		\includegraphics[height=\myfactor\textwidth]{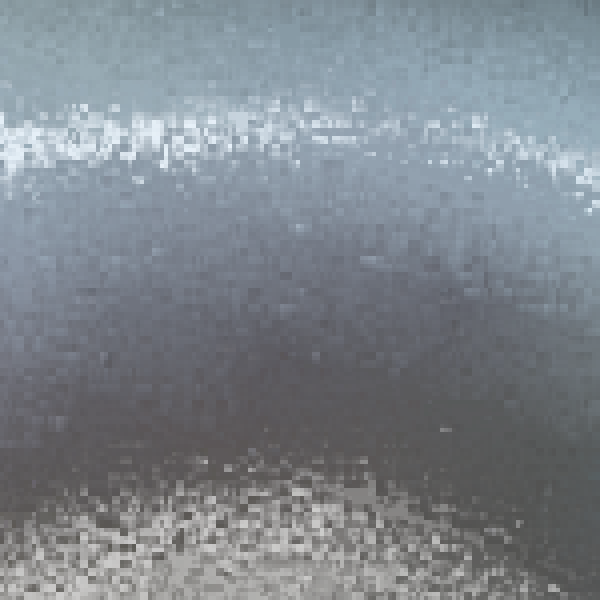} &
		\includegraphics[height=\myfactor\textwidth]{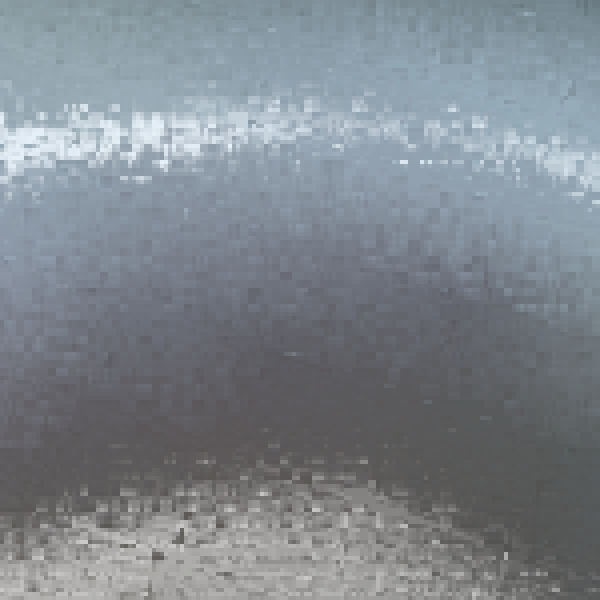} &
		\includegraphics[height=\myfactor\textwidth]{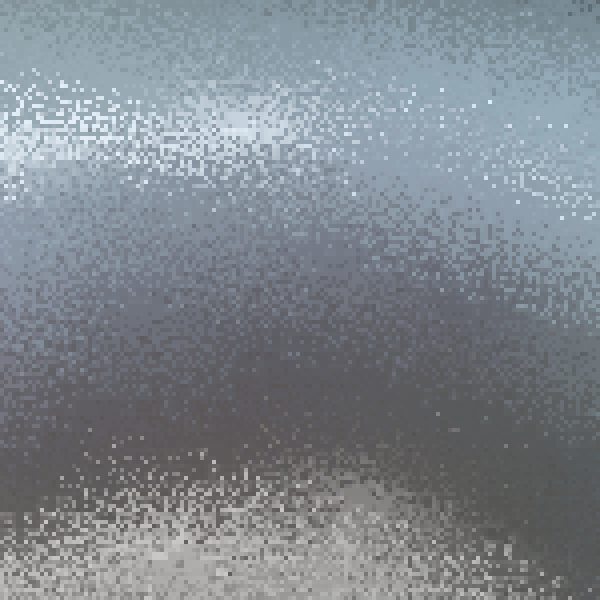} &
		\includegraphics[height=\myfactor\textwidth]{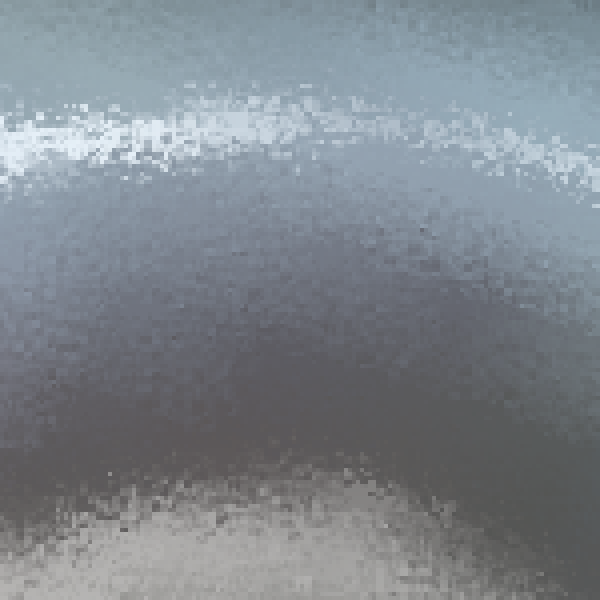}
		\\
		ground truth & our weighted & IS clamped & MIS clamped & PMIS clamped & regression clamped
		\end{tabular}
	}
	\caption{Visual results from the estimators with bilinear filtering. See the caption in
		Figure~\ref{fig_estimators_vs_SPP} for an explanation of the abbreviations.
		STF produces a poor highlight, IS and MIS
		are substantially more noisy than our weighted estimator, and PMIS
		misses part of the highlight. The regression estimator works well but has higher
		runtime cost than ours and only guarantees results within the range of the texel values if clamping is used.
	}
	\label{fig_estimator_images}
\end{figure}

We have evaluated the error in rendered images for a number of candidate estimators
(Section~\ref{sec_existing_estimators} and~\ref{sec:estimators}); the results are summarized
in Figure~\ref{fig_estimators_vs_SPP}.
The baseline is classic one-tap STF. We have included clamped and non-clamped versions of
standard importance sampling (IS), multiple importance sampling (MIS),
pairwise MIS (PMIS), and the regression estimator. Those are compared against our novel
weighted estimator (Equation~\ref{eq_bart_weighted_estimator}).
The left part of Figure~\ref{fig_estimators_vs_SPP}
shows PSNR (computed in linear RGB space) as a function of samples per pixel (SPP).
Figure~\ref{fig_estimator_images} shows visual
results from these estimators for the left part of Figure~\ref{fig_estimators_vs_SPP}.
Our weighted estimator consistently has both lowest numeric error and the best visual appearance with respect to the reference.

We have also evaluated how the degree of texture magnification affects error for the various estimators; 
results are shown on the right side of Figure~\ref{fig_estimators_vs_SPP}.
At zoom factors of one and slightly greater, there is little texture-space overlap of each pixel's filter.
This leads to the problematic case for importance sampling discussed in Section~\ref{sec_prevwork_mc} where one pixel
may sample a texel with low probability but then share it with a pixel where that texel has a high filter weight;
as a result, that estimator has high variance.
In contrast, methods like our weighted estimator essentially revert to one-tap STF's performance in that case,
which is all that can be expected when sharing is rarely successful.
As the zoom factor increases, the mismatch of PMFs between nearby pixels decreases and all methods give lower error
as sharing is successful more frequently.
This provides an empirical demonstration of our main insight---with increasing magnification factors,
adjacent pixels tend to share the same texel values and draw them from the same distributions, 
dramatically improving the reuse probability.
Throughout this range, our estimator has the lowest error and the clamped regression estimator is a close second.

In practice our estimator has lower computational cost than the regression estimator as it requires fewer arithmetic
operations and only a single loop over the samples rather than three.
Therefore, we will focus on our weighted estimator alone for the remainder of this section.

\begin{table}
	\caption{Quality measures, PSNR (in dB, where higher is better) and 
		 FLIP (lower is better), for different
		filter footprint configurations at 1~SPP.  `*' means \textit{exact filtering} (Section~\ref{sec:exact_filtering}) is being used;
		`q' means quad intrinsics are used; and `w' means wave intrinsics are used.
                All variants of our method have lower error than one-tap STF, with error decreasing more with larger sharing footprints
                and when exact filtering is used.}
	\footnotesize{
	\begin{tabular}{|c||c|c|c|c|c|c|c|c|c|c|c|}
		\hline
		& STF & $2\times 2$q  & $2\times 2$q* &$2\times 2$w  & $2\times 2$w* &
		$3\times 3$ & $3\times 3$* & PS $9\times$ & PS $9\times$* & $4\times 4$ & $4\times 4$*  \\
		\hline\hline
		\textbf{PSNR} ($\uparrow$) & 27.82 & 36.76 & 36.89 & 34.94 & 35.09 & 40.14 & 41.60 & 38.07 & 39.47 & 42.29 & 44.87 \\ \hline
		\textbf{FLIP} ($\downarrow$) & 0.0480 & 0.0311 & 0.0309 & 0.0342 & 0.0338 & 0.0258 & 0.0214 & 0.0268 & 0.0233 & 0.0243 & 0.0157
		\\
		\hline
	\end{tabular}
	}
	\label{table_filter_sizes}
\end{table}

\subsection{Evaluation of Sharing Footprints}
\label{sec_results_footprints}
Table~\ref{table_filter_sizes} shows the effect of varying the texel sharing footprint with a bilinear filter.
It includes square texel sharing footprints,  
$2\times 2$ quad and wave intrinsics,
$3\times 3$, $4\times 4$, as well as
a pseudorandom sparse footprint with $9$ texels (PS $9\times$) of sharing.
We also present results for \textit{exact filtering} (EF),
described in Section~\ref{sec:exact_filtering}. Our optimized quad STBN masks
were used for all results except $3\times 3$, which uses FAST EMA product
(see Section~\ref{sec_results_blue_noise}).

As expected, larger footprints tend to give lower error.
For $2\times 2$, there are rarely enough samples for EF to be applicable.
However, for larger footprints ($\geq 9$), EF reduces the error.
The pseudorandom sparse footprint with size 9 (PS $9\times$), is not quite
as good as a $3\times 3$ footprint, which is expected since it spreads out the
filter footprint compared to the square $3\times 3$.
However, this pattern works better when spatiotemporal denoising is used (Section~\ref{sec_results_denoising}).

\begin{figure}[t]
	\centering
	\includegraphics[width=0.55\columnwidth]{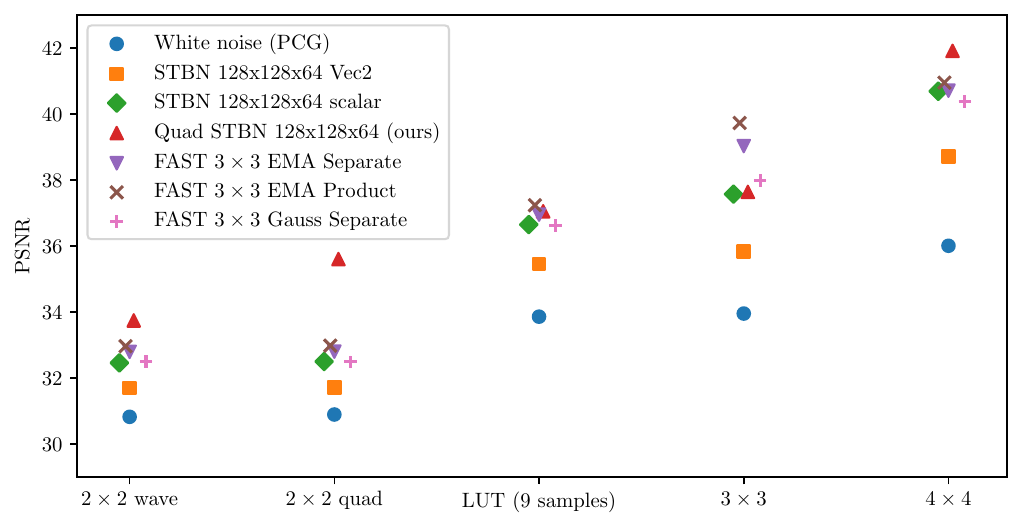}
	\caption{Impact of the pseudorandom number source on the PSNR score of our algorithm (non-denoised) for different footprints when used for bilinear filtering.}
	\label{fig_stbn}
\end{figure}

\subsection{Blue Noise Mask Evaluation}
\label{sec_results_blue_noise}
We have measured the effect of our modifications to the STBN pattern proposed in Section~\ref{sec_blue_noise}
to error in images and compared to FAST noise~\cite{donnelly:2023:filter}.
We present the quantitative results in Figure~\ref{fig_stbn}.
The improvement from our modified STBN pattern is dramatic for the quad sharing variants
and it also provides a marginally better PSNR than the alternatives for the $4 \times 4$ footprint.
FAST noise significantly improves the results for the $3 \times 3$ footprint, as it was optimized for a
$3 \times 3$ blurring kernel.
Therefore, we recommend the $3 \times 3$ FAST noise for the $3 \times 3$ footprint, 
while for all the other footprints we suggest using our modified STBN pattern.

\subsection{Bicubic B-Spline Filtering}
\label{sec_results_bicubic}
Our analysis so far has focused on the bilinear filter due to its ubiquity in real-time rendering
and to facilitate the adoption of STF as a drop-in replacement for the traditional hardware magnification
filter.
However, one of the benefits of STF is that higher-quality filters like bicubic or Gaussian filters may be used at no additional texture sampling cost.

\begin{figure}[t]
	\centering
	\includegraphics[width=\columnwidth]{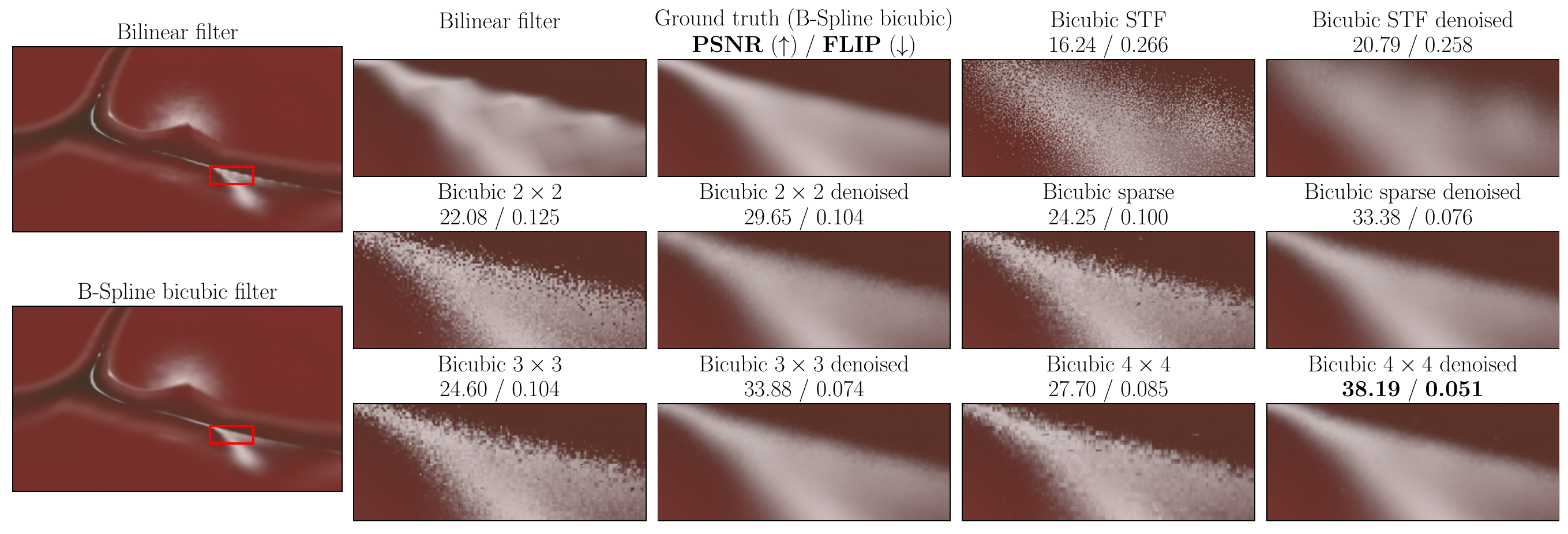}
	\caption{Use of our method with the B-spline bicubic filter, both for single frames and denoised with DLSS.
    We provide the bilinear-filtered image for comparison. 
    The B-spline bicubic filter removes unpleasant diagonal specular aliasing. The original STF method fails to reproduce sharp specular peaks.
    With increasing sharing footprint size the error decreases, especially on specular peaks.
    However, with large sharing footprints, starting with the $3\times 3$ square, visible square or streak structures start to appear.}
	\label{fig_bicubic}
\end{figure}

We evaluate our method with a bicubic B-spline filter. This filter is commonly used in offline rendering applications
due to its pleasant appearance and removing some diagonal aliasing of the bilinear filter.
Figure~\ref{fig_bicubic} shows results with different sharing pattern configurations, without and with DLSS spatiotemporal denoising.
In this case, one-tap STF fails to reproduce the shape of the specular highlight.
Our method, even with the $2\times 2$ square sharing footprint, significantly reduces the error and better preserves the highlight's appearance.
Increasing the footprint size further reduces the error, but turns noise into visible structured and square or streak artifacts.
We analyze this effect in the following section.

\subsection{Spatiotemporal Denoising}
\label{sec_results_denoising}
Our technique can have non-obvious interactions with spatiotemporal denoisers.
On the one hand, it significantly reduces the variance of individual pixels, making denoising easier.
Conversely, it introduces correlation between adjacent pixels due to texture sample reuse.
Denoisers typically assume uncorrelated (or inversely correlated, in the case of blue noise sampling) signals
and struggle with distinguishing between image structure and correlated noise.
Numerous temporal filtering techniques are used in production game engines~\cite{Yang:2020:Survey}, so for our evaluation we focus on two ends
of the complexity spectrum---a basic implementation of TAA in Falcor inspired by the Unreal Engine 4 TAA~\cite{Karis:2014:High},
as well as a modern machine-learning based approach of DLSS~\cite{Liu:2022:DLSS}.
DLSS comes with different quality profiles, from pure spatiotemporal denoising and antialiasing
(called \emph{DLAA}), to temporal frame upsampling.
Those additional profiles are commonly used to improve the frame rate at the cost of some visual quality.
To make our evaluation complete, we analyze behavior of our method combined with both DLAA and DLSS (with the Balanced profile, rendering at 58\% resolution).
While we describe our observations and include visual examples,
we encourage the reader to review our video in the supplementary material to see these effects in motion. Our video shows results using both TAA and the DLAA version of DLSS.

\begin{figure}[t]
	\centering
	\includegraphics[width=0.8\columnwidth]{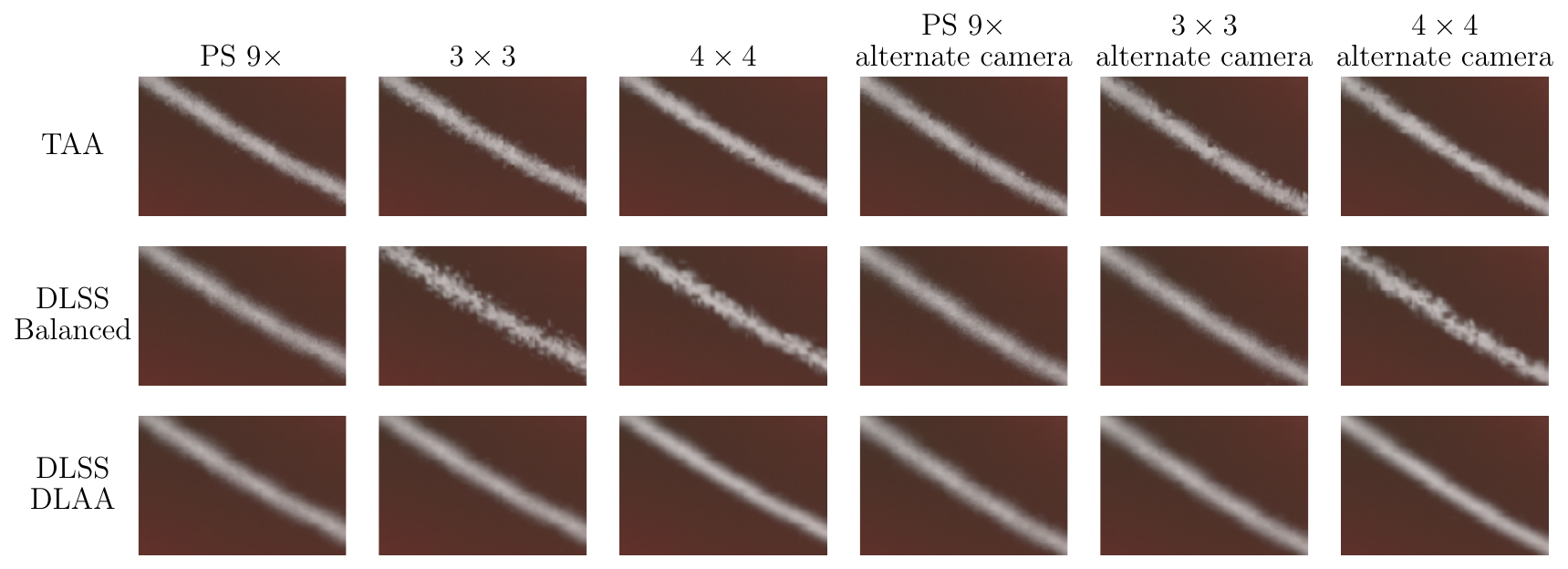}
	\caption{Results of \textbf{DLSS DLAA}, \textbf{DLSS Balanced}, and \textbf{TAA} with different footprints.
    TAA is consistently the most noisy but shows acceptable results with sparse footprints (PS $9\times$).
    DLSS Balanced struggles with the $3\times 3$ footprint at some camera angles and $4\times 4$ at all tested camera angles.
    DLAA consistently denoises all footprint configurations well and closely matches reference images.
    Across the compared denoisers, the sparse footprint shows consistently good results at all tested camera angles.
    }
	\label{fig_dlss_3x3_4x4_sparse_different_camera_angles}
\end{figure}

Our main finding is that with any sample reuse footprint, DLSS in the DLAA profile is able
to remove the residual noise effectively and with a good temporal stability.
On the other hand, the Balanced profile can struggle with larger regular footprints, such as $3\times 3$ or $4\times 4$.
While the $4\times 4$ footprint is consistently challenging, the $3\times 3$ footprint shows different behaviors depending
on the camera position---both in static and the dynamic scenes.
Addressing those artifacts was our main motivation for pseudorandom sparse footprints (Section~\ref{sec_stochastic_footprints});
we show their efficiency in Figure~\ref{fig_dlss_3x3_4x4_sparse_different_camera_angles}.
While sparse footprints are marginally noisier than the $3\times 3$ footprint, they do not show view-dependent inconsistency.

By comparison, TAA tends to consistently retain small amounts of residual noise of the bilinear filter with all footprints and camera trajectories.
Sparse footprints tend to help its noise reduction, but on average, it tends to be much noisier than DLSS.
With the bicubic filter, TAA produces severe banding-like visual artifacts caused by the local color bounding box clamping~\cite{Karis:2014:High}.

\subsection{Performance}
\label{sec_results_performance}
To test the computational performance of our method in one of the targeted use cases,
we measured frame time in a stand-alone renderer supporting neural
texture decompression~\cite{Vaidyanathan2023}. The rendering resolution was
$3840\times 2160$ and we zoomed in on a piece of geometry so that every pixel
had magnified textures. The material consisted of 9 texture channels split between three textures
(base color, normal map, material parameters). 
The baseline is one-tap STF. %
Our method with a $2\times 2$ footprint took just 0.04~ms longer than the baseline,
while $3\times 3$  and $4\times 4$ took 0.11~ms and 0.14~ms longer, respectively.
As a reference, we also attempted to decode $2\times 2$ texels needed
for a classic full bilinear filter. A full bilinear filter took more than $5\times$ the cost of a single sample,
instead of expected $4\times$ linear scaling, most likely due to high register pressure of NTC decompression.

\section{Conclusions and Future Work}
\label{sec_conclusions}

We have presented a significant improvement to STF under
magnification based on sharing texels among nearby pixels,
allowing a more accurate estimate of the filtered value.
The additional cost of our method is small and limited to arithmetic and wave register swizzling instructions,
i.e., there is no additional texture sampling cost or memory traffic.
Our method not only reduces stochastic noise, making post-rendering denoisers more effective,
but it also reduces the visual difference between filtering after shading and filtering before shading.
These properties make STF more attractive for existing game engines, and thus ease the adoption of novel compression algorithms
like NTC~\cite{Vaidyanathan2023} where the number of texels accessed directly affects performance.

Under magnification, our approach produces results closer to filtering before shading than filtering after shading.
While this is beneficial for many applications, Pharr et al.~\cite{Pharr2024} discuss cases where filtering
after shading is nevertheless preferred.
Our technique can be used selectively and not applied to those specific textures,
though it would be worthwhile in future work to investigate if we can get other benefits of our approach, 
such as noise and variance reduction, by applying some of our insights to already-shaded pixel values under magnification.

Developing and evaluating our algorithm, we found that a custom blue noise pattern can improve the quality
dramatically, up to 6dB PSNR difference. This result and insight encourages joint research of novel custom blue noise masks
together with other low-sample rendering techniques.

With our approach, wave lanes independently sample texels; within a sharing footprint, this may lead to some texels being sampled repeatedly
while other useful ones are not sampled at all.
It would be useful to find an efficient way to coordinate pixel texture samples to avoid such cases,
though we note it might be challenging since each lane has a unique set of other lanes that it draws texels from.
However, the benefits of a solution to this problem could go beyond improved image quality---especially under high magnification,
it may be possible to sample less than one texel per lane on average, thus improving performance of multi-texture sampling.

\begin{acks}
Many thanks to Pontus Ebelin for help with image \& video metrics and reviewing
of the paper.
Thanks also to Benedikt Bitterli for reading multiple drafts of the paper and
offering many helpful suggestions, especially related to how we presented
the connection between discrete and continuous MC estimators.
We would also like to thank Marco Salvi for thoughtful comments and feedback as well as for suggesting the randomization of sharing footprints and
Chris Wyman for proof-reading and suggestions on improving the structure
of the manuscript. 
We had many fruitful discussions during the development of this work Peter Morley, Johannes Deligiannis, Alexey Panteleev, Mike Songy, Nathan Hoobler,
and Homam Bahnassi, including a suggestion to use larger wave sharing footprints and
their evaluation of the covered techniques in existing production renderers and on real-world
game scenes.
We also thank Aaron Lefohn for continuous support of this work.
The Bricks 090 texture set was retrieved from \url{https://ambientcg.com/}
and the gravel texture set from \url{https://kaimoisch.com/free-textures/}.
\end{acks}

\bibliographystyle{ACM-Reference-Format}
\bibliography{references}


\begin{thebibliography}{32}


\ifx \showCODEN    \undefined \def \showCODEN     #1{\unskip}     \fi
\ifx \showDOI      \undefined \def \showDOI       #1{#1}\fi
\ifx \showISBNx    \undefined \def \showISBNx     #1{\unskip}     \fi
\ifx \showISBNxiii \undefined \def \showISBNxiii  #1{\unskip}     \fi
\ifx \showISSN     \undefined \def \showISSN      #1{\unskip}     \fi
\ifx \showLCCN     \undefined \def \showLCCN      #1{\unskip}     \fi
\ifx \shownote     \undefined \def \shownote      #1{#1}          \fi
\ifx \showarticletitle \undefined \def \showarticletitle #1{#1}   \fi
\ifx \showURL      \undefined \def \showURL       {\relax}        \fi
\providecommand\bibfield[2]{#2}
\providecommand\bibinfo[2]{#2}
\providecommand\natexlab[1]{#1}
\providecommand\showeprint[2][]{arXiv:#2}

\bibitem[Anderson and Moore(1979)]%
        {Anderson:1979:Optimal}
\bibfield{author}{\bibinfo{person}{Brian D.~O. Anderson} {and}
  \bibinfo{person}{John~B. Moore}.} \bibinfo{year}{1979}\natexlab{}.
\newblock \bibinfo{booktitle}{\emph{Optimal Filtering}}.
\newblock \bibinfo{publisher}{Prentice-Hall}.
\newblock


\bibitem[Bekaert et~al\mbox{.}(2000)]%
        {bekaert2000weighted}
\bibfield{author}{\bibinfo{person}{Philippe Bekaert}, \bibinfo{person}{Mateu
  Sbert}, {and} \bibinfo{person}{Yves~D Willems}.}
  \bibinfo{year}{2000}\natexlab{}.
\newblock \showarticletitle{{Weighted Importance Sampling Techniques for
  {M}onte {C}arlo radiosity}}. In \bibinfo{booktitle}{\emph{Eurographics
  Workshop on Rendering}}. Springer, \bibinfo{pages}{35--46}.
\newblock


\bibitem[Bitterli(2022)]%
        {Bitterli2022}
\bibfield{author}{\bibinfo{person}{Benedikt Bitterli}.}
  \bibinfo{year}{2022}\natexlab{}.
\newblock \bibinfo{title}{Correlations and Reuse for Fast and Accurate
  Physically Based Light Transport}.
\newblock
\newblock


\bibitem[Bitterli et~al\mbox{.}(2020)]%
        {bitterli2020spatiotemporal}
\bibfield{author}{\bibinfo{person}{Benedikt Bitterli}, \bibinfo{person}{Chris
  Wyman}, \bibinfo{person}{Matt Pharr}, \bibinfo{person}{Peter Shirley},
  \bibinfo{person}{Aaron Lefohn}, {and} \bibinfo{person}{Wojciech Jarosz}.}
  \bibinfo{year}{2020}\natexlab{}.
\newblock \showarticletitle{Spatiotemporal Reservoir Resampling for Real-time
  Ray Tracing with Dynamic Direct Lighting}.
\newblock \bibinfo{journal}{\emph{ACM Transactions on Graphics (SIGGRAPH)}}
  \bibinfo{volume}{39}, \bibinfo{number}{4} (\bibinfo{date}{July}
  \bibinfo{year}{2020}), \bibinfo{pages}{148:1--17}.
\newblock
\urldef\tempurl%
\url{https://doi.org/10.1145/3386569.3392481}
\showDOI{\tempurl}


\bibitem[Dongarra(2022)]%
        {Dongarra2022}
\bibfield{author}{\bibinfo{person}{Jack Dongarra}.}
  \bibinfo{year}{2022}\natexlab{}.
\newblock \bibinfo{title}{{A Not So Simple Matter of Software}}.
\newblock \bibinfo{howpublished}{Turing Award Keynote}.
\newblock
\newblock
\shownote{\url{https://www.hpcwire.com/2022/11/16/jack-dongarra-a-not-so-simple-matter-of-software/}}.


\bibitem[Donnelly et~al\mbox{.}(2024)]%
        {donnelly:2023:filter}
\bibfield{author}{\bibinfo{person}{William Donnelly}, \bibinfo{person}{Alan
  Wolfe}, \bibinfo{person}{Judith B{\"u}tepage}, {and} \bibinfo{person}{Jon
  Vald{\'e}s}.} \bibinfo{year}{2024}\natexlab{}.
\newblock \showarticletitle{Filter-Adapted Spatiotemporal Sampling for
  Real-Time Rendering}.
\newblock \bibinfo{journal}{\emph{Proceedings of the ACM on Computer Graphics
  and Interactive Techniques}} \bibinfo{volume}{7}, \bibinfo{number}{1}
  (\bibinfo{year}{2024}), \bibinfo{pages}{13:1--16}.
\newblock


\bibitem[Georgiev and Fajardo(2016)]%
        {georgiev:2016:blue}
\bibfield{author}{\bibinfo{person}{Iliyan Georgiev} {and}
  \bibinfo{person}{Marcos Fajardo}.} \bibinfo{year}{2016}\natexlab{}.
\newblock \showarticletitle{Blue-Noise Dithered Sampling}.
\newblock In \bibinfo{booktitle}{\emph{ACM SIGGRAPH 2016 Talks}}.
  \bibinfo{pages}{1--1}.
\newblock


\bibitem[Haar and Aaltonen(2015)]%
        {Haar:2015:gpu}
\bibfield{author}{\bibinfo{person}{Ulrich Haar} {and}
  \bibinfo{person}{Sebastian Aaltonen}.} \bibinfo{year}{2015}\natexlab{}.
\newblock \showarticletitle{GPU-driven Rendering Pipelines}.
\newblock \bibinfo{journal}{\emph{SIGGRAPH Advances in Real-Time Rendering in
  Games course}} (\bibinfo{year}{2015}).
\newblock


\bibitem[Handscomb(1964)]%
        {Handscomb1964}
\bibfield{author}{\bibinfo{person}{David~Christopher Handscomb}.}
  \bibinfo{year}{1964}\natexlab{}.
\newblock \showarticletitle{Remarks on a Monte Carlo Integration Method}.
\newblock \bibinfo{journal}{\emph{Numer. Math.}} \bibinfo{volume}{6},
  \bibinfo{number}{1} (\bibinfo{year}{1964}), \bibinfo{pages}{261--268}.
\newblock
\urldef\tempurl%
\url{https://doi.org/10.1007/BF01386074}
\showDOI{\tempurl}


\bibitem[Hesterberg(1995)]%
        {Hesterberg1995WeightedAI}
\bibfield{author}{\bibinfo{person}{Tim Hesterberg}.}
  \bibinfo{year}{1995}\natexlab{}.
\newblock \showarticletitle{Weighted Average Importance Sampling and Defensive
  Mixture Distributions}.
\newblock \bibinfo{journal}{\emph{Technometrics}}  \bibinfo{volume}{37}
  (\bibinfo{year}{1995}), \bibinfo{pages}{185--194}.
\newblock
\urldef\tempurl%
\url{https://api.semanticscholar.org/CorpusID:122839484}
\showURL{%
\tempurl}


\bibitem[Hofmann et~al\mbox{.}(2021)]%
        {Hofmann2021}
\bibfield{author}{\bibinfo{person}{Nikolai Hofmann}, \bibinfo{person}{Jon
  Hasselgren}, \bibinfo{person}{Petrik Clarberg}, {and} \bibinfo{person}{Jacob
  Munkberg}.} \bibinfo{year}{2021}\natexlab{}.
\newblock \showarticletitle{Interactive Path Tracing and Reconstruction of
  Sparse Volumes}.
\newblock \bibinfo{journal}{\emph{Proceedings of the ACM on Computer Graphics
  and Interactive Techniques}} \bibinfo{volume}{4}, \bibinfo{number}{1}
  (\bibinfo{date}{April} \bibinfo{year}{2021}), \bibinfo{pages}{5:1--19}.
\newblock
\urldef\tempurl%
\url{https://doi.org/10.1145/3451256}
\showDOI{\tempurl}


\bibitem[Kallweit et~al\mbox{.}(2022)]%
        {Kallweit2022}
\bibfield{author}{\bibinfo{person}{Simon Kallweit}, \bibinfo{person}{Petrik
  Clarberg}, \bibinfo{person}{Craig Kolb}, \bibinfo{person}{Tom{\'a}{\v s}
  Davidovi{\v c}}, \bibinfo{person}{Kai-Hwa Yao}, \bibinfo{person}{Theresa
  Foley}, \bibinfo{person}{Yong He}, \bibinfo{person}{Lifan Wu},
  \bibinfo{person}{Lucy Chen}, \bibinfo{person}{Tomas Akenine{-}M{\"{o}}ller},
  \bibinfo{person}{Chris Wyman}, \bibinfo{person}{Cyril Crassin}, {and}
  \bibinfo{person}{Nir Benty}.} \bibinfo{year}{2022}\natexlab{}.
\newblock \bibinfo{title}{The Falcor Rendering Framework}.
\newblock \bibinfo{howpublished}{BSD-Licensed Github Repository}.
\newblock


\bibitem[Karis(2014)]%
        {Karis:2014:High}
\bibfield{author}{\bibinfo{person}{Brian Karis}.}
  \bibinfo{year}{2014}\natexlab{}.
\newblock \showarticletitle{High-quality Temporal Supersampling}.
\newblock \bibinfo{journal}{\emph{Advances in Real-Time Rendering in Games,
  SIGGRAPH Courses}} \bibinfo{volume}{1}, \bibinfo{number}{10.1145}
  (\bibinfo{year}{2014}), \bibinfo{pages}{2614028--2615455}.
\newblock


\bibitem[Kim et~al\mbox{.}(2024)]%
        {Kim2024neuralvdb}
\bibfield{author}{\bibinfo{person}{Doyub Kim}, \bibinfo{person}{Minjae Lee},
  {and} \bibinfo{person}{Ken Museth}.} \bibinfo{year}{2024}\natexlab{}.
\newblock \showarticletitle{NeuralVDB: High-resolution Sparse Volume
  Representation using Hierarchical Neural Networks}.
\newblock \bibinfo{journal}{\emph{ACM Transactions on Graphics}}
  \bibinfo{volume}{43}, \bibinfo{number}{2} (\bibinfo{year}{2024}),
  \bibinfo{pages}{20:1--21}.
\newblock
\showISSN{0730-0301}
\urldef\tempurl%
\url{https://doi.org/10.1145/3641817}
\showDOI{\tempurl}


\bibitem[Liu(2022)]%
        {Liu:2022:DLSS}
\bibfield{author}{\bibinfo{person}{Edward Liu}.}
  \bibinfo{year}{2022}\natexlab{}.
\newblock \showarticletitle{{DLSS 2.0 - Image Reconstruction for Real-Time
  Rendering with Deep learning}}. In \bibinfo{booktitle}{\emph{Game Developers
  Conference}}.
\newblock


\bibitem[McGuire et~al\mbox{.}(2012)]%
        {mcguire:2012:scalable}
\bibfield{author}{\bibinfo{person}{Morgan McGuire}, \bibinfo{person}{Michael
  Mara}, {and} \bibinfo{person}{David Luebke}.}
  \bibinfo{year}{2012}\natexlab{}.
\newblock \showarticletitle{Scalable Ambient Obscurance}. In
  \bibinfo{booktitle}{\emph{High Performance Graphics}}.
  \bibinfo{pages}{97--103}.
\newblock


\bibitem[Microsoft(2021)]%
        {Microsoft2021}
\bibfield{author}{\bibinfo{person}{Microsoft}.}
  \bibinfo{year}{2021}\natexlab{}.
\newblock \bibinfo{title}{HLSL Shader Model 6.0}.
\newblock
  \bibinfo{howpublished}{\url{https://learn.microsoft.com/en-us/windows/win32/direct3dhlsl/hlsl-shader-model-6-0-features-for-direct3d-12}}.
\newblock
\newblock
\shownote{[Online; accessed 2024-09-11]}.


\bibitem[Misso et~al\mbox{.}(2022)]%
        {Misso:2022:Unbiased}
\bibfield{author}{\bibinfo{person}{Zackary Misso}, \bibinfo{person}{Benedikt
  Bitterli}, \bibinfo{person}{Iliyan Georgiev}, {and} \bibinfo{person}{Wojciech
  Jarosz}.} \bibinfo{year}{2022}\natexlab{}.
\newblock \showarticletitle{Unbiased and Consistent Rendering using Biased
  Estimators}.
\newblock \bibinfo{journal}{\emph{ACM Transactions on Graphics (Proceedings of
  SIGGRAPH)}} \bibinfo{volume}{41}, \bibinfo{number}{4} (\bibinfo{date}{July}
  \bibinfo{year}{2022}).
\newblock
\urldef\tempurl%
\url{https://doi.org/10/gqjn66}
\showDOI{\tempurl}


\bibitem[Owen(2013)]%
        {Owen2013mcbook}
\bibfield{author}{\bibinfo{person}{Art~B. Owen}.}
  \bibinfo{year}{2013}\natexlab{}.
\newblock \bibinfo{booktitle}{\emph{Monte Carlo Theory, Methods and Examples}}.
\newblock \bibinfo{publisher}{\url{https://artowen.su.domains/mc/}}.
\newblock


\bibitem[Penner(2011)]%
        {penner:2011:shader}
\bibfield{author}{\bibinfo{person}{Eric Penner}.}
  \bibinfo{year}{2011}\natexlab{}.
\newblock \showarticletitle{Shader Amortization Using Pixel Quad Message
  Passing}.
\newblock In \bibinfo{booktitle}{\emph{GPU Pro 2}}. \bibinfo{publisher}{CRC
  Press}, \bibinfo{pages}{349--366}.
\newblock


\bibitem[Pharr et~al\mbox{.}(2023)]%
        {Pharr2023}
\bibfield{author}{\bibinfo{person}{Matt Pharr}, \bibinfo{person}{Wenzel Jakob},
  {and} \bibinfo{person}{Greg Humphreys}.} \bibinfo{year}{2023}\natexlab{}.
\newblock \bibinfo{booktitle}{\emph{{Physically Based Rendering: From Theory to
  Implementation}} (\bibinfo{edition}{4th} ed.)}.
\newblock \bibinfo{publisher}{The MIT Press}.
\newblock


\bibitem[Pharr et~al\mbox{.}(2024)]%
        {Pharr2024}
\bibfield{author}{\bibinfo{person}{Matt Pharr}, \bibinfo{person}{Bartlomiej
  Wronski}, \bibinfo{person}{Marco Salvi}, {and} \bibinfo{person}{Marcos
  Fajardo}.} \bibinfo{year}{2024}\natexlab{}.
\newblock \showarticletitle{{Filtering After Shading with Stochastic Texture
  Filtering}}.
\newblock \bibinfo{journal}{\emph{Proceedings of the ACM on Computer Graphics
  and Interactive Techniques}} \bibinfo{volume}{7}, \bibinfo{number}{1}
  (\bibinfo{year}{2024}), \bibinfo{pages}{14:1--20}.
\newblock


\bibitem[Powell and Swann(1966)]%
        {Powell1966weighted}
\bibfield{author}{\bibinfo{person}{Michael~JD. Powell} {and}
  \bibinfo{person}{J. Swann}.} \bibinfo{year}{1966}\natexlab{}.
\newblock \showarticletitle{Weighted Uniform Sampling—a Monte Carlo Technique
  for Reducing Variance}.
\newblock \bibinfo{journal}{\emph{IMA Journal of Applied Mathematics}}
  \bibinfo{volume}{2}, \bibinfo{number}{3} (\bibinfo{year}{1966}),
  \bibinfo{pages}{228--236}.
\newblock


\bibitem[Reinhard et~al\mbox{.}(2001)]%
        {Reinhard2001natural}
\bibfield{author}{\bibinfo{person}{Erik Reinhard}, \bibinfo{person}{Peter
  Shirley}, {and} \bibinfo{person}{Tom Troscianko}.}
  \bibinfo{year}{2001}\natexlab{}.
\newblock \showarticletitle{Natural Image Statistics for Computer Graphics}.
\newblock \bibinfo{journal}{\emph{Univ. Utah Tech Report UUCS-01-002}}
  (\bibinfo{date}{March} \bibinfo{year}{2001}).
\newblock


\bibitem[Smith et~al\mbox{.}(1962)]%
        {Smith:1962:Application}
\bibfield{author}{\bibinfo{person}{Gerald~L Smith}, \bibinfo{person}{Stanley~F
  Schmidt}, {and} \bibinfo{person}{Leonard~A McGee}.}
  \bibinfo{year}{1962}\natexlab{}.
\newblock \bibinfo{booktitle}{\emph{Application of Statistical Filter Theory to
  the Optimal Estimation of Position and Velocity on Board a Circumlunar
  Vehicle}}. Vol.~\bibinfo{volume}{135}.
\newblock \bibinfo{publisher}{National Aeronautics and Space Administration}.
\newblock


\bibitem[Spanier(1979)]%
        {Spanier1979new}
\bibfield{author}{\bibinfo{person}{Jerome Spanier}.}
  \bibinfo{year}{1979}\natexlab{}.
\newblock \showarticletitle{A New Family of Estimators for Random Walk
  Problems}.
\newblock \bibinfo{journal}{\emph{IMA Journal of Applied Mathematics}}
  \bibinfo{volume}{23}, \bibinfo{number}{1} (\bibinfo{year}{1979}),
  \bibinfo{pages}{1--31}.
\newblock


\bibitem[Szirmay-Kalos and Szecsi(2003)]%
        {SzirmayKalos2023wisphoton}
\bibfield{author}{\bibinfo{person}{Laszlo Szirmay-Kalos} {and}
  \bibinfo{person}{Laszl Szecsi}.} \bibinfo{year}{2003}\natexlab{}.
\newblock \showarticletitle{{Improved Indirect Photon Mapping with Weighted
  Importance Sampling}}. In \bibinfo{booktitle}{\emph{Eurographics 2003 - Short
  Presentations}}.
\newblock
\showISSN{1017-4656}
\urldef\tempurl%
\url{https://doi.org//10.2312/egs.20031068}
\showDOI{\tempurl}


\bibitem[Talbot et~al\mbox{.}(2005)]%
        {talbot2005resampling}
\bibfield{author}{\bibinfo{person}{Justin Talbot}, \bibinfo{person}{David
  Cline}, {and} \bibinfo{person}{Parris Egbert}.}
  \bibinfo{year}{2005}\natexlab{}.
\newblock \showarticletitle{{Importance Resampling for Global Illumination}}.
  In \bibinfo{booktitle}{\emph{Eurographics Symposium on Rendering}},
  \bibfield{editor}{\bibinfo{person}{Kavita Bala} {and} \bibinfo{person}{Philip
  Dutre}} (Eds.).
\newblock
\showISBNx{3-905673-23-1}
\showISSN{1727-3463}
\urldef\tempurl%
\url{https://doi.org//10.2312/EGWR/EGSR05/139-146}
\showDOI{\tempurl}


\bibitem[Vaidyanathan et~al\mbox{.}(2023)]%
        {Vaidyanathan2023}
\bibfield{author}{\bibinfo{person}{Karthik Vaidyanathan},
  \bibinfo{person}{Marco Salvi}, \bibinfo{person}{Bartlomiej Wronski},
  \bibinfo{person}{Tomas Akenine{-}M{\"{o}}ller}, \bibinfo{person}{Pontus
  Ebelin}, {and} \bibinfo{person}{Aaron Lefohn}.}
  \bibinfo{year}{2023}\natexlab{}.
\newblock \showarticletitle{{Random-Access Neural Compression of Material
  Textures}}.
\newblock \bibinfo{journal}{\emph{ACM Transactions on Graphics}}
  \bibinfo{volume}{42}, \bibinfo{number}{4} (\bibinfo{year}{2023}),
  \bibinfo{pages}{88:1--25}.
\newblock
\urldef\tempurl%
\url{https://doi.org/10/gsk4fz}
\showDOI{\tempurl}


\bibitem[Veach and Guibas(1995)]%
        {Veach1995MIS}
\bibfield{author}{\bibinfo{person}{Eric Veach} {and}
  \bibinfo{person}{Leonidas~J. Guibas}.} \bibinfo{year}{1995}\natexlab{}.
\newblock \showarticletitle{Optimally Combining Sampling Techniques for Monte
  Carlo Rendering}. In \bibinfo{booktitle}{\emph{Proceedings of SIGGRAPH}}.
  \bibinfo{pages}{419–428}.
\newblock
\showISBNx{0897917014}
\urldef\tempurl%
\url{https://doi.org/10.1145/218380.218498}
\showDOI{\tempurl}


\bibitem[Wolfe et~al\mbox{.}(2022)]%
        {Wolfe2022}
\bibfield{author}{\bibinfo{person}{Alan Wolfe}, \bibinfo{person}{Nathan
  Morrical}, \bibinfo{person}{Tomas Akenine-M\"oller}, {and}
  \bibinfo{person}{Ravi Ramamoorthi}.} \bibinfo{year}{2022}\natexlab{}.
\newblock \showarticletitle{{Spatiotemporal Blue Noise Masks}}. In
  \bibinfo{booktitle}{\emph{Eurographics Symposium on Rendering}}.
  \bibinfo{pages}{117--126}.
\newblock


\bibitem[Yang et~al\mbox{.}(2020)]%
        {Yang:2020:Survey}
\bibfield{author}{\bibinfo{person}{Lei Yang}, \bibinfo{person}{Shiqiu Liu},
  {and} \bibinfo{person}{Marco Salvi}.} \bibinfo{year}{2020}\natexlab{}.
\newblock \showarticletitle{{A Survey of Temporal Antialiasing Techniques}}.
\newblock \bibinfo{journal}{\emph{Computer Graphics Forum}}
  \bibinfo{volume}{39}, \bibinfo{number}{2} (\bibinfo{year}{2020}),
  \bibinfo{pages}{607--621}.
\newblock


\end{thebibliography}



\begin{thebibliography}{4}


\ifx \showCODEN    \undefined \def \showCODEN     #1{\unskip}     \fi
\ifx \showDOI      \undefined \def \showDOI       #1{#1}\fi
\ifx \showISBNx    \undefined \def \showISBNx     #1{\unskip}     \fi
\ifx \showISBNxiii \undefined \def \showISBNxiii  #1{\unskip}     \fi
\ifx \showISSN     \undefined \def \showISSN      #1{\unskip}     \fi
\ifx \showLCCN     \undefined \def \showLCCN      #1{\unskip}     \fi
\ifx \shownote     \undefined \def \shownote      #1{#1}          \fi
\ifx \showarticletitle \undefined \def \showarticletitle #1{#1}   \fi
\ifx \showURL      \undefined \def \showURL       {\relax}        \fi
\providecommand\bibfield[2]{#2}
\providecommand\bibinfo[2]{#2}
\providecommand\natexlab[1]{#1}
\providecommand\showeprint[2][]{arXiv:#2}

\bibitem[Anderson and Moore(1979)]%
        {Anderson:1979:Optimal}
\bibfield{author}{\bibinfo{person}{Brian D.~O. Anderson} {and}
  \bibinfo{person}{John~B. Moore}.} \bibinfo{year}{1979}\natexlab{}.
\newblock \bibinfo{booktitle}{\emph{Optimal Filtering}}.
\newblock \bibinfo{publisher}{Prentice-Hall}.
\newblock


\bibitem[Misso et~al\mbox{.}(2022)]%
        {Misso:2022:Unbiased}
\bibfield{author}{\bibinfo{person}{Zackary Misso}, \bibinfo{person}{Benedikt
  Bitterli}, \bibinfo{person}{Iliyan Georgiev}, {and} \bibinfo{person}{Wojciech
  Jarosz}.} \bibinfo{year}{2022}\natexlab{}.
\newblock \showarticletitle{Unbiased and Consistent Rendering using Biased
  Estimators}.
\newblock \bibinfo{journal}{\emph{ACM Transactions on Graphics (Proceedings of
  SIGGRAPH)}} \bibinfo{volume}{41}, \bibinfo{number}{4} (\bibinfo{date}{July}
  \bibinfo{year}{2022}).
\newblock
\urldef\tempurl%
\url{https://doi.org/10/gqjn66}
\showDOI{\tempurl}


\bibitem[Pharr et~al\mbox{.}(2024)]%
        {Pharr2024}
\bibfield{author}{\bibinfo{person}{Matt Pharr}, \bibinfo{person}{Bartlomiej
  Wronski}, \bibinfo{person}{Marco Salvi}, {and} \bibinfo{person}{Marcos
  Fajardo}.} \bibinfo{year}{2024}\natexlab{}.
\newblock \showarticletitle{{Filtering After Shading with Stochastic Texture
  Filtering}}.
\newblock \bibinfo{journal}{\emph{Proceedings of the ACM on Computer Graphics
  and Interactive Techniques}} \bibinfo{volume}{7}, \bibinfo{number}{1}
  (\bibinfo{year}{2024}), \bibinfo{pages}{14:1--20}.
\newblock


\bibitem[Smith et~al\mbox{.}(1962)]%
        {Smith:1962:Application}
\bibfield{author}{\bibinfo{person}{Gerald~L Smith}, \bibinfo{person}{Stanley~F
  Schmidt}, {and} \bibinfo{person}{Leonard~A McGee}.}
  \bibinfo{year}{1962}\natexlab{}.
\newblock \bibinfo{booktitle}{\emph{Application of Statistical Filter Theory to
  the Optimal Estimation of Position and Velocity on Board a Circumlunar
  Vehicle}}. Vol.~\bibinfo{volume}{135}.
\newblock \bibinfo{publisher}{National Aeronautics and Space Administration}.
\newblock


\end{thebibliography}

\SupplementaryMaterials

\vfill\eject

\pagebreak
\begin{center}
\textbf{\large Supplemental Material for Improved Stochastic Texture Filtering Through Sample Reuse}
\end{center}

This supplemental material includes further discussion of Monte Carlo estimates with stochastic texture filtering (STF),
a variance analysis of filtering after shading, further examples of texel sharing footprints
as well as a description of our optimization algorithm for generating sparse footprints,
an example implementation of the {\small\textcolor{violet}{\texttt{GetFilterPMF}}} function used in the example implementation
in Section~\ref{sec_method_details},
and some results and discussion about the use of our sample sharing techniques with volumetric ray marching.

\section{STF and Monte Carlo Integral Estimates}
\label{app:stf-mc-connection}

Pharr et al.'s paper on stochastic texture filtering did not make a direct connection between
integral Monte Carlo estimators and the stochastic texture filtering algorithms introduced there
but rather derived STF algorithms by framing them as stochastic evaluation of sums of weighted
texel values~\cite{Pharr2024}. Because our weighted STF estimator, Equation~\ref{eq_bart_weighted_estimator},
is an integral estimator, here we make the straightforward connection between integral estimators and stochastic
texture filtering for completeness.

With traditional STF, we are applying Monte Carlo integration to the integral texture filtering
equation, \ref{eq:texture-continuous-integral}:
\begin{align}
  t_\mathrm{r}(u,v) &= \int \! \! \! \! \int \left( \sum_i^n \, \delta(u'-u_i) \delta(v'-v_i) \, \mathbf{T}_{u_i,v_i} \right) \,
     f_\mathrm{r}(u-u', v-v') \, \mathrm{d}u' \, \mathrm{d}v' \\
     &= \int \! \! \! \! \int \sum_i^n \delta(u'-u_i) \delta(v'-v_i) \, \mathbf{T}_{u_i,v_i} \, f_\mathrm{r}(u_i-u', v_i-v') \, \mathrm{d}u' \, \mathrm{d}v'.
\end{align}
We will define the PDF
\begin{equation}
p(u, v) = \sum_i^n \delta(u-u_i) \delta(v-v_i) \, f_\mathrm{r}(u_i-u, v_i-v).
\end{equation}
Under the assumption that the texture reconstruction filter $f_\mathrm{r}$ is normalized, it is easy to see that this is a valid PDF.

Samples from the PDF can be taken by selecting a term $i$ of the sum with probability proportional to $f_\mathrm{r}(u_i-u,v_i-v)$.
In turn, we have a sample point $(u_i,v_i)$.

If we apply the importance sampling Monte Carlo estimator, we have
\begin{equation}
t_\mathrm{r}(u,v) \approx \frac{\cancel{\delta(u-u_i)} \cancel{\delta(v-v_i)} \, \mathbf{T}_{u_i,v_i} \, \cancel{f_\mathrm{r}(u_i-u, v_i-v)}}%
  {\cancel{\delta(u-u_i)} \cancel{\delta(v-v_i)} \, \cancel{f_\mathrm{r}(u_i-u, v_i-v)}} =
\mathbf{T}_{u_i,v_i},
\end{equation}
giving the one-tap STF estimator.

\section{Variance and Bias with Filtering After Shading}\label{sec:app-nonlinear-taylor}
For cases where filtering before shading is the desired result, it is useful to be able to characterize the error
from STF and filtering after shading in order to evaluate and design STF estimators.
This is challenging in general, as a wide variety of nonlinearities are present in shading functions.
We therefore propose a simple approach based on statistical analysis of nonlinear
transformations of random variables.
For a more complete treatment of the statistical analysis of error resulting from nonlinearities applied to random variables,
we refer the reader to the statistical and control theory literature and extended Kalman
filters~\cite{Anderson:1979:Optimal,Smith:1962:Application} as well as recent advances in unbiasing rendering algorithms
using telescoping Taylor series~\cite{Misso:2022:Unbiased}.

Consider a shading function $f$ where the true filtered texture value is $\mu$:
with filtering before shading, we filter the texture using Equation~\ref{eq:texture-continuous-sum} to compute $\mu$ and then return $f(\mu)$.
With filtering after shading and one-tap STF~\cite{Pharr2024}, the texture is represented by a random variable $X$ corresponding to a single texel
that is sampled according to the texture filter, $X \sim f_\mathrm{r}$.
The estimator is $f(X)$.
We would like to understand how the expected value of filtering after shading, $\mathbb{E}\left[f(X)\right]$, relates to $f(\mu)$.

To approximate this error, we can use the Taylor expansion of $f$ around $\mu$.
For example, consider the second-order expansion:\footnote{In practice, many functions used in rendering, such as specular shading,
	are nonlinear and have slowly decaying higher-order derivatives, so a second order expansion is insufficient.
	However, we can apply this expansion to multi-variate functions and include
	higher-order derivatives and thus, higher-order statistical moments.}
\begin{align}
	\mathbb{E}\left[f(X)\right] & {} = \mathbb{E}\left[f\left(\mu + \left(X - \mu\right)\right)\right] \\
	& {} \approx \mathbb{E}\left[f(\mu) + f'(\mu)\left(X-\mu\right) + \frac{1}{2}f''(\mu) \left(X - \mu\right)^2 \right] \\
	& {} = f(\mu) + f'(\mu) \, \mathbb{E} \left[ X-\mu \right] + \frac{1}{2}f''(\mu) \, \mathbb{E} \left[ \left(X - \mu\right)^2 \right].
\end{align}
Because one-tap STF gives an unbiased estimate of $\mu$, $\mathbb{E}\left[X - \mu\right] = 0$ and $\mathbb{E} \left[ \left(X - \mu\right)^2 \right]$ is $X$'s variance,
which we will denote by $\sigma_X$. Dropping $f(\mu)$, the result of filtering before shading, we are left with
\begin{equation}
	\frac{f''(\mu)}{2}\sigma_X^2
\end{equation}
as the error due to filtering after shading.
In other words, the error depends on the second derivative of the shading function and the squared variance of $X$ (and the higher-order
terms we have neglected); we see how the variance of $X$ directly contributes to the final error and the resulting bias.

This analysis fits with our earlier error analysis in Section~\ref{sec:stferror}:
when the variation in the filtered texel values is small, STF yields a small error and filtering after shading gives results that are close to filtering before shading.
With constant signals with zero variance, the error is zero and the two approaches are equivalent.

Unlike one-tap STF, our method uses an estimator that has a small bias (Section~\ref{sec:estimators}).
Thus, $\mathbb{E}\left[X - \mu\right] \ne 0$ and 
$\mathbb{E} \left[ \left(X - \mu\right)^2 \right]$ includes both bias and variance.
The error is approximated as:
\begin{equation}
  f'(\mu) \, \mathbb{E} \left[ X-\mu \right] + \frac{1}{2}f''(\mu) \, \mathbb{E} \left[ \left(X - \mu\right)^2 \right].
\end{equation}
The error includes terms that depend on both the first and second derivatives of the shading function $f$,
though in practice the overall error is lower than with one-tap STF since $X$ is closer to $\mu$ with our approach.

From these results, we can see why the requirement of not introducing any variation in regions of constant texture values is so important---even
a small amount of error may introduce a large error in the shaded result.
Furthermore, estimators like standard importance sampling that may have unbounded weights (recall Section~\ref{sec_existing_estimators})
result in not only much higher variance, but also high further statistical error moments.

While Taylor expansion formally does not apply to every function used in rendering (for example, a step comparison operator used in shadow mapping is not differentiable),
this analysis still provides an insight and intuition for evaluating
different estimators: that the better the variance reduction of an estimator
(or, generalizing to higher-order moments, smaller amounts of noise and tighter distributions),
the closer the result is to filtering before shading.

\section{Illustration of Larger Footprints}

\begin{figure}[t]
	\centering
	\includegraphics[width=0.7\columnwidth]{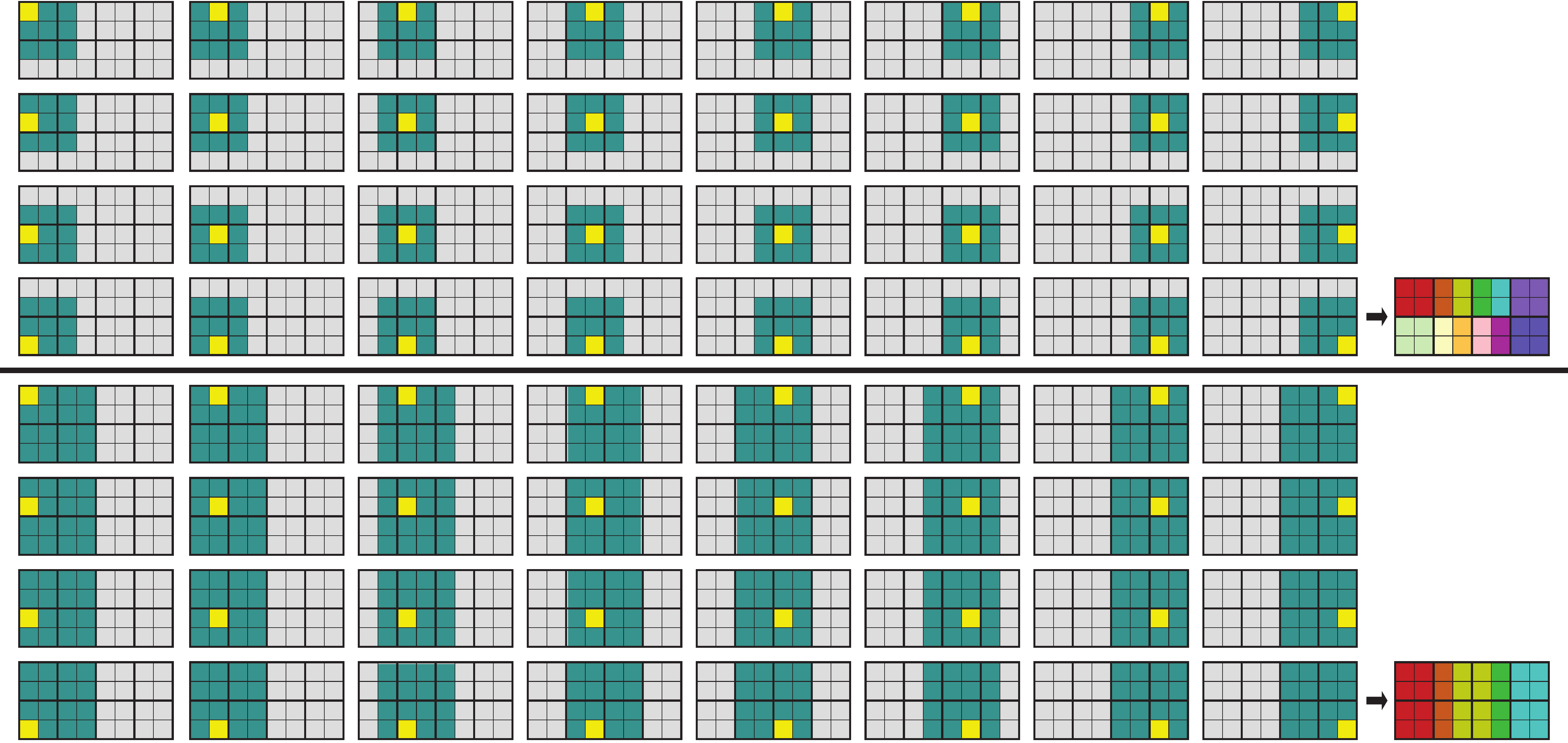}
	\caption{A set of possible $3\times 3$ footprints (top) and 
	  $4\times 4$ footprints (bottom). As in Figure~\ref{fig_footprint2x2},
          for each footprint, yellow signifies the lane in the wave that the footprint
          is associated with and green signifies the other lanes that it draws texel
          values from. The colored illustrations on the right shows which groups
          of lanes all end up with the same texel values to filter after sharing.
	}
	\label{fig_footprint3x3_4x4}
\end{figure}

Figure~\ref{fig_footprint3x3_4x4} illustrates the footprint
for deterministic $3\times 3$ and $4 \times 4$ square footprints in a 32-lane wave. 
Note that as the footprints become larger, more lanes in the wave all filter
the same set of texel values; these texel sharing sets are illustrated in the lower right.
However, since the filters are larger, we find that the error is also lower in general,
as discussed in Section~\ref{sec_results_footprints}.

\section{Wave Intrinsics in Other Shader Stages}
While the specific wave mapping for other shading stages, such as ray tracing shaders, is undefined,
we have successfully tested our method with general wave intrinsics in those stages.
We have verified that our method works with shadow rays that sample alpha masks from textures and is able to
achieve some visual quality improvements over the original STF technique.
However, we note that lanes can map to arbitrary and distant rays and thus the
resulting filtering quality is not guaranteed.
We also advise additional caution and checking whether other lanes are active, as behavior of 
\texttt{WaveReadLaneAt(value, laneId)} is undefined for inactive lanes.

\section{Pseudorandom Sparse Footprint Generation Algorithm}\label{sec:app-stochastic-footprint}

\begin{figure}[tb]
	\centering
	\includegraphics[width=0.9\columnwidth]{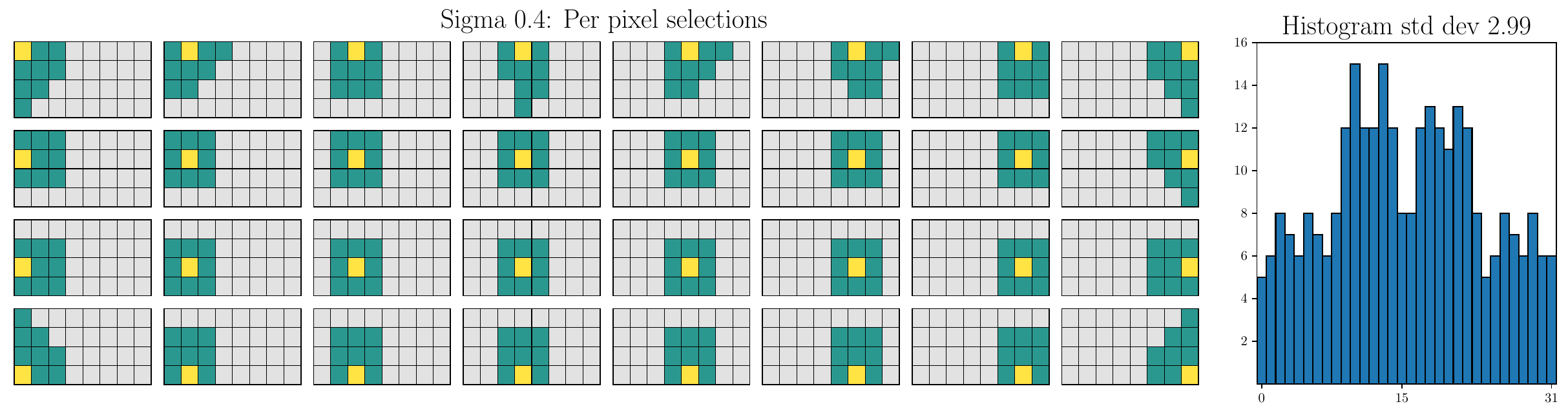}\\[1ex]
    \includegraphics[width=0.9\columnwidth]{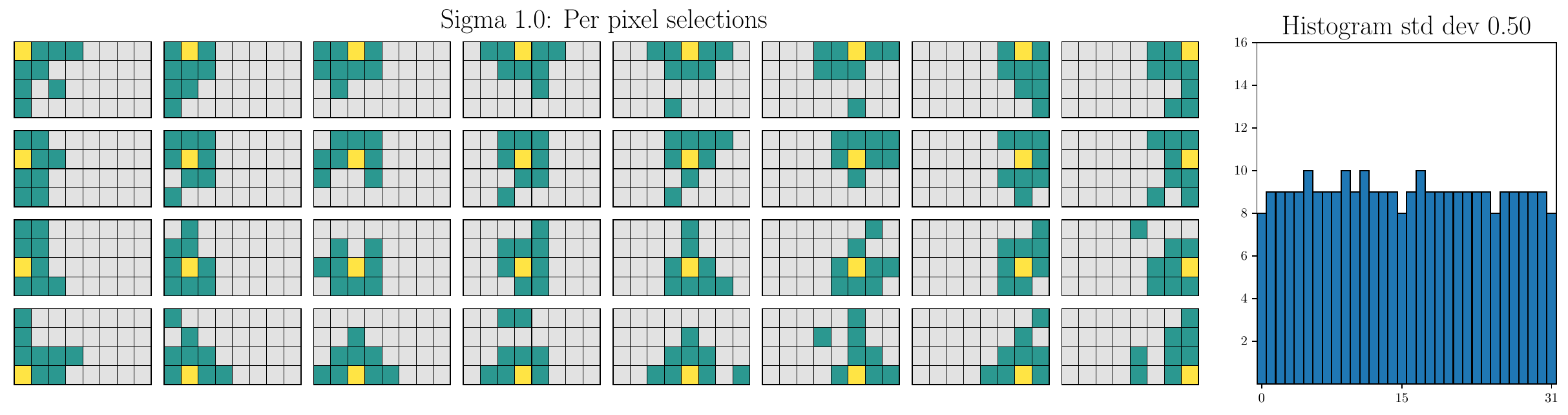}\\[1ex]
    \includegraphics[width=0.9\columnwidth]{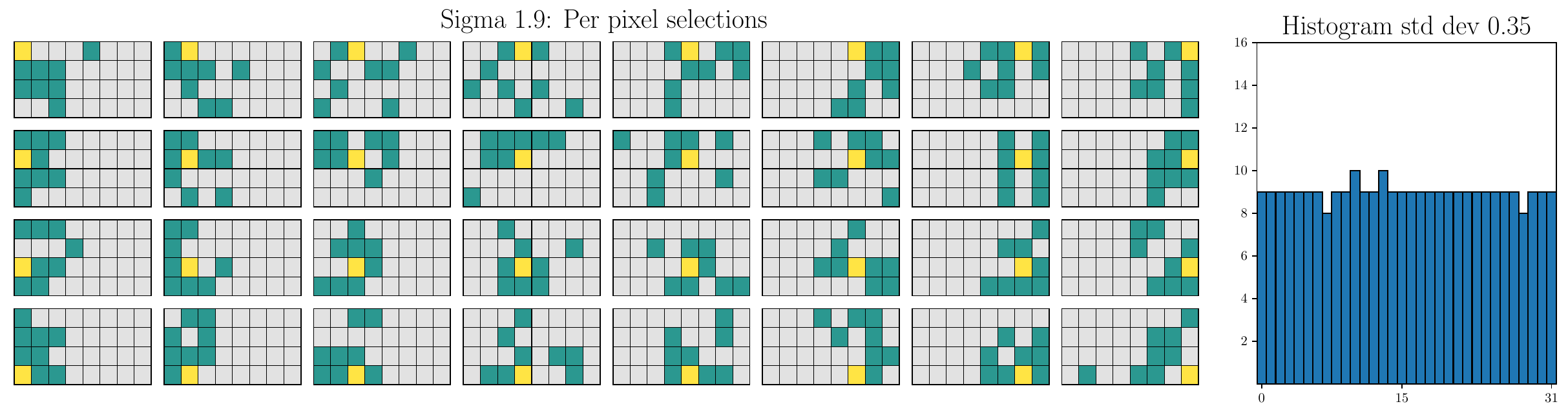}
    \caption{Three sets of sample sharing footprints, with increasing values of the $\sigma$ parameter.
      \textbf{Top}: with a small $\sigma$, most of the patterns are the same as the deterministic $3 \times 3$ footprint;
      as shown by the histogram on the right, some lanes (the ones in the center of the wave) are used for sharing much more
      than others (the ones along the edges and the corners.)
      \textbf{Middle}: increasing the $\sigma$ gives more irregular patterns and a more uniform histogram.
      \textbf{Bottom}: a further increase of $\sigma$ gives a histogram with slightly lower standard deviation but
      with irregular patterns that may sample far-away lanes.  In general, the farther away the lanes used
      for sharing are, the less effective reuse will be, since shared texels may be outside of the current lane's texture filter footprint.
	}
	\label{fig_sparse_different_sigma}
\end{figure}

To generate the sparse wave sharing footprints introduced in Section~\ref{sec_stochastic_footprints},
we use a simple three-stage global optimization algorithm.

In the first stage, for each lane in a wave, we generate a set of 16--32 candidate configurations.
Candidate elements to share are generated by sampling from a normal distribution with a fixed standard deviation $\sigma$.
Close to the wave borders and corners, we relax the $\sigma$ to allow considering farther-away candidates.
We continue taking samples, discarding repeated lanes, until the number of candidates equals
the desired wave sharing element count.
Increasing values of $\sigma$ reduce the footprints' locality but make it easier to have each lane be used for sharing the same number of times
and have irregular shapes that do not lead to visible structure in images (Figure~\ref{fig_structured_patterns}).
Figure~\ref{fig_sparse_different_sigma} shows results for three different $\sigma$ values;
the bottom part of Figure~\ref{fig_stochastic_footprints} was generated with a $\sigma$ of 1.4.

Having generated candidates for each pixel, we proceed to the second stage of the algorithm.
We randomly select a candidate for each pixel, count how many times each wave lane has been selected for sharing,
and then find the standard deviation of these counts.
We use the standard deviation as a score, where lower standard deviations are better, corresponding to
greater uniformity in how often each lane is used.
We repeat this process 10,000 times and select the configuration with the best score.

In the third stage, we proceed with a variation of the \textit{coordinate descent} algorithm, where we attempt to improve
the configuration selected after the second stage.
We exhaustively iterate through all the lanes and check if using any of the other candidates from the first stage for the lane
would improve the configuration's overall score.
We continue this process until no better candidate is found for any of the lanes.

Finally, because this approach does not guarantee convergence to a global minimum, we repeat it from scratch 30 times and retain the
pattern with the best score.

Although our optimization algorithm is brute-force,
the search space of waves of 32--64 elements is relatively small and our algorithm still runs quickly.
Our naive Python implementation running on a single CPU core 
can generate a complete set of sharing footprints in less than 40 seconds of CPU time.
This has allowed for quick iterations and experiments as well as generating different patterns for different frames to break up temporal artifacts.
For the results in the paper, we run ten times as many iterations of the first two stages, with a corresponding increase in pattern generation time
by a factor of $\sim\!\!10$.
We find this time to be acceptable for a preprocess;
if higher performance was necessary, the iterations of the first two stages could run in parallel and a higher-performance language like C++
could be used for the implementation.
We include our implementation in the supplementary material.

\section{Filter PMF Implementation}
\label{app:filter-pdf}

Each time through the loop in the code listing in Section~\ref{sec_method_details}, we consider
a texel sampled by one of the lanes in the sharing footprint for the current lane.
To evaluate its contribution to the weighted importance sampling estimator, Equation~\ref{eq_bart_weighted_estimator},
we need to compute the probability that the current lane would have sampled that texel;
this is handled by the call to {\textcolor{violet}{\texttt{GetFilterPMF}}}.

Below is an example implementation of this function for a bilinear filter.
Because the filter is normalized, the PMF for a given texel is simply its bilinear filter weight,
which is easily computed in a few lines of code. We note that depending on the rendering technique used,
this function might need to aditionally verify if other lanes sample from the same texture set
and return a zero PMF on any mismatch.

\lstdefinestyle{mystyle}
{
	language = C,
	keywordstyle = [1]{\color{magenta}},
	keywordstyle = [2]{\color{violet}},
	morekeywords = [1]{float4, uint, int2},
	morekeywords = [2]{WaveReadLaneAt, GetFilterPMF},
	basicstyle=\scriptsize\ttfamily, %
	frame=single,
}
\begin{lstlisting}[language=C,  style=mystyle]
float GetFilterPMF(float2 texelFloatCoords, int2 texelIntCoords)
{
    float2 texelDistance = texelFloatCoords - texelIntCoords;
    float2 filterPdf = clamp(1 - abs(texelDistance), 0, 1);
    return filterPdf.x * filterPdf.y;
}
\end{lstlisting}

\section{Tricubic Reconstruction for Volumetric Rendering}
\label{app:tricubic-volumetric}

\begin{figure}[t]
	\centering
	\newcommand{\myfactor}{0.3}
	\begin{tabular}{ccc}
		\centering\raisebox{-14.5pt}[0pt][0pt]{\includegraphics[width=0.3525\textwidth]{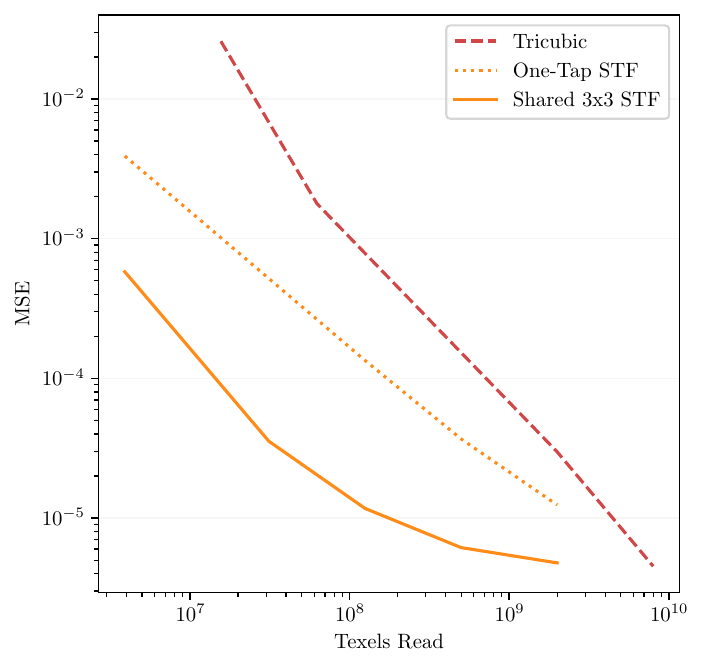}} &
		\centering\includegraphics[width=0.275\textwidth]{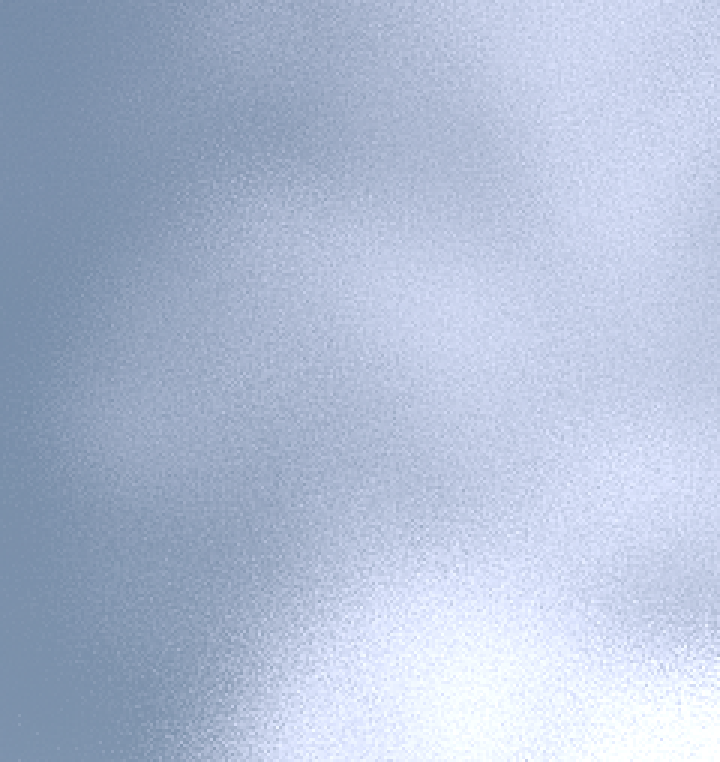} &
		\centering\includegraphics[width=0.275\textwidth]{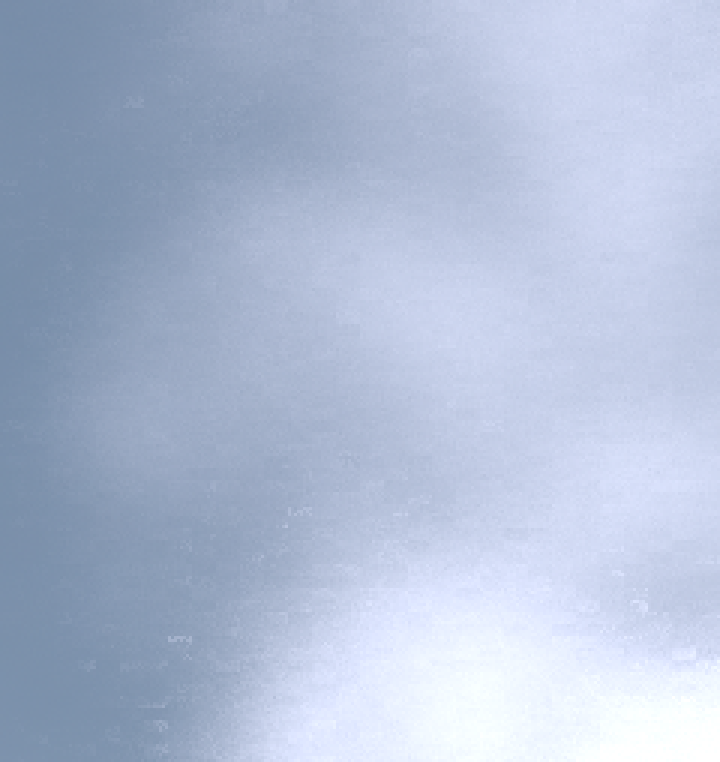}
        \end{tabular}
	\caption{Sample reuse for volumetric ray marching of a cloud data set.
          \textbf{Left}: Plot comparing mean squared error for full tricubic filtering, one-tap STF, and STF with
          $3 \times 3$ sparse sharing footprints. Sharing significantly reduces MSE, roughly in proportion
          to the number of successfully shared texel lookups.
          \textbf{Middle}: with 64 ray marching steps and one-tap STF, error manifests as high-frequency noise.
          \textbf{Right}: with 64 steps with texel sharing, numeric error is much lower and noise is reduced, but block artifacts appear.
          }
	\label{fig_cloud_error}
\end{figure}

We have also investigated the effect of sample sharing for STF with ray marched volumetric rendering;
our results are summarized in Figure~\ref{fig_cloud_error}.
As shown in the plot, texel sharing significantly reduces the numeric error compared to one-tap STF for a given number of texel lookups.
We also note that one-tap STF has significantly lower error than full tricubic filtering given an equal number of samples;
we attribute this to the benefit of importance sampling---full tricubic filtering accesses all the 64 texels under the filter,
many of which make a relatively small contribution to the final result.

However, as shown by the images, texel sharing leads to block-structured artifacts in the image with sharing and 64 ray marching steps.
We believe that the high-frequency noise from one-tap STF is likely to be preferable in this case
as it would be easier to remove with post-rendering filtering like TAA or DLSS.
(These artifacts do disappear at higher sampling rates, i.e., more steps along each ray.)
We attribute these artifacts to the extent of the tricubic filter: with $3 \times 3$ sharing in a wave, even if all neighboring pixels
provide unique useful texels, less than $15\%$ of the texels under the filter will be available.
Further, nearby pixels will generally filter similar incomplete sets of texels, leading to correlation between pixels in each wave.

\afterdoc

\end{document}